\pdfoutput=1

\documentclass[11pt,twoside,a4paper,cmspaper,final,collab]{cms-tdr}
\def\svnVersion{482866P}\def\cmsCernNoTag{CERN-EP-2018-071}\def\cmsCernDate{\today}\def\cmsMessage{Published in the European Physical Journal C as \href{http://dx.doi.org/10.1140/epjc/s10052-018-6373-0}{\doi{10.1140/epjc/s10052-018-6373-0.}}}

\begin{document}\cmsNoteHeader{SMP-16-015}

\hyphenation{had-ron-i-za-tion}
\hyphenation{cal-or-i-me-ter}
\hyphenation{de-vices}
\RCS$HeadURL: svn+ssh://svn.cern.ch/reps/tdr2/papers/SMP-16-015/trunk/SMP-16-015.tex $
\RCS$Id: SMP-16-015.tex 473688 2018-09-03 09:07:02Z pgras $

\newlength\cmsFigWidth
\ifthenelse{\boolean{cms@external}}{\setlength\cmsFigWidth{0.85\columnwidth}}{\setlength\cmsFigWidth{0.4\textwidth}}
\ifthenelse{\boolean{cms@external}}{\providecommand{\cmsLeft}{upper\xspace}}{\providecommand{\cmsLeft}{left\xspace}}
\ifthenelse{\boolean{cms@external}}{\providecommand{\cmsRight}{lower\xspace}}{\providecommand{\cmsRight}{right\xspace}}
\cmsNoteHeader{SMP-16-015}

\newcommand{\MGaMC}{{\textsc{MG5}\_a\textsc{MC}}\xspace}
\newcommand{\Nj}{\ensuremath{N_{\text{jets}}}}
\newcommand{\citefigfour}{Other details are as mentioned in the Fig.~\ref{fig:sigNjet} caption}
\newcommand{\ptbal}{\ensuremath{\pt^{\text{bal}}}\xspace}
\newcommand{\JZB}{\ensuremath{\text{JZB}}\xspace}
\newcommand{\ppZnj}{\ensuremath{\Pp\Pp \to \cPZ + N \text{ jets}}\xspace}
\newcommand{\Njetti}{N$_{\text{jetti}}$\xspace}
\newcommand{\PYTHIAeight}{{\PYTHIA}8\xspace}
\newcommand{\GENEVA}{\textsc{geneva}\xspace}
\providecommand{\FloatBarrier}{}
\providecommand{\cmsTable}[1]{\resizebox{\textwidth}{!}{#1}}

\title{Measurement of differential cross sections for $\cPZ$ boson production in association with jets in proton-proton collisions at $\sqrt{s} = 13\TeV$}

\titlerunning{$\cPZ$ boson production in association with jets in proton-proton collisions at $\sqrt{s} = 13\TeV$}

\date{\today}

\abstract{The production of a $\cPZ$ boson, decaying to two charged leptons, in association with jets in proton-proton collisions at a centre-of-mass energy of 13\TeV is measured. Data recorded with the CMS detector at the LHC are used that correspond to an integrated luminosity of 2.19\fbinv. The cross section is measured as a function of the jet multiplicity and its dependence on the transverse momentum of the $\cPZ$ boson, the jet kinematic variables (transverse momentum and rapidity), the scalar sum of the jet momenta, which quantifies the hadronic activity, and the balance in transverse momentum between the reconstructed jet recoil and the $\cPZ$ boson. The measurements are compared with predictions from four different calculations. The first two merge matrix elements with different parton multiplicities in the final state and parton showering, one of which includes one-loop corrections. The third is a fixed-order calculation with next-to-next-to-leading order accuracy for the process with a $\cPZ$ boson and one parton in the final state. The fourth combines the fully differential next-to-next-to-leading order calculation of the process with no parton in the final state with next-to-next-to-leading logarithm resummation and parton showering.}

\hypersetup{%
pdfauthor={CMS Collaboration},%
pdftitle={Measurement of the differential cross section for Z boson production in association with jets in proton-proton collisions at sqrt s = 13 TeV},%
pdfsubject={CMS},%
pdfkeywords={CMS, physics, standard model, cross section, Z boson, jets, proton, LHC}}

\maketitle

\section{Introduction}
\label{introduction}

Measurements of vector boson production in association with jets provide fundamental tests of quantum chromodynamics (QCD). The high centre-of-mass energy at the CERN LHC allows the production of an electroweak boson along with a large number of jets with large transverse momenta. A precise knowledge of the kinematic distributions in processes with large jet multiplicity is essential to exploit the potential of the LHC experiments. Comparison of the measurements with predictions motivates additional Monte Carlo (MC) generator development and  improves our understanding of the prediction uncertainties. Furthermore, the production of a massive vector boson together with jets is an important background to a number of standard model (SM) processes  (production of a single top quark, $\ttbar$, and Higgs boson as well as vector boson fusion and WW scattering) as well as to searches for physics beyond the SM, \eg supersymmetry. Leptonic decay modes of the vector bosons are often used in the measurement of SM processes and searches for physics beyond the SM since they have a sufficiently high branching fraction and clean signatures that provide a strong rejection of backgrounds. Differential cross sections for the associated production of a $\cPZ$ boson with hadronic jets have been previously measured by the ATLAS, CMS, and LHCb Collaborations in proton-proton collisions at centre-of-mass energies of 7 \cite{Aad:2013ysa,Aad:2011qv,Chatrchyan:2011ne,Khachatryan:2014zya}, 8~\cite{Khachatryan:2015ira,Khachatryan:2016crw,AbellanBeteta:2016ugk} and 13 \cite{Aaboud:2017hbk} \TeV, and by the CDF and D0 Collaborations in proton-antiproton collisions at 1.96\TeV \cite{Aaltonen:2007ae,Abazov:2008ez}.

In this paper, we present measurements of the cross section multiplied by the branching fraction for the production of a $\cPZ/\gamma^*$ boson in association with jets and its subsequent decay into a pair of oppositely charged leptons ($\ell^+\ell^-$) in proton-proton collisions at a centre-of-mass energy of 13\TeV. The measurements from the two final states, with an electron--positron pair (electron channel) and with a muon--antimuon pair (muon channel), are combined. The measurements are performed with data from the CMS detector recorded in 2015 at the LHC corresponding to 2.19\fbinv of integrated luminosity. For convenience, $\cPZ/\gamma^*$ is denoted as $\cPZ$. In this paper a $\cPZ$ boson is defined as a pair of oppositely charged muons or electrons with invariant mass in the range $91\pm20\GeV$. This range is chosen to have a good balance between the signal acceptance, the rejection of background processes, and the ratio of $\cPZ$ boson to $\gamma^*$ event yields. It is also consistent with previous measurements~\cite{Khachatryan:2014zya,Khachatryan:2015ira,Khachatryan:2016crw} and eases comparisons.

The cross section is measured as a function of the jet multiplicity ($\Nj$), transverse momentum (\pt) of the $\cPZ$ boson, and of the jet transverse momentum and rapidity ($y$) of the first, second, and third jets, where the jets are ordered by decreasing \pt. Furthermore, the cross section is measured as a function of the scalar sum of the jet transverse momenta (\HT) for event samples with at least one, two, and three jets. These observables have been studied in previous measurements. In addition, we study the balance in transverse momentum between the reconstructed jet recoil and the $\cPZ$ boson for the different jet multiplicities and two $\cPZ$ boson \pt regions ($\pt(\cPZ) < 50\GeV$ and $\pt(\cPZ) > 50\GeV$).

\section{The CMS detector}
\label{cms}
The central feature of the CMS apparatus is a superconducting solenoid of 6\unit{m} internal diameter, providing a magnetic field of 3.8\unit{T}. Within the solenoid volume are a silicon pixel and strip tracker, a lead tungstate crystal electromagnetic calorimeter (ECAL), and a brass and scintillator hadron calorimeter (HCAL), each composed of a barrel and two endcap sections. Forward calorimeters extend the pseudorapidity coverage provided by the barrel and endcap detectors up to $\abs{\eta}=5$. The electron momentum is estimated by combining the energy measurement in the ECAL with the momentum measurement in the tracker. The momentum resolution for electrons with $\pt \approx 45\GeV$ from $\Z \rightarrow \Pe \Pe$ decays ranges from 1.7\% for nonshowering electrons in the barrel region ($\abs{\eta}< 1.444$) to 4.5\% for showering electrons in the endcaps ($1.566 < \abs{\eta} < 3$)~\cite{Khachatryan:2015hwa}. When combining information from the entire detector, the jet energy resolution is 15\% at 10\GeV, 8\% at 100\GeV, and 4\% at 1\TeV, to be compared to about 40, 12, and 5\% obtained when only the ECAL and HCAL calorimeters are used. Muons are measured in the pseudorapidity range $\abs{\eta} < 2.4$, with detection planes made using three technologies: drift tubes, cathode strip chambers, and resistive plate chambers. Matching muons to tracks measured in the silicon tracker results in a relative transverse momentum resolution for muons with $20 < \pt < 100\GeV$ of 1.3--2.0\% in the barrel and better than 6\% in the endcaps. The \pt resolution in the barrel is better than 10\% for muons with \pt up to 1\TeV~\cite{Chatrchyan:2012xi}.

Events of interest are selected using a two-tiered trigger system~\cite{Khachatryan:2016bia}. The first level (L1), composed of custom hardware processors, uses information from the calorimeters and muon detectors to select events at a rate of around 100\unit{kHz} within a time interval of less than 4\mus. The second level, known as the high-level trigger (HLT), consists of a farm of processors running a version of the full event reconstruction software optimized for fast processing, and reduces the event rate to around 1\unit{kHz} before data storage.

\section{Observables}
\label{mobs}

The cross section is measured for jet multiplicities up to 6 and differentially as a function of the transverse momentum of the $\cPZ$ boson and as a function of several jet kinematic variables, including the jet transverse momentum, rapidity, and the scalar sum of jet transverse momenta.

Jet kinematic variables are measured for event samples with at least one, two, and three jets. In the following, the jet multiplicity will be referred to as ``inclusive'' to designate events with at least $N$ jets and as ``exclusive'' for events with exactly $N$ jets.

The balance between the $\cPZ$ boson and jet transverse momenta is also studied via the \pt balance observable $\ptbal = \lvert \ptvec(\cPZ) + \sum_{\text{jets}} \ptvec(\text{j}_i) \rvert$, and the so-called jet-$\cPZ$ balance $\JZB = \lvert \sum_{\text{jets}} \ptvec(\text{j}_i) \rvert - \lvert \ptvec(\cPZ) \rvert$, where the sum runs over jets with $\pt > 30\GeV$ and $\abs{y} < 2.4$ \cite{Dias:2011zf,Chatrchyan:2012qka}. The hadronic activity not included in the jets will lead to an imbalance that translates into $\pt^{\text{bal}}$ and \JZB values different from zero. It includes the activity in the forward region ($\abs{y} > 2.4$), which is the dominant contribution according to simulation. Gluon radiation in the central region that is not clustered in a jet with $\pt>30\GeV$ will also contribute to the imbalance. Hadronic activity not included in the jets will lead to a shift of the $\pt^{\text{bal}}$ distribution peak to larger values. The \JZB variable distinguishes between two configurations, one where transverse momentum due to the unaccounted hadronic activity is in the direction of the $\cPZ$ boson and another where it is in the opposite direction. Events in the first configuration that have a large imbalance will populate the positive tail of the \JZB distribution, while those in the second configuration populate the negative tail.

The distribution of \ptbal is measured for events with minimum jet multiplicities of 1, 2, and~3. To separate low and high jet multiplicity events without \pt and y constraints on the jets, the \JZB variable is also studied for $\pt(\cPZ)$ below and above 50\GeV.

The $\cPZ$ boson transverse momentum $\pt(\cPZ)$ can be described via fixed-order calculations in perturbative QCD at high values, while at small transverse momentum this requires resummation of multiple soft-gluon emissions to all orders in perturbation theory~\cite{Dokshitzer:1978yd,Collins:1984kg}. The measurement of the distribution of $\pt(\cPZ)$ for events with at least one jet, due to the increased phase space for soft gluon radiation, leads to an understanding of the balance in transverse momentum between the jets and the $\cPZ$ boson, and can be used for comparing theoretical predictions that treat multiple soft-gluon emissions in different ways.

\section{Phenomenological models and theoretical calculations}
\label{theory}

The measured $\cPZ + \text{ jets}$ cross section is compared to four different calculations: two merging matrix elements (MEs) with various final-state parton multiplicities together with parton showering; a third with a combination of next-to-next-to-leading order (NNLO) calculation with next-to-next-to-leading logarithmic (NNLL) resummation and with parton showering; and a fourth with fixed-order calculation.

The first two calculations use \MGvATNLO version 2.2.2 (denoted \MGaMC)~\cite{Alwall:2014hca}, which is interfaced with \PYTHIAeight (version 8.212)~\cite{Sjostrand:2014zea}. \PYTHIAeight is used to include initial- and final-state parton showers and hadronisation. Its settings are defined by the CUETP8M1 tune~\cite{Khachatryan:2015pea}, in particular the NNPDF 2.3~\cite{Ball:2012cx} leading order (LO) parton distribution function (PDF) is used and the strong coupling $\alpS(m_{\cPZ})$ is set to 0.130. The first calculation includes MEs computed at LO for the five processes \ppZnj, $N=0\ldots4$ and matched to the parton shower using the \kt-MLM~\cite{Alwall:2007fs,Alwall:2008qv} scheme with the matching scale set at 19\GeV. In the ME calculation, the NNPDF~3.0 LO PDF~\cite{Ball:2014uwa} is used and $\alpS(m_{\cPZ})$ is set to 0.130 at the $\cPZ$ boson mass scale. The second calculation includes MEs computed at NLO for the three processes \ppZnj, $N=0\ldots2$ and merged with the parton shower using the FxFx~\cite{Frederix:2012ps} scheme with the merging scale set at 30\GeV. The NNPDF~3.0 next-to-leading order (NLO) PDF is used and $\alpS(m_{\cPZ})$ is set to 0.118. This second calculation is also employed to derive nonperturbative corrections for the fixed-order prediction discussed in the following.

The third calculation uses the \GENEVA 1.0-RC2 MC program (GE), where an NNLO calculation for Drell--Yan production is combined with higher-order resummation~\cite{Alioli:2015toa,Alioli:2012fc}. Logarithms of the 0-jettiness resolution variable, ${\tau}$, also known as beam thrust and defined in Ref.~\cite{Stewart:2010tn}, are resummed at NNLL including part of the next-to-NNLL (N$^{3}$LL) corrections. The accuracy refers to the $\tau$ dependence of the cross section and is denoted NNLL'$_\tau$. The PDF set PDF4LHC15 NNLO~\cite{Butterworth:2015oua} is used for this calculation and $\alpS(m_{\cPZ})$ is set to 0.118. The resulting parton-level events are further combined with parton showering and hadronisation provided by \PYTHIAeight using the same tune as for \MGaMC.

Finally, the distributions measured for $\Nj\ge1$ are compared with the fourth calculation performed at NNLO accuracy for $\cPZ +1 $ jet using the $N$-jettiness subtraction scheme ($N_{\text{jetti}}$)~\cite{Boughezal:2016isb,Boughezal:2015ded}. The PDF set CT14~\cite{Dulat:2015mca} is used for this calculation. The nonperturbative correction obtained from \MGaMC and \PYTHIAeight is applied.  It is calculated for each bin of the measured distributions from the ratio of the cross section values obtained with and without multiple parton interactions and hadronisation. This correction is less than 7\%.

Given the large uncertainty in the LO calculation for the total cross section, the prediction with LO MEs is rescaled to match the $\Pp\Pp\to \cPZ$ cross section calculated at NNLO in \alpS and includes NLO quantum electrodynamics (QED) corrections with \FEWZ~\cite{Melnikov:2006kv} (version 3.1b2). The values used to normalise the cross section of the \MGaMC predictions are given in Table~\ref{tab:theory_xsec}. All the numbers correspond to a 50\GeV dilepton mass threshold applied before QED final-state radiation (FSR). With \FEWZ, the cross section is computed in the dimuon channel, using a mass threshold applied after QED FSR, but including the photons around the lepton at a distance $R = \sqrt{(\Delta \eta)^2+(\Delta \phi)^2}$ smaller than 0.1. The number given in the table includes a correction computed with the LO sample to account for the difference in the mass definition. This correction is small, $+0.35\%$. When the mass threshold is applied before FSR, the cross section is assumed to be the same for the electron and muon channels.

\begin{table*}[ht]
\centering
 \topcaption{Values of the $\Pp\Pp \to \ell^+\ell^-$ total cross section used for the calculation in data-theory comparison plots. The cross section used, the cross section from the MC generator (``native''), and the ratio of the two ($k$) are provided. The phase space of the sample to which the cross section values correspond is indicated in the second column.}

\cmsTable{
\begin{tabular}{lccccc}
                                        &                    & Native cross     &             & Used cross  &\\
Prediction                              & Phase space        & section [pb]     & Calculation & section [pb]& $k$\\
  \hline
  \MGaMC+\PYTHIAeight, ${\le} 4$ j LO+PS       & $m_{\ell^+\ell^-}>50\GeV$ & 1652             & \FEWZ NNLO   & 1929        & 1.17 \\
  \MGaMC+\PYTHIAeight, ${\le }2$ j NLO+PS      & $m_{\ell^+\ell^-}>50\GeV$ & 1977             & native      & 1977        & 1     \\
  \GENEVA                          & $m_{\ell^+\ell^-} \in [50, 150\GeV]$ & 1980 & native & 1980 & 1 \\
\end{tabular}
}
\label{tab:theory_xsec}
\end{table*}

Uncertainties in the ME calculation (denoted {\em theo. unc.} in the figure legends) are estimated for the NLO \MGaMC, NNLO, and \GENEVA calculations following the prescriptions recommended by the authors of the respective generators. The uncertainty coming from missing terms in the fixed-order calculation is estimated by varying the normalisation ($\mu_{\mathrm{R}}$) and factorisation ($\mu_{\mathrm{F}}$) scales by factors 0.5 and 2. In the case of the FxFx-merged sample, the envelope of six combinations of the variations is considered, the two combinations where one scale is varied by a factor 0.5 and the other by a factor 2 are excluded. In the case of the NNLO and \GENEVA samples the two scales are varied by the same factor, leading to only two combinations. For \GENEVA, the uncertainty is symmetrised  by using the maximum of the up and down uncertainties for both cases. The uncertainty from the resummation is also estimated and added in quadrature. It is estimated using six profile scales~\cite{Abbate:2010xh,Ligeti:2008ac}, as described in Ref.~\cite{Alioli:2015toa}. Uncertainties in PDF and \alpS values are also estimated in the case of the FxFx-merged sample. The PDF uncertainty is estimated using the set of 100 replicas of the NNPDF~3.0 NLO PDF, and the uncertainty in the \alpS value used in the ME calculation is estimated by varying it by $\pm0.001$. These two uncertainties are added in quadrature to the ME calculation uncertainties. For \GENEVA and NLO \MGaMC all these uncertainties are obtained using the reweighting method~\cite{Frederix:2011ss,Alioli:2015toa} implemented in these generators.

\section{Simulation}
\label{samples}

MC event generators are used to simulate proton-proton interactions and produce events from signal and background processes.
The response of the detector is modeled with \GEANTfour \cite{Allison:2006ve}.
The $\cPZ (\to \ell^+ \ell^-) + \text{ jets}$ process is generated with NLO \MGaMC interfaced with \PYTHIAeight, using the FxFx merging scheme as described in Section~\ref{theory}. The sample includes the $\cPZ\to \Pgt^+\Pgt^-$ process, which is considered a background.
Other processes that can give a final state with two oppositely charged same-flavour leptons and jets are $\PW\PW$, $\PW\cPZ$, $\cPZ\cPZ$, $\PQt\PAQt$ pairs, and single top quark production. The $\PQt\PAQt$ and single top quark backgrounds are generated using \POWHEG version~2~\cite{Nason:2004rx,Frixione:2007vw,Alioli:2010xd,Frixione:2007nw} interfaced with \PYTHIAeight. Background samples corresponding to diboson electroweak production (denoted VV in the figure legends)~\cite{Nason:2013ydw} are generated at NLO with \POWHEG interfaced to \PYTHIAeight ($\PW\PW$), \MGaMC interfaced to \PYTHIAeight or \PYTHIAeight alone ($\PW\cPZ$ and $\cPZ\cPZ$). The background sample corresponding to $\PW + \text{ jets}$ production ($\PW$) is generated at NLO using \MGaMC interfaced with \PYTHIAeight, utilizing the FxFx merging scheme.

The events collected at the LHC contain multiple superimposed proton-proton collisions within a single beam crossing, an effect known as pileup. Samples of simulated pileup are generated with a distribution of proton-proton interactions per beam bunch crossing close to that observed in data. The number of pileup interactions, averaging around 20, varies with the beam conditions. The correct description of pileup is ensured by reweighting the simulated sample to match the number of interactions measured in data.

\section{Object reconstruction and event selection}
\label{eventselection}

The particle-flow (PF) algorithm~\cite{Sirunyan:2017ulk} is used to reconstruct the events. It combines the information from the various elements of the CMS detector to reconstruct and identify each particle in the event. The reconstructed particles are called PF candidates. If several primary vertices are reconstructed, we use the one with the largest quadratic sum of associated track transverse momenta as the vertex of the hard scattering and the other vertices are assumed to be pileup.

The online trigger selects events with two isolated electrons (muons) with transverse momenta of at least 17 and 12 (17 and 8) \GeV. After offline reconstruction, the leptons are required to satisfy $\pt > 20 \GeV$ and $\abs{\eta} < 2.4$. We require that the two electrons (muons) with highest transverse momenta form a pair of oppositely charged leptons with an invariant mass in the range $91\pm20\GeV$. The transition region between the ECAL barrel and endcap ($1.444 < \abs{\eta}< 1.566$) is excluded in the reconstruction of electrons and the missing acceptance is corrected to the full $\abs{\eta} < 2.4$ region. The reconstruction of electrons and muons is described in detail in Refs.~\cite{Khachatryan:2015hwa,Chatrchyan:2012xi}. The identification criteria applied for electrons and muons are identical to those described in the Ref.~\cite{Khachatryan:2016crw} except for the thresholds of the isolation variables, which are optimised for 13\TeV centre-of-mass energy in our analysis. Electrons (muons) are considered isolated based on the scalar $\pt$ sum of the nearby PF candidates with a distance $R = \sqrt{(\Delta \eta)^2+(\Delta \phi)^2} < 0.3$ (0.4). The scalar \pt sum must be less than 15 (25)\% of the electron (muon) transverse momentum. We also correct the simulation for differences from data in the trigger, and the lepton identification, reconstruction and isolation efficiencies. These corrections, which depend on the run conditions, are derived using data taken during the run period, and they typically amount to 1--2\% for the reconstruction and identification efficiency and 3--5\% for the trigger efficiency.

Jets at the generator level are defined from the stable particles ($c\tau > 1\cm$), neutrinos excluded, clustered with the anti-$\kt$ algorithm~\cite{Cacciari:2008gp} using a radius parameter of 0.4. The jet four-momentum is obtained according to the E-scheme~\cite{fastjetmanual} (vector sum of the four-momenta of the constituents). In the reconstructed data, the algorithm is applied to the PF candidates. The technique of charged-hadron subtraction~\cite{Sirunyan:2017ulk} is used to reduce the pileup contribution by removing charged particles that originate from pileup vertices. The jet four-momentum is corrected for the difference observed in the simulation between jets built from PF candidates and generator-level particles. The jet mass and direction are kept constant for the corrections, which are functions of the jet $\eta$ and \pt, as well as the energy density and jet area quantities defined in Ref.~\cite{Cacciari:2007fd,Chatrchyan:2011ds}. The latter are used in the correction of the energy offset introduced by the pileup interactions. Further jet energy corrections are applied for differences between data and simulation in the pileup in zero-bias events and in the \pt balance in dijet, $\cPZ +\text{ jet}$, and $\gamma+\text{ jet}$ events. Since the \pt balance in  $\cPZ+\text{ jet}$ events is one of the observables we are measuring in this paper, it is important to understand how it is used in the jet calibration. The balance is measured for events with two objects (jet, $\gamma$, or $\cPZ$ boson) back-to-back in the transverse plane ($\abs{\Delta\phi - \pi} < 0.34$) associated with a possible third object, a soft jet. The measurement is made for various values of $\rho=\pt^{\text{soft jet}}/\pt^{\text{ref}}$, running from 0.1 to 0.3, and extrapolated to $\rho = 0$. In the case the back-to-back objects are a jet and a boson,  $\pt^{\text{ref}}$ is defined as the transverse momentum of the boson, while in the case of two jets it is defined as the average of their transverse momenta. All jets down to $\pt = 5$ or 10\GeV, including jets reconstructed in the forward calorimeter, are considered for the soft jet. The data-simulation adjustment is therefore done for ideal topologies with only two objects, whose transverse momenta must be balanced. The jet calibration procedure is detailed in the Ref.~\cite{Khachatryan:2016kdb}. In this measurement, jets are further required to satisfy the loose identification criteria defined in Ref.~\cite{CMS:2016jetID}. Despite the vertex requirement used in the jet clustering some jets are reconstructed from pileup candidates; these jets are suppressed using the multivariate technique described in Ref.~\cite{CMS:2013wea}.  Jets with $\pt>30\GeV$ and $\abs{y}<2.4$ are used in this analysis.

\section{Backgrounds estimation}
\label{background}

The contributions from background processes are estimated using the simulation samples described in Section~\ref{samples} and are subtracted from the measured distributions. The dominant background, $\ttbar$, is also measured from data. This $\ttbar$ background contributes mainly due to events with two same-flavour leptons. The production cross sections for $\Pep\Pem$ and $\Pgmp\Pgmm$ events from $\ttbar$ are identical to the cross section of $\Pep\Pgmm$ and $\Pem\Pgmp$ and can therefore be estimated from the latter. We select events in the $\ttbar$ control sample using the same criteria as for the measurement, but requiring the two leptons to have different flavours. This requirement rejects the signal and provides a sample enriched in $\PQt\PAQt$ events. Each of the distributions that we are measuring is derived from this sample and compared with the simulation. This comparison produces a discrepancy for events with at least one jet that we correct by applying a correction factor $\mathcal{C}$ to the simulation depending on the event jet multiplicity. These factors, together with their uncertainties, are given in Table~\ref{tab:TTSF}.

After applying this correction to the simulation, all the distributions considered in this measurement agree with data in the $\ttbar$ control sample. The agreement is demonstrated with a $\chi^2$-test. We conclude that a parametrization as a function of the jet multiplicity is sufficient to capture the dependency on the event topology. Remaining sources of uncertainties are the estimate of the lepton reconstruction and selection efficiencies and of the yield of events from processes other than \ttbar entering in the control region. This yield is estimated from the simulation. Based on the sizes of the statistical uncertainties and background contributions, both these uncertainties are negligible. Therefore, the uncertainty in the correction factor is reduced to the statistical uncertainties in the data and simulation samples.

\begin{table}[h]
  \centering
  \topcaption{The correction factors ($\mathcal{C}$) applied to the simulated $\ttbar$ sample with their uncertainties, which are derived from the statistical uncertainties in the data and simulation samples.}
\begin{tabular}{cc}
  $\Nj$ & $\mathcal{C}$ \\
  \hline
  $=$0 & 1 \\
  $=$1 & 0.94 $\pm$ 0.04 \\
  $=$2 & 0.97 $\pm$ 0.03 \\
  $=$3 & 1.01 $\pm$ 0.04 \\
  $=$4 & 0.86 $\pm$ 0.06 \\
  $=$5 & 0.61 $\pm$ 0.09 \\
  $=$6 & 0.68 $\pm$ 0.17 \\
\end{tabular}
\label{tab:TTSF}
\end{table}

\begin{figure*}
  \centering
  \includegraphics[width=0.45\textwidth]{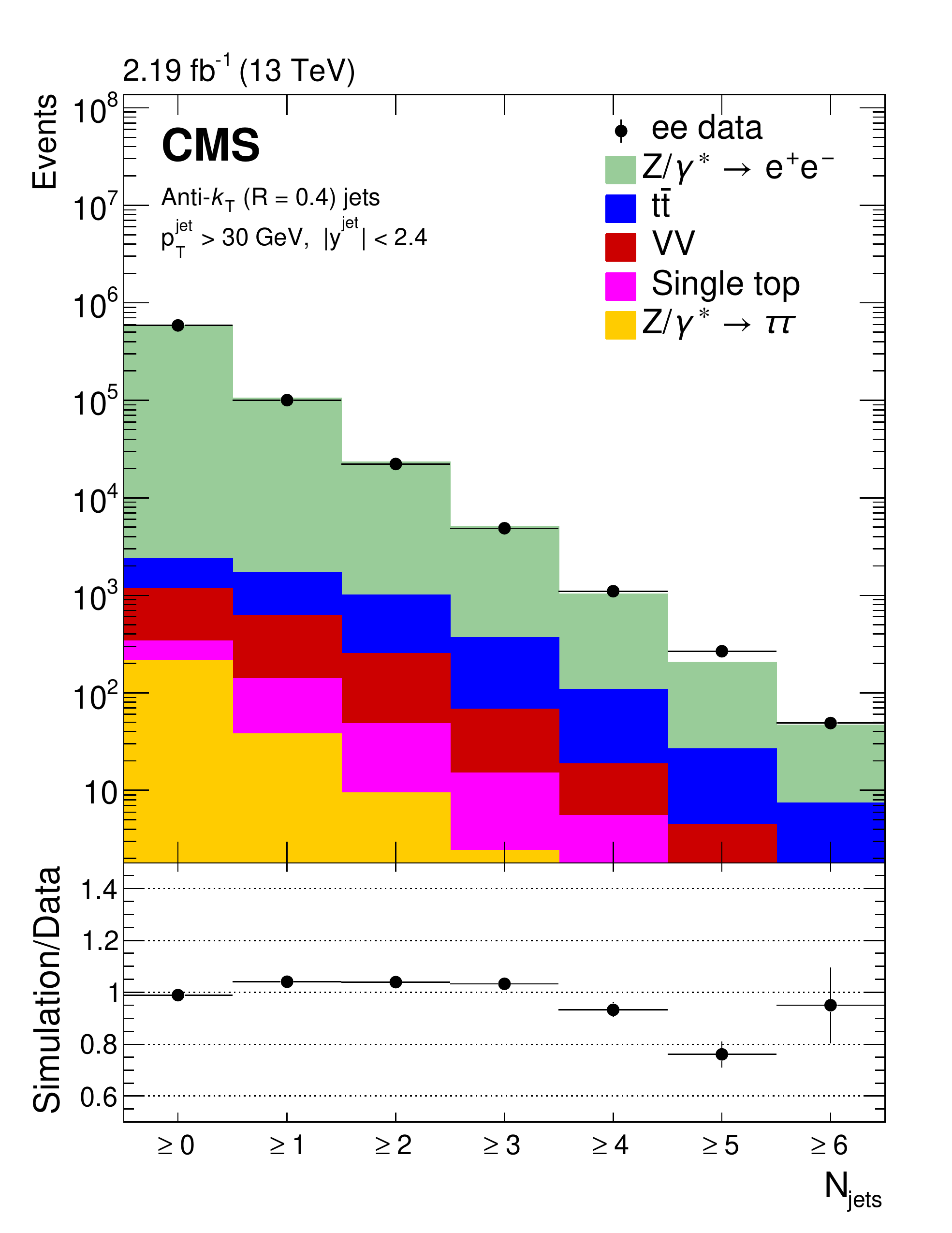}%
  \includegraphics[width=0.45\textwidth]{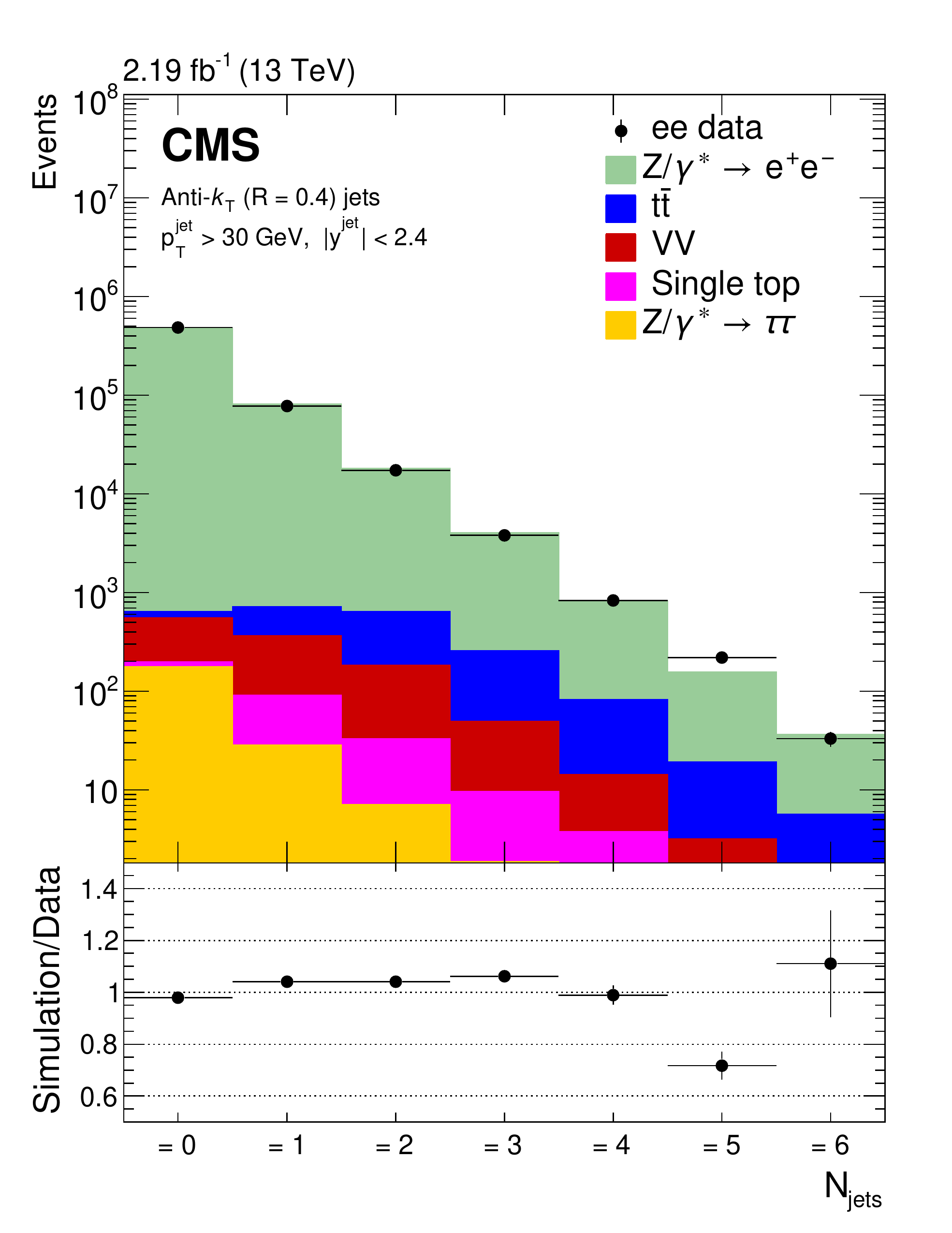}\\
  \includegraphics[width=0.45\textwidth]{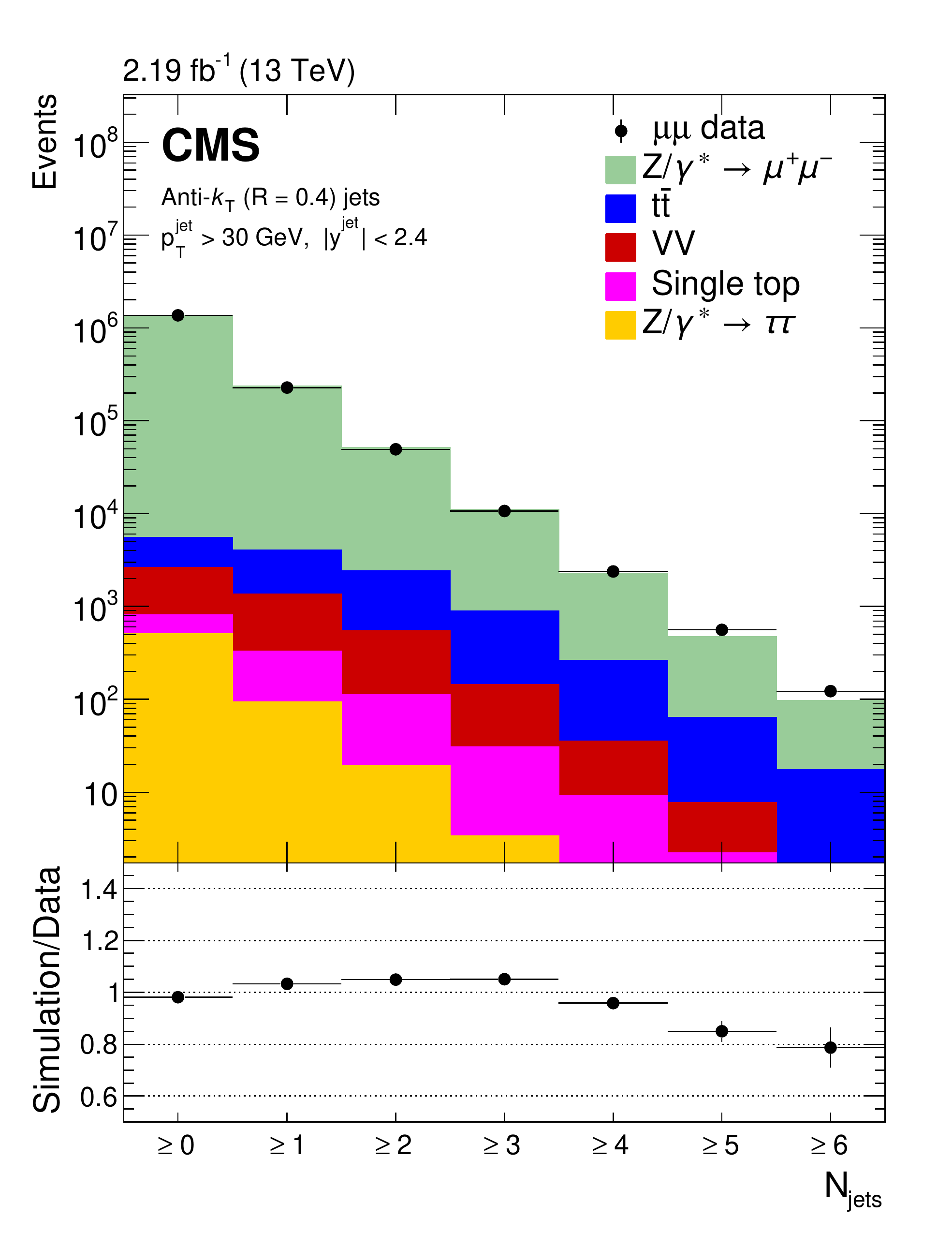}%
  \includegraphics[width=0.45\textwidth]{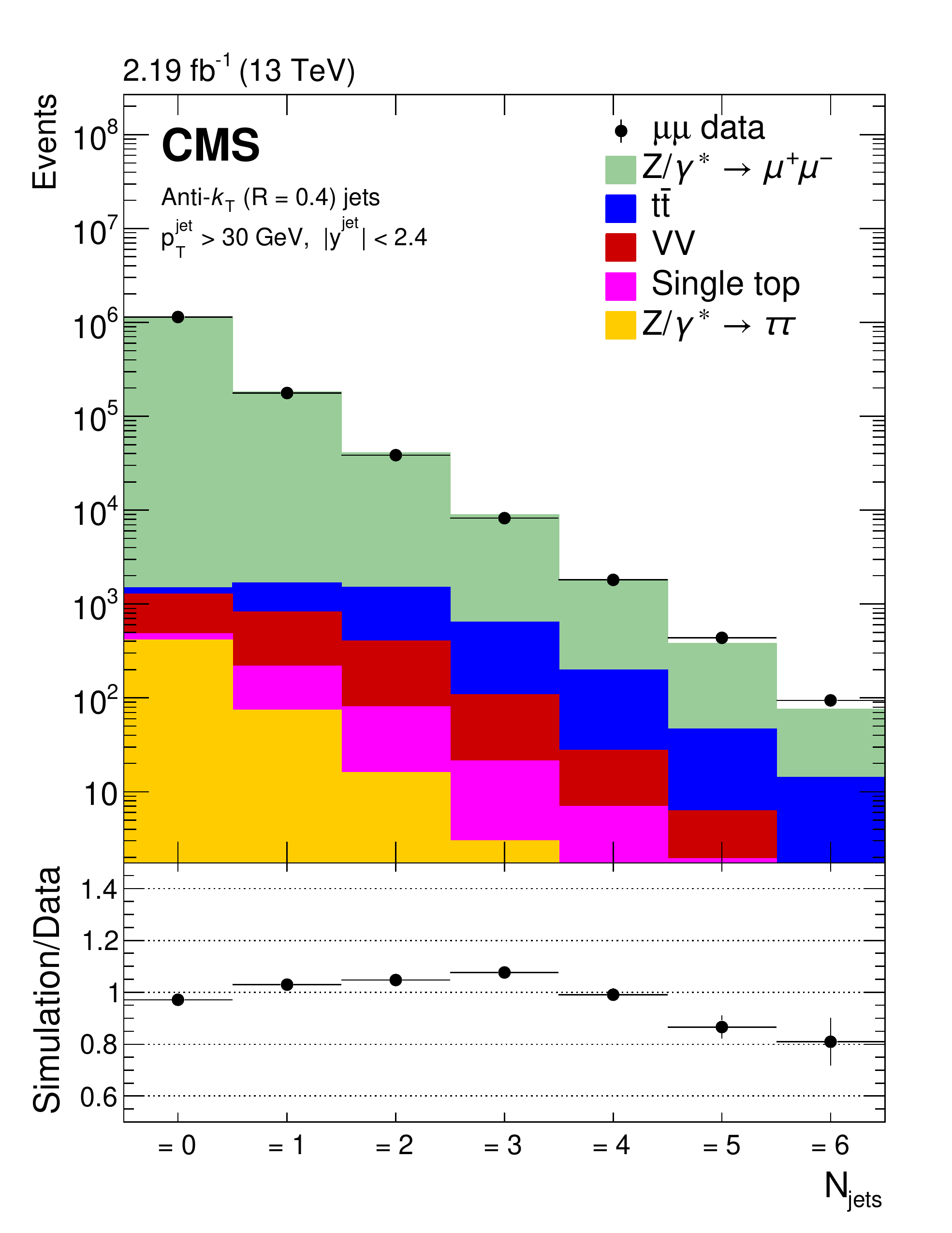}
  \caption{Reconstructed data, simulated signal, and background distributions of the inclusive (left) and exclusive (right) jet multiplicity for the electron (upper) and muon (lower) channels. The background distributions are obtained from the simulation, except for the $\PQt\PAQt$ contribution which is estimated from the data as explained in the text. The error bars correspond to the statistical uncertainty. In the ratio plots, they include both the uncertainties from data and from simulation. The set of generators described in Section~\ref{samples} has been used for the simulation.}
  \label{fig:njet}
\end{figure*}

\begin{figure*}
  \centering
  \includegraphics[width=0.45\textwidth]{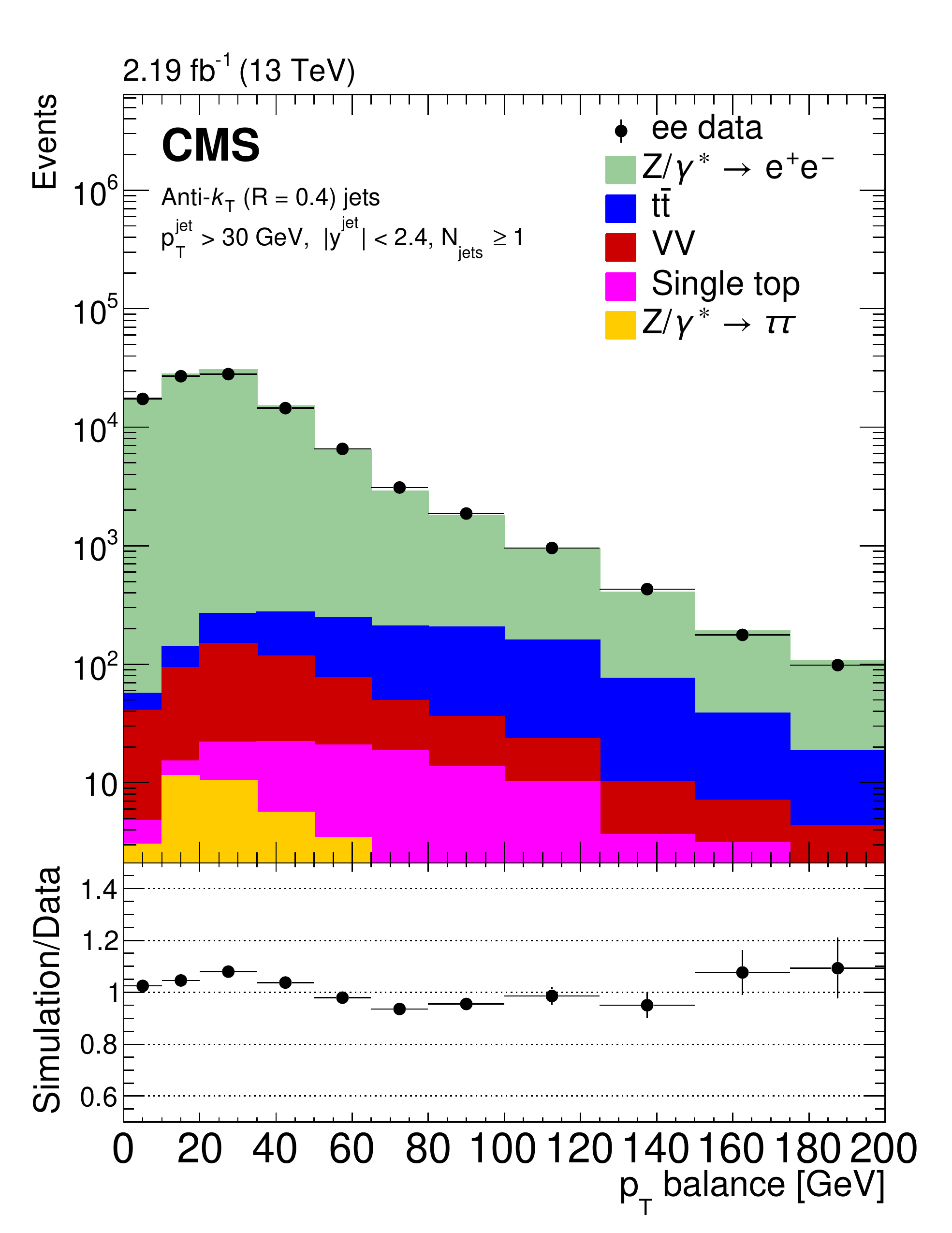}%
  \includegraphics[width=0.45\textwidth]{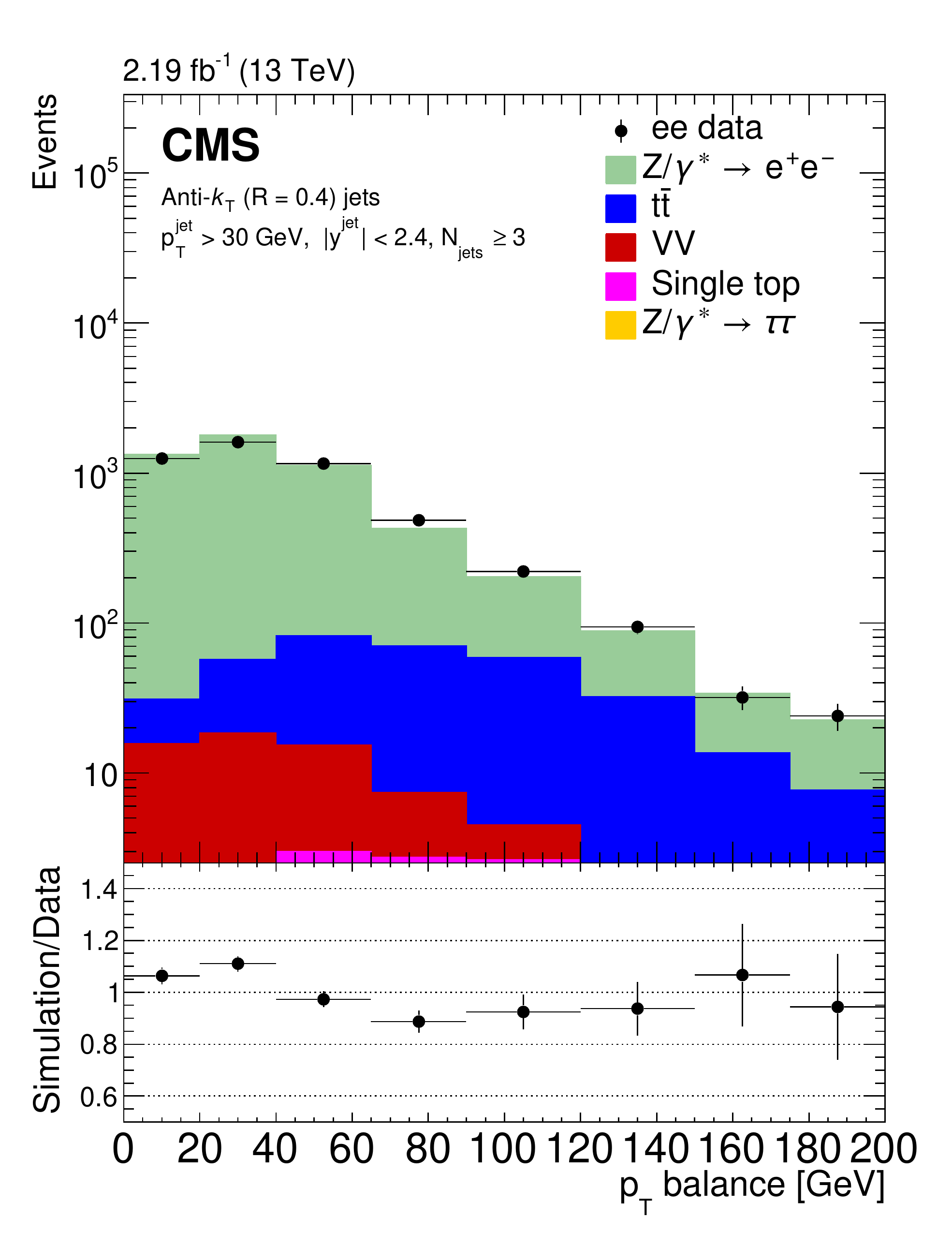}\\
  \includegraphics[width=0.45\textwidth]{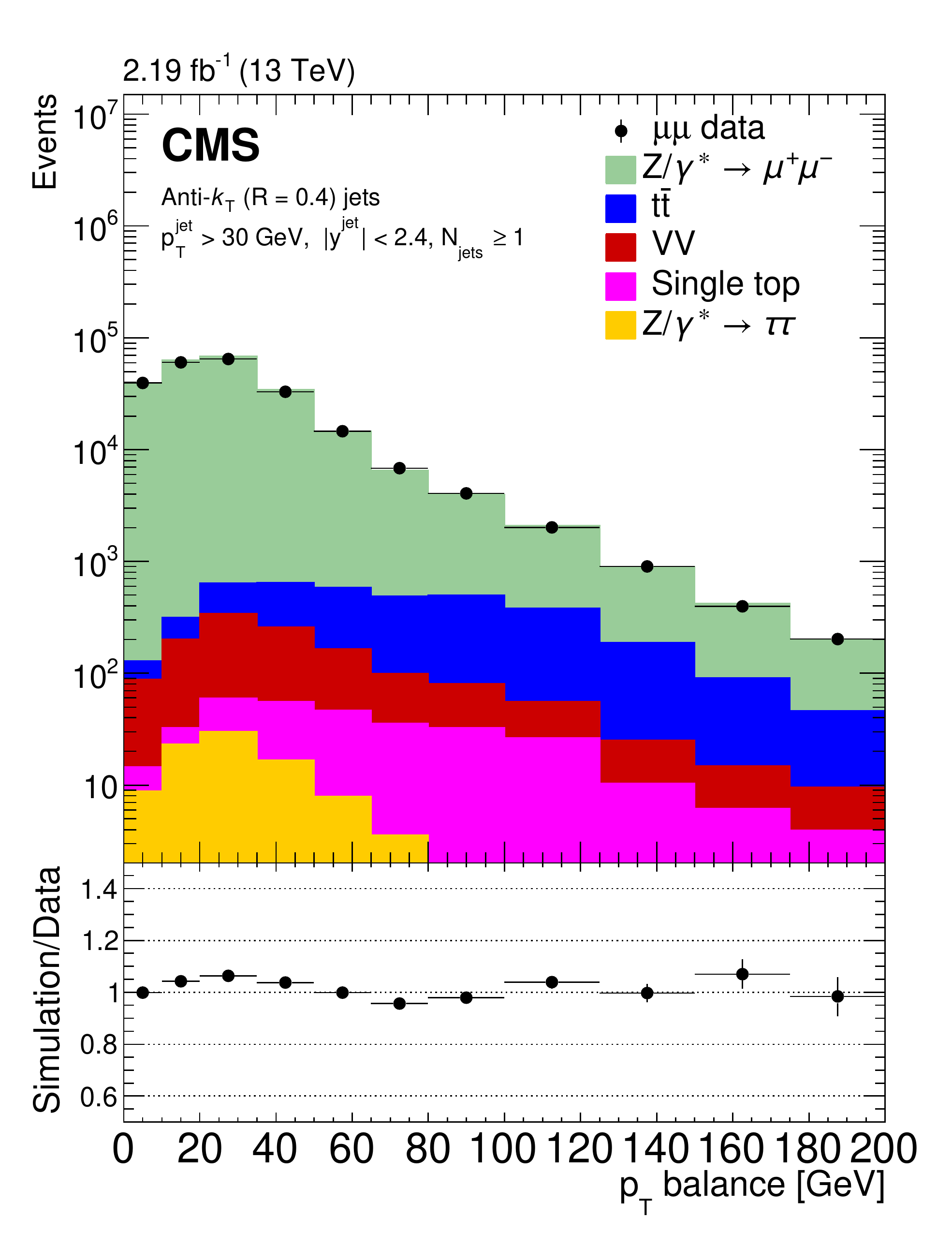}%
  \includegraphics[width=0.45\textwidth]{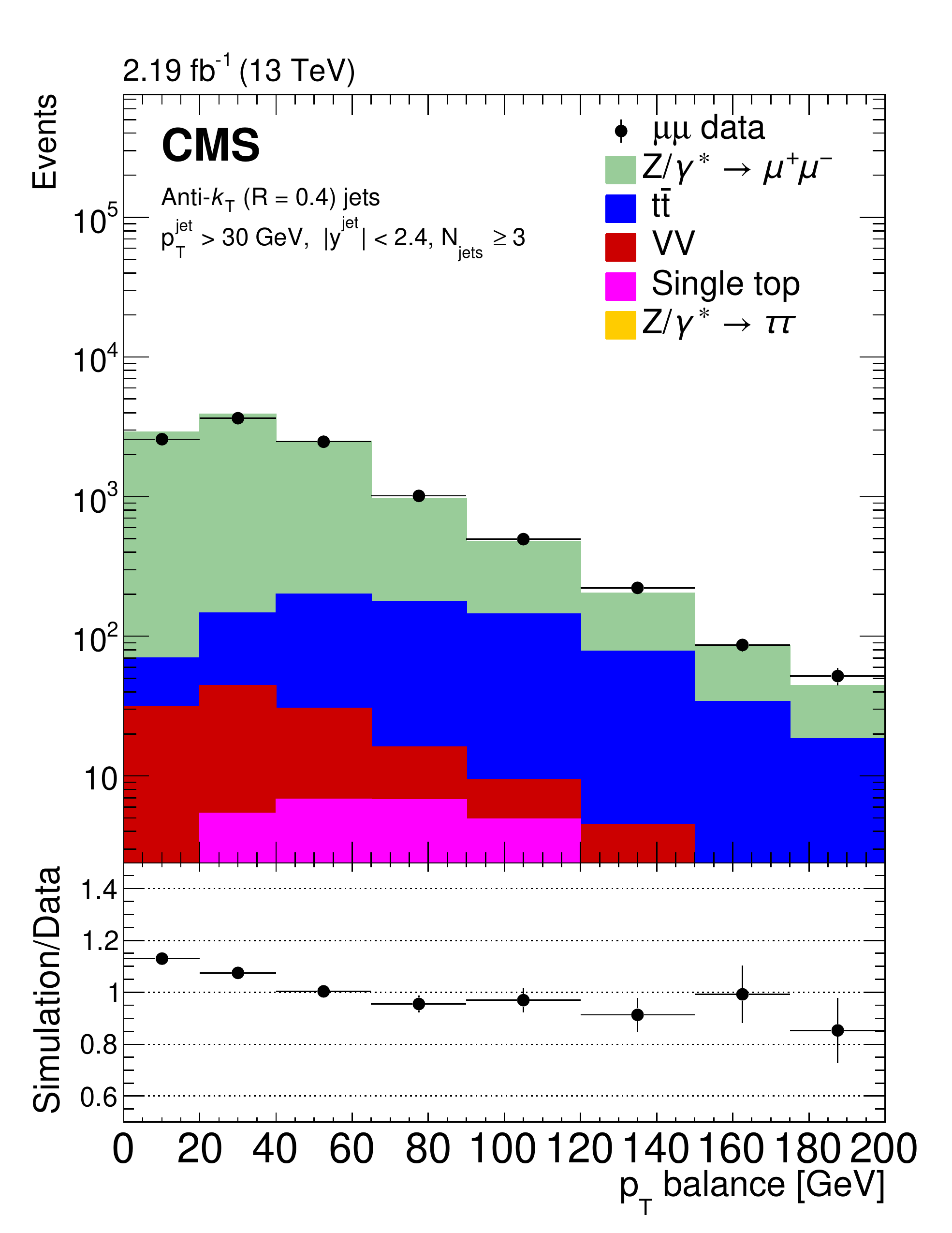}
  \caption{Reconstructed data, simulated signal, and background distributions of the transverse momentum balance between the $\cPZ$ boson and the sum of the jets with at least one jet (left) and three jets (right) for the electron (upper) and muon (lower) channels. The background distributions are obtained from the simulation, except for the $\PQt\PAQt$ contribution which is estimated from the data as explained in the text. The error bars correspond to the statistical uncertainty. In the ratio plots, they include both the uncertainties from data and from simulation. The set of generators described in Section~\ref{samples} has been used for the simulation.}
  \label{fig:reco-ptbal}
\end{figure*}

\begin{figure*}
  \centering               \includegraphics[width=0.5\textwidth]{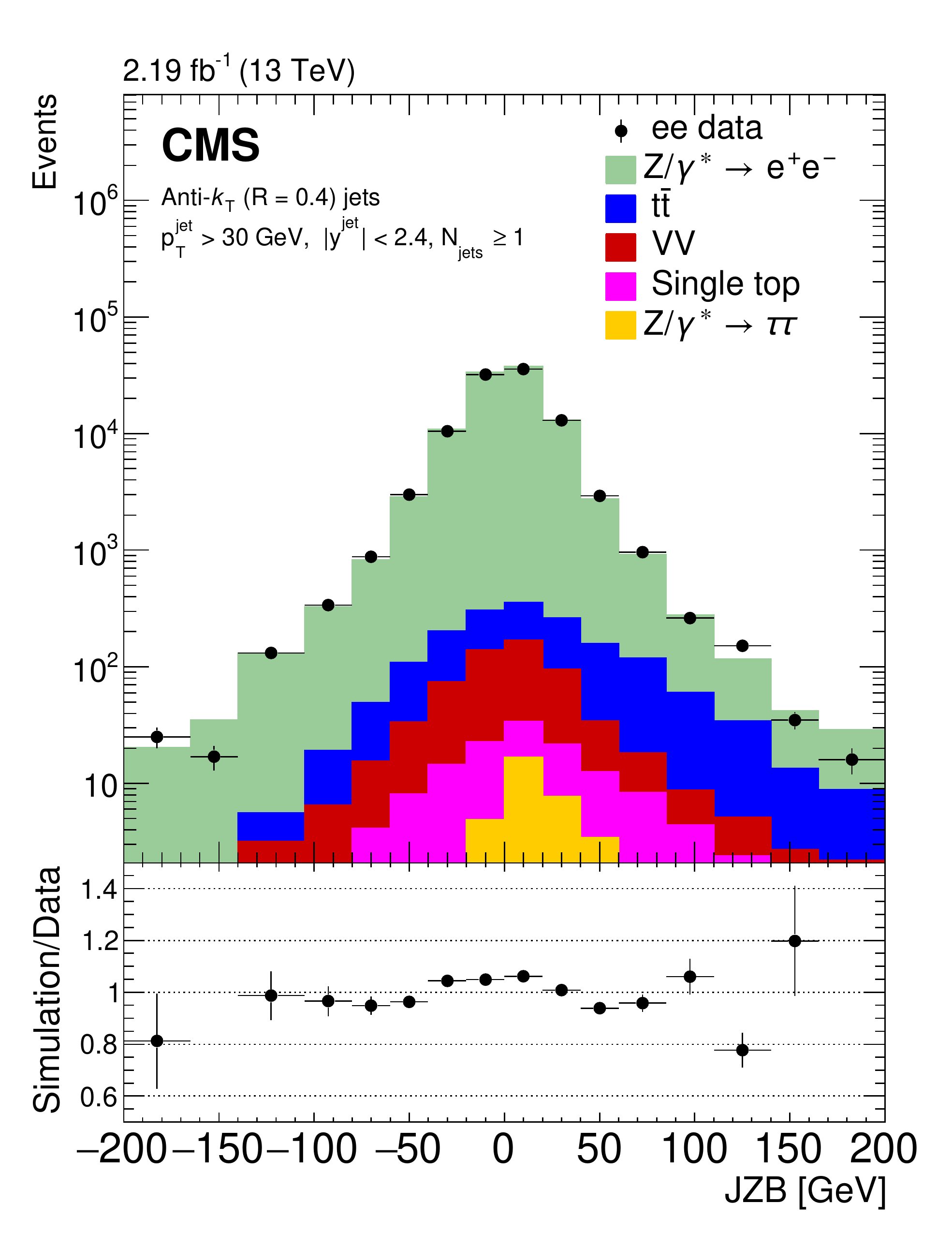}\includegraphics[width=0.5\textwidth]{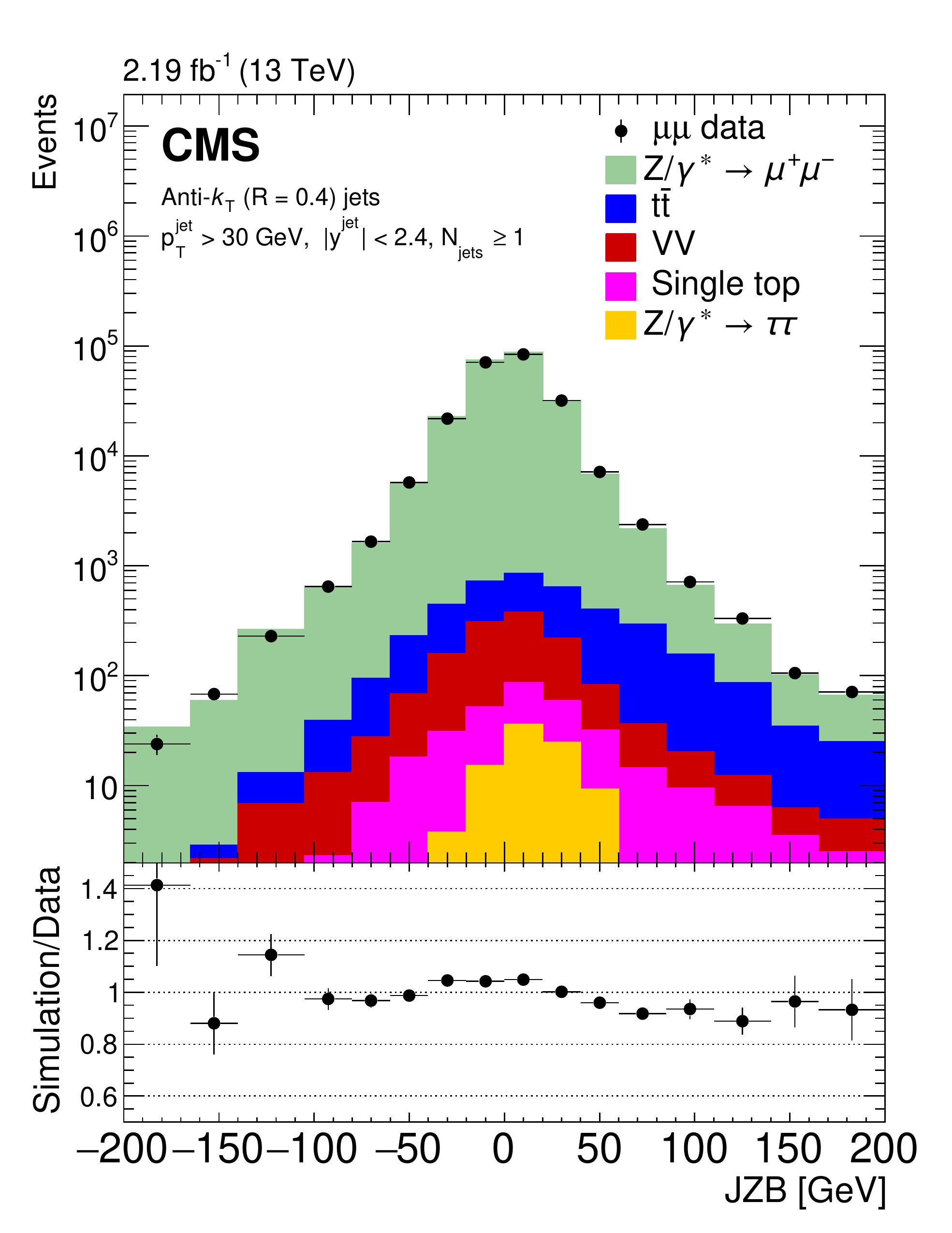}
  \caption{Reconstructed data, simulated signal, and background distributions of the \JZB variable for the electron (left) and muon (right) channels. The background distributions are obtained from the simulation, except for the $\PQt\PAQt$ contribution which is estimated from the data as explained in the text. The error bars correspond to the statistical uncertainty. In the ratio plots, they include both the uncertainties from data and from simulation. The set of generators described in Section~\ref{samples} has been used for the simulation.}
  \label{fig:reco-jzb}
\end{figure*}

The jet multiplicity distributions in data and simulation are presented in Fig.~\ref{fig:njet}. The background contamination is below 1\% for the inclusive cross section, and increases with the number of jets to close to 10\% for a jet multiplicity of three and above due to $\PQt\PAQt$ production. Multijet and $\PW$ events could pass the selection if one or two jets are misidentified as leptons. The number of multijet events is estimated from data using a control sample obtained by requiring two same-sign same-flavour lepton candidates, whereas the number of $\PW$ events is estimated from simulation. Both contributions are found to be negligible. Fig.~\ref{fig:reco-ptbal} shows the \ptbal distribution separately for electron and muon channels. The $\ttbar$ background does not peak at the same $\pt$ balance as the signal, and has a broader spectrum. The \JZB distribution is shown in Fig.~\ref{fig:reco-jzb}. The $\ttbar$ background is asymmetric, making a larger contribution to the positive side of the distribution because transverse energy is carried away by neutrinos from $\PW$  boson decays, leading to a reduction in the negative term of the \JZB expression. Overall the agreement between data and simulation before the background subtraction is good and differences are within about 10\%.

\section{Unfolding procedure}
\label{unfolding}

The fiducial cross sections are obtained by subtracting the simulated backgrounds from the data distributions and correcting the background-subtracted data distributions back to the particle level using an unfolding procedure, which takes into account detector effects such as detection efficiency and resolution. The unfolding is performed using the D'Agostini iterative method with early stopping~\cite{D'Agostini:1994zf} implemented in the RooUnfold toolkit~\cite{Adye:2011gm}. The response matrix describes the migration probability between the particle- and reconstructed-level quantities, including the overall reconstruction efficiencies. It is computed using a $\cPZ + \text{ jets}$ sample simulated with \MGaMC interfaced with \PYTHIAeight, using the FxFx merging scheme as described in Section~\ref{theory}. The optimal number of iterations is determined separately for each distribution by studying the fluctuations introduced by the unfolding with toy MC experiments generated at each step of the iteration. Final unfolded results have also been checked to be consistent with data-simulation comparisons on detector-folded distributions.

Because of the steep slope at the lower boundary of the jet transverse momentum distributions and in order to improve its accuracy, the unfolding is performed for these distributions using histograms with two additional bins, $[20, 24]$ and $[24, 30]\GeV$, below the nominal \pt threshold. The additional bins are discarded after the unfolding

The particle-level values refer to the stable leptons from the decay of the $\cPZ$ boson and to the jets built from the stable particles (c$\tau>1\cm$) other than neutrinos using the same algorithm as for the measurements. The momenta of all the photons whose $R$ distance to the lepton axis is smaller than 0.1 are added to the lepton momentum to account for the effects of the final-state radiation; the leptons are said to be ``dressed''.
The momentum of the $\cPZ$ boson is taken to be the sum of the momenta of the two highest-\pt electrons (or muons). The phase space for the cross section measurement is restricted to events with a $\cPZ$ boson mass between 71 and 111\GeV and both leptons with $\pt > 20\GeV$ and $\abs{\eta} < 2.4$. Jets are required to have $\pt > 30\GeV$, $\abs{y} < 2.4$ and a spatial separation from the dressed leptons of $R > 0.4$.

\section{Systematic uncertainties}
\label{systematics}

The systematic uncertainties are propagated to the measurement by varying the corresponding simulation parameters by one standard deviation up and down when computing the response matrix. The uncertainty sources are independent, and the resulting uncertainties are therefore added in quadrature. Tables~\ref{tab:combZNGoodJets_Zexc} to~\ref{tab:combJZB_ptHigh} present the uncertainties for each differential cross section.

The dominant uncertainty comes from the jet energy scale (JES). It typically amounts to 5\% for a jet multiplicity of one and increases with the number of reconstructed jets. The uncertainty in the jet resolution (JER), which is responsible for the bin-to-bin migrations that is corrected by the unfolding, is estimated and the resulting uncertainty is typically 1\%.

The most important uncertainty after the JES arises from the measured efficiency (Eff) of trigger, lepton reconstruction, and lepton identification, which results in a measurement uncertainty of about 2\% up to 4\% for events with leptons of large transverse momenta. The uncertainty in the measurement of the integrated luminosity (Lumi) is 2.3\%~\cite{CMS:2016eto}. The resulting uncertainty on the measured distributions is 2.3\%, although the uncertainty is slightly larger in regions that contain background contributions that are estimated from simulation.

The largest background contribution to the uncertainty (Bkg) comes from the reweighting procedure for the $\ttbar$ simulation, which is estimated to be less than 1\% for jet multiplicity below 4. Theoretical contributions come from the accuracy of the predicted cross sections, and include the uncertainties from PDFs, \alpS and the fixed-order calculation. Three other small sources of uncertainty are: (1) the lepton energy scale (LES) and resolution (LER), which are below 0.3\% in every bin of the measured distributions; (2) the uncertainty in the pileup model, where the 5\% uncertainty in the average number of pileup events results in an uncertainty in the measurement smaller than 1\%; and (3) the uncertainty in the input distribution used to build the response matrix used in the unfolding and described as follows.

Because of the finite binning a different distribution will lead to a different response matrix. This uncertainty is estimated by weighting the simulation to agree with the data in each distribution and building a new response matrix. The weighting is done using a finer binning than for the measurement. The difference between the nominal results and the results unfolded using the alternative response matrix is taken as the systematic uncertainty, denoted {\em Unf model}. An additional uncertainty comes from the finite size of the simulation sample used to build the response matrix. This source of uncertainty is denoted {\em Unf stat} in the table and is included in the systematic uncertainty of the measurement.

\begin{table*}
\centering
\topcaption{Cross section in exclusive jet multiplicity for the combination of both decay channels and breakdown of the uncertainties.}
\footnotesize{
\begin{tabular}{cccccccccccc}
$\Nj$ & $\dd{\sigma}{\Nj}$ & Tot. unc & Stat & JES & JER & Eff & Lumi & Bkg & Pileup & Unf model & Unf stat\\
 & [\text{pb]} & [\%]& [\%] & [\%] & [\%] & [\%] & [\%] & [\%] & [\%] & [\%] & [\%]\\
\hline
$=$ 0 & 652. & 3.0 & 0.090 & 1.1 & 0.046 & 1.5 & 2.3 & $<$0.01 & 0.22 & \NA  & 0.026 \\
$=$ 1 & 98.0 & 5.1 & 0.27 & 4.3 & 0.18 & 1.5 & 2.3 & 0.012 & 0.30 & \NA  & 0.10 \\
$=$ 2 & 22.3 & 7.3 & 0.62 & 6.7 & 0.20 & 1.6 & 2.3 & 0.026 & 0.43 & \NA  & 0.26 \\
$=$ 3 & 4.68 & 10. & 1.3 & 9.8 & 0.39 & 1.7 & 2.3 & 0.13 & 0.29 & \NA  & 0.54 \\
$=$ 4 & 1.01 & 11. & 3.4 & 10. & 0.24 & 1.7 & 2.3 & 0.42 & 0.56 & \NA  & 1.4 \\
$=$ 5 & 0.274 & 14. & 5.0 & 12. & 0.076 & 2.0 & 2.3 & 1.2 & 0.30 & \NA  & 2.2 \\
$=$ 6 & 0.045 & 24. & 15. & 17. & 0.35 & 1.8 & 2.4 & 3.5 & 1.7 & \NA  & 6.6 \\
\end{tabular}}
\label{tab:combZNGoodJets_Zexc}
\end{table*}

\begin{table*}
\centering
\topcaption{Cross section in inclusive jet multiplicity for the combination of both decay channels and breakdown of the uncertainties.}
\footnotesize{
\begin{tabular}{cccccccccccc}
$\Nj$ & $\dd{\sigma}{\Nj}$ & Tot. unc & Stat & JES & JER & Eff & Lumi & Bkg & Pileup & Unf model & Unf stat\\
 & [\text{pb]} & [\%]& [\%] & [\%] & [\%] & [\%] & [\%] & [\%] & [\%] & [\%] & [\%]\\
\hline
$\geq$ 0 & 778. & 2.8 & 0.080 & 0.079 & $<$0.01 & 1.5 & 2.3 & $<$0.01 & 0.24 & \NA  & 0.025 \\
$\geq$ 1 & 126.3 & 5.7 & 0.22 & 5.0 & 0.19 & 1.5 & 2.3 & $<$0.01 & 0.32 & \NA  & 0.086 \\
$\geq$ 2 & 28.3 & 7.9 & 0.51 & 7.4 & 0.22 & 1.6 & 2.3 & 0.072 & 0.41 & \NA  & 0.21 \\
$\geq$ 3 & 6.02 & 11. & 1.1 & 10. & 0.29 & 1.7 & 2.3 & 0.25 & 0.35 & \NA  & 0.46 \\
$\geq$ 4 & 1.33 & 12. & 2.7 & 11. & 0.16 & 1.7 & 2.3 & 0.65 & 0.54 & \NA  & 1.1 \\
$\geq$ 5 & 0.319 & 14. & 4.8 & 13. & 0.097 & 1.9 & 2.3 & 1.5 & 0.50 & \NA  & 2.2 \\
$\geq$ 6 & 0.045 & 24. & 15. & 17. & 0.35 & 1.8 & 2.4 & 3.5 & 1.7 & \NA  & 6.6 \\
\end{tabular}}
\label{tab:combZNGoodJets_Zinc}
\end{table*}

\begin{table*}
\centering
\topcaption{Differential cross section in $\pt(\cPZ)$  ($\Nj \geq 1$) for the combination of both decay channels and breakdown of the uncertainties.}
\cmsTable{
\begin{tabular}{cccccccccccccc}
  $\pt(\cPZ)$ & $\dd{\sigma}{\pt(\cPZ)}$ & Tot. & Stat & JES & JER & Eff & Lumi & Bkg & LES & LER & Pileup & Unf   & Unf \vspace{-0.4em}\\
              &                          & unc  &      &     &     &     &      &     &     &     &        & model & stat\\
{[{\GeV}]} & ${\scriptstyle [\frac{\text{pb}}{{\GeV}}]}$ & [\%]& [\%] & [\%] & [\%] & [\%] & [\%] & [\%] & [\%] & [\%] & [\%] & [\%] & [\%]\\
\hline
$0 \ldots 1.25$ & 0.073 & 18. & 5.4 & 16. & 0.81 & 1.6 & 2.3 & $<$0.01 & 1.2 & 0.93 & 0.22 & 5.5 & 2.2 \\
$1.25 \ldots 2.5$ & 0.212 & 14. & 3.2 & 13. & 0.89 & 1.6 & 2.3 & $<$0.01 & 0.67 & 0.37 & 0.34 & 1.9 & 1.3 \\
$2.5 \ldots 3.75$ & 0.309 & 13. & 2.7 & 13. & 0.82 & 1.5 & 2.3 & $<$0.01 & 0.55 & 0.30 & 0.17 & 1.7 & 1.1 \\
$3.75 \ldots 5$ & 0.377 & 13. & 2.4 & 13. & 0.86 & 1.6 & 2.3 & $<$0.01 & 0.73 & 0.18 & 0.43 & 1.2 & 1.0 \\
$5 \ldots 6.25$ & 0.422 & 14. & 2.3 & 13. & 0.85 & 1.5 & 2.3 & $<$0.01 & 0.55 & 0.085 & 0.50 & 1.7 & 1.1 \\
$6.25 \ldots 7.5$ & 0.487 & 13. & 2.2 & 12. & 0.88 & 1.5 & 2.3 & $<$0.01 & 0.51 & 0.11 & 0.34 & 1.8 & 1.0 \\
$7.5 \ldots 8.75$ & 0.537 & 13. & 2.1 & 12. & 0.85 & 1.5 & 2.3 & $<$0.01 & 0.57 & 0.073 & 0.30 & 2.0 & 1.0 \\
$8.75 \ldots 10$ & 0.580 & 12. & 1.9 & 12. & 0.81 & 1.6 & 2.3 & $<$0.01 & 0.62 & 0.040 & 0.24 & 2.7 & 0.93 \\
$10 \ldots 11.25$ & 0.631 & 13. & 1.9 & 12. & 0.74 & 1.6 & 2.3 & $<$0.01 & 0.67 & 0.030 & 0.29 & 3.1 & 0.91 \\
$11.25 \ldots 12.5$ & 0.697 & 12. & 1.8 & 11. & 0.81 & 1.6 & 2.3 & $<$0.01 & 0.55 & 0.11 & 0.20 & 3.2 & 0.91 \\
$12.5 \ldots 15$ & 0.757 & 12. & 1.4 & 11. & 0.89 & 1.6 & 2.3 & $<$0.01 & 0.48 & 0.098 & 0.18 & 2.8 & 0.71 \\
$15 \ldots 17.5$ & 0.87 & 12. & 1.4 & 11. & 0.86 & 1.5 & 2.3 & $<$0.01 & 0.98 & 0.093 & 0.058 & 2.2 & 0.68 \\
$17.5 \ldots 20$ & 0.98 & 12. & 1.3 & 12. & 0.87 & 1.5 & 2.3 & $<$0.01 & 0.81 & 0.085 & 0.43 & 1.1 & 0.66 \\
$20 \ldots 25$ & 1.15 & 11. & 0.87 & 11. & 0.79 & 1.6 & 2.3 & $<$0.01 & 0.67 & 0.044 & 0.19 & 1.4 & 0.43 \\
$25 \ldots 30$ & 1.47 & 11. & 0.79 & 10. & 0.54 & 1.6 & 2.3 & $<$0.01 & 0.63 & 0.017 & 0.30 & 1.4 & 0.36 \\
$30 \ldots 35$ & 1.80 & 9.3 & 0.75 & 8.6 & 0.32 & 1.5 & 2.3 & $<$0.01 & 0.50 & 0.035 & 0.45 & 1.9 & 0.32 \\
$35 \ldots 40$ & 2.03 & 7.3 & 0.69 & 6.4 & 0.11 & 1.6 & 2.3 & $<$0.01 & 0.26 & 0.055 & 0.35 & 1.7 & 0.28 \\
$40 \ldots 45$ & 2.04 & 6.0 & 0.72 & 5.0 & 0.061 & 1.6 & 2.3 & $<$0.01 & 0.11 & 0.046 & 0.38 & 1.5 & 0.29 \\
$45 \ldots 50$ & 1.908 & 4.9 & 0.74 & 3.8 & 0.028 & 1.6 & 2.3 & $<$0.01 & 0.18 & 0.034 & 0.39 & 1.0 & 0.29 \\
$50 \ldots 60$ & 1.617 & 3.9 & 0.59 & 2.5 & 0.025 & 1.5 & 2.3 & 0.012 & 0.22 & 0.039 & 0.41 & 0.74 & 0.23 \\
$60 \ldots 70$ & 1.204 & 3.4 & 0.68 & 1.6 & 0.023 & 1.6 & 2.3 & 0.018 & 0.51 & 0.031 & 0.23 & 0.53 & 0.26 \\
$70 \ldots 80$ & 0.881 & 3.2 & 0.77 & 1.0 & 0.017 & 1.6 & 2.3 & 0.024 & 0.65 & 0.055 & 0.38 & 0.52 & 0.30 \\
$80 \ldots 90$ & 0.634 & 3.3 & 0.87 & 0.64 & 0.011 & 1.6 & 2.3 & 0.028 & 0.93 & $<$0.01 & 0.25 & 0.63 & 0.35 \\
$90 \ldots 100$ & 0.444 & 3.3 & 1.0 & 0.38 & 0.022 & 1.6 & 2.3 & 0.031 & 0.80 & 0.081 & 0.36 & 0.74 & 0.42 \\
$100 \ldots 110$ & 0.333 & 3.3 & 1.2 & 0.34 & $<$0.01 & 1.6 & 2.3 & 0.026 & 0.66 & $<$0.01 & 0.25 & 0.77 & 0.48 \\
$110 \ldots 130$ & 0.2212 & 3.3 & 1.0 & 0.22 & $<$0.01 & 1.6 & 2.3 & 0.021 & 0.87 & 0.019 & 0.20 & 0.79 & 0.41 \\
$130 \ldots 150$ & 0.1308 & 3.4 & 1.3 & 0.16 & 0.010 & 1.7 & 2.3 & 0.021 & 0.88 & 0.023 & 0.073 & 0.88 & 0.54 \\
$150 \ldots 170$ & 0.0813 & 3.6 & 1.6 & 0.18 & 0.013 & 1.7 & 2.3 & 0.016 & 0.75 & 0.027 & 0.11 & 1.0 & 0.67 \\
$170 \ldots 190$ & 0.0516 & 3.9 & 2.0 & 0.13 & 0.015 & 1.8 & 2.3 & 0.022 & 0.87 & 0.017 & 0.17 & 1.1 & 0.84 \\
$190 \ldots 220$ & 0.0317 & 4.0 & 2.1 & 0.11 & $<$0.01 & 1.8 & 2.3 & 0.034 & 0.69 & 0.033 & 0.10 & 1.1 & 0.90 \\
$220 \ldots 250$ & 0.01835 & 4.5 & 2.8 & 0.028 & $<$0.01 & 1.8 & 2.3 & 0.041 & 0.82 & 0.020 & 0.11 & 1.4 & 1.2 \\
$250 \ldots 400$ & 0.00508 & 4.5 & 2.5 & 0.055 & $<$0.01 & 2.0 & 2.3 & 0.065 & 0.80 & $<$0.01 & 0.12 & 1.4 & 1.1 \\
$400 \ldots 1000$ & 0.000187 & 7.8 & 6.1 & $<$0.01 & $<$0.01 & 1.7 & 2.4 & 0.11 & 1.7 & 0.062 & 0.58 & 2.6 & 2.4 \\
\end{tabular}}
\label{tab:combZPt_Zinc1jet}
\end{table*}

\begin{table*}
\centering
\topcaption{Differential cross section in $1^{\text{st}}$ jet \pt ($\Nj \geq 1$) for the combination of both decay channels and breakdown of the uncertainties.}
\cmsTable{
\begin{tabular}{cccccccccccc}
$\pt(j_1)$ & $\dd{\sigma}{\pt(j_1)}$ & Tot. unc & Stat & JES & JER & Eff & Lumi & Bkg & Pileup & Unf model & Unf stat\\
{[{\GeV}]} & ${\scriptstyle [\frac{\text{pb}}{{\GeV}}]}$ & [\%]& [\%] & [\%] & [\%] & [\%] & [\%] & [\%] & [\%] & [\%] & [\%]\\
\hline
$30 \ldots 41$ & 3.99 & 5.9 & 0.28 & 5.1 & 0.17 & 1.5 & 2.3 & $<$0.01 & 0.39 & 0.34 & 0.11 \\
$41 \ldots 59$ & 2.07 & 5.4 & 0.35 & 4.5 & 0.18 & 1.5 & 2.3 & 0.011 & 0.33 & 0.35 & 0.13 \\
$59 \ldots 83$ & 0.933 & 5.1 & 0.45 & 4.2 & 0.17 & 1.6 & 2.3 & 0.015 & 0.25 & 0.26 & 0.18 \\
$83 \ldots 118$ & 0.377 & 5.1 & 0.59 & 4.1 & 0.20 & 1.6 & 2.3 & 0.051 & 0.28 & 0.24 & 0.24 \\
$118 \ldots 168$ & 0.1300 & 5.1 & 0.92 & 4.1 & 0.22 & 1.6 & 2.3 & 0.070 & 0.057 & 0.30 & 0.38 \\
$168 \ldots 220$ & 0.0448 & 4.9 & 1.4 & 3.8 & 0.21 & 1.6 & 2.3 & 0.077 & 0.21 & 0.30 & 0.59 \\
$220 \ldots 300$ & 0.01477 & 6.4 & 2.0 & 5.3 & 0.32 & 1.6 & 2.3 & 0.065 & 0.30 & 0.37 & 0.86 \\
$300 \ldots 400$ & 0.00390 & 7.0 & 3.4 & 5.2 & 0.24 & 1.7 & 2.3 & 0.096 & 0.28 & 0.72 & 1.4 \\
\end{tabular}}
\label{tab:combFirstJetPt_Zinc1jet}
\end{table*}

\begin{table*}
\centering
\topcaption{Differential cross section in $2^{\text{nd}}$ jet \pt ($\Nj \geq 2$) for the combination of both decay channels and breakdown of the uncertainties.}
\cmsTable{
\begin{tabular}{cccccccccccc}
$\pt(j_2)$ & $\dd{\sigma}{\pt(j_2)}$ & Tot. unc & Stat & JES & JER & Eff & Lumi & Bkg & Pileup & Unf model & Unf stat\\
{[{\GeV}]} & ${\scriptstyle [\frac{\text{pb}}{{\GeV}}]}$ & [\%]& [\%] & [\%] & [\%] & [\%] & [\%] & [\%] & [\%] & [\%] & [\%]\\
\hline
$30 \ldots 41$ & 1.125 & 8.5 & 0.56 & 7.9 & 0.22 & 1.6 & 2.3 & 0.020 & 0.51 & 0.38 & 0.24 \\
$41 \ldots 59$ & 0.457 & 7.4 & 0.73 & 6.8 & 0.13 & 1.6 & 2.3 & 0.049 & 0.33 & 0.34 & 0.31 \\
$59 \ldots 83$ & 0.173 & 6.5 & 1.1 & 5.7 & 0.16 & 1.6 & 2.3 & 0.15 & 0.31 & 0.39 & 0.44 \\
$83 \ldots 118$ & 0.0590 & 5.6 & 1.7 & 4.4 & 0.16 & 1.6 & 2.3 & 0.22 & 0.48 & 0.21 & 0.66 \\
$118 \ldots 168$ & 0.0187 & 6.0 & 2.3 & 4.7 & 0.20 & 1.7 & 2.3 & 0.25 & 0.19 & 0.13 & 0.89 \\
$168 \ldots 250$ & 0.00518 & 6.6 & 3.4 & 4.6 & 0.33 & 1.7 & 2.3 & 0.22 & 0.21 & 0.19 & 1.3 \\
\end{tabular}}
\label{tab:combSecondJetPt_Zinc2jet}
\end{table*}

\begin{table*}
\centering
\topcaption{Differential cross section in $3^{\text{rd}}$ jet \pt ($\Nj \geq 3$) for the combination of both decay channels and breakdown of the uncertainties.}
\cmsTable{
\begin{tabular}{cccccccccccc}
$\pt(j_3)$ & $\dd{\sigma}{\pt(j_3)}$ & Tot. unc & Stat & JES & JER & Eff & Lumi & Bkg & Pileup & Unf model & Unf stat\\
{[{\GeV}]} & ${\scriptstyle [\frac{\text{pb}}{{\GeV}}]}$ & [\%]& [\%] & [\%] & [\%] & [\%] & [\%] & [\%] & [\%] & [\%] & [\%]\\
\hline
$30 \ldots 41$ & 0.289 & 11. & 1.2 & 10. & 0.26 & 1.6 & 2.3 & 0.12 & 0.42 & 0.93 & 0.50 \\
$41 \ldots 59$ & 0.0972 & 9.3 & 1.8 & 8.6 & 0.14 & 1.7 & 2.3 & 0.28 & 0.41 & 1.0 & 0.72 \\
$59 \ldots 83$ & 0.0306 & 7.9 & 2.9 & 6.5 & 0.31 & 1.7 & 2.3 & 0.48 & 0.69 & 1.2 & 1.1 \\
$83 \ldots 118$ & 0.00756 & 11. & 4.7 & 8.7 & 0.46 & 1.9 & 2.3 & 0.83 & 0.74 & 0.83 & 1.7 \\
$118 \ldots 168$ & 0.00180 & 10. & 8.1 & 3.7 & 0.40 & 1.8 & 2.4 & 0.82 & 0.50 & 1.3 & 3.0 \\
$168 \ldots 250$ & 0.000342 & 17. & 14. & 6.1 & 0.20 & 1.8 & 2.3 & 0.71 & 1.5 & 2.2 & 5.3 \\
\end{tabular}}
\label{tab:combThirdJetPt_Zinc3jet}
\end{table*}

\begin{table*}
  \centering
  \topcaption{Differential cross section in $1^{\text{st}}$ jet $\vert y \vert$ ($\Nj \geq 1$) for the combination of both decay channels and breakdown of the uncertainties.}
\cmsTable{
\begin{tabular}{cccccccccccc}
$\abs{y(j_1)}$ & $\dd{\sigma}{\abs{y(j_1)}}$ & Tot. unc & Stat & JES & JER & Eff & Lumi & Bkg & Pileup & Unf model & Unf stat\\
 & [\text{pb]} & [\%]& [\%] & [\%] & [\%] & [\%] & [\%] & [\%] & [\%] & [\%] & [\%]\\
\hline
$0 \ldots 0.2$ & 70.4 & 4.9 & 0.62 & 4.0 & 0.089 & 1.5 & 2.3 & 0.015 & 0.23 & 0.11 & 0.25 \\
$0.2 \ldots 0.4$ & 69.5 & 5.0 & 0.63 & 4.1 & 0.097 & 1.5 & 2.3 & 0.015 & 0.29 & 0.14 & 0.26 \\
$0.4 \ldots 0.6$ & 66.7 & 5.0 & 0.65 & 4.1 & 0.12 & 1.5 & 2.3 & 0.014 & 0.20 & 0.14 & 0.26 \\
$0.6 \ldots 0.8$ & 64.7 & 5.2 & 0.64 & 4.3 & 0.18 & 1.6 & 2.3 & 0.014 & 0.30 & 0.15 & 0.26 \\
$0.8 \ldots 1$ & 62.3 & 5.2 & 0.68 & 4.3 & 0.087 & 1.5 & 2.3 & 0.013 & 0.20 & 0.17 & 0.28 \\
$1 \ldots 1.2$ & 57.3 & 5.1 & 0.71 & 4.2 & 0.19 & 1.5 & 2.3 & 0.012 & 0.28 & 0.24 & 0.29 \\
$1.2 \ldots 1.4$ & 52.0 & 5.4 & 0.75 & 4.6 & 0.16 & 1.5 & 2.3 & $<$0.01 & 0.29 & 0.25 & 0.31 \\
$1.4 \ldots 1.6$ & 47.8 & 6.1 & 0.77 & 5.4 & 0.087 & 1.5 & 2.3 & $<$0.01 & 0.32 & 0.31 & 0.32 \\
$1.6 \ldots 1.8$ & 43.5 & 6.3 & 0.80 & 5.6 & 0.21 & 1.5 & 2.3 & $<$0.01 & 0.34 & 0.21 & 0.34 \\
$1.8 \ldots 2$ & 38.9 & 6.7 & 0.84 & 6.0 & 0.38 & 1.5 & 2.3 & $<$0.01 & 0.41 & 0.32 & 0.36 \\
$2 \ldots 2.2$ & 34.3 & 7.2 & 0.90 & 6.5 & 0.44 & 1.5 & 2.3 & $<$0.01 & 0.62 & 0.40 & 0.39 \\
$2.2 \ldots 2.4$ & 29.5 & 7.2 & 1.0 & 6.4 & 0.66 & 1.5 & 2.3 & $<$0.01 & 0.66 & 0.36 & 0.44 \\
\end{tabular}}
\label{tab:combFirstJetAbsRapidity_Zinc1jet}
\end{table*}

\begin{table*}
\centering
  \topcaption{Differential cross section in $2^{\text{nd}}$ jet $\vert y \vert$ ($\Nj \geq 2$) for the combination of both decay channels and breakdown of the uncertainties.}
\cmsTable{
\begin{tabular}{cccccccccccc}
$\abs{y(j_2)}$ & $\dd{\sigma}{\abs{y(j_2)}}$ & Tot. unc & Stat & JES & JER & Eff & Lumi & Bkg & Pileup & Unf model & Unf stat\\
 & [\text{pb]} & [\%]& [\%] & [\%] & [\%] & [\%] & [\%] & [\%] & [\%] & [\%] & [\%]\\
\hline
$0 \ldots 0.2$ & 15.1 & 7.2 & 1.4 & 6.4 & 0.11 & 1.6 & 2.3 & 0.078 & 0.30 & 0.26 & 0.62 \\
$0.2 \ldots 0.4$ & 14.4 & 7.3 & 1.5 & 6.6 & 0.041 & 1.6 & 2.3 & 0.082 & 0.15 & 0.33 & 0.64 \\
$0.4 \ldots 0.6$ & 14.4 & 7.4 & 1.4 & 6.6 & 0.13 & 1.6 & 2.3 & 0.074 & 0.49 & 0.35 & 0.64 \\
$0.6 \ldots 0.8$ & 13.7 & 7.5 & 1.5 & 6.7 & 0.25 & 1.6 & 2.3 & 0.071 & 0.35 & 0.27 & 0.68 \\
$0.8 \ldots 1$ & 13.9 & 7.5 & 1.5 & 6.7 & 0.17 & 1.6 & 2.3 & 0.065 & 0.17 & 0.093 & 0.70 \\
$1 \ldots 1.2$ & 12.43 & 7.4 & 1.6 & 6.6 & 0.11 & 1.6 & 2.3 & 0.065 & 0.42 & 0.13 & 0.70 \\
$1.2 \ldots 1.4$ & 11.89 & 8.1 & 1.5 & 7.4 & 0.082 & 1.6 & 2.3 & 0.062 & 0.23 & 0.10 & 0.68 \\
$1.4 \ldots 1.6$ & 11.00 & 7.7 & 1.7 & 6.9 & 0.15 & 1.6 & 2.3 & 0.052 & 0.51 & 0.11 & 0.76 \\
$1.6 \ldots 1.8$ & 10.09 & 8.6 & 1.7 & 7.8 & 0.25 & 1.6 & 2.3 & 0.049 & 0.48 & 0.19 & 0.78 \\
$1.8 \ldots 2$ & 9.35 & 8.2 & 1.8 & 7.4 & 0.33 & 1.6 & 2.3 & 0.043 & 0.65 & 0.44 & 0.84 \\
$2 \ldots 2.2$ & 8.48 & 8.6 & 1.8 & 7.8 & 0.48 & 1.6 & 2.3 & 0.035 & 0.50 & 0.67 & 0.85 \\
$2.2 \ldots 2.4$ & 7.04 & 9.3 & 2.0 & 8.4 & 0.37 & 1.6 & 2.3 & 0.037 & 0.93 & 1.2 & 0.96 \\
\end{tabular}}
\label{tab:combSecondJetAbsRapidity_Zinc2jet}
\end{table*}

\begin{table*}
\centering
\topcaption{Differential cross section in $3^{\text{rd}}$ jet $\abs{y}$ ($\Nj \geq 3$) for the combination of both decay channels and breakdown of the uncertainties.}
\cmsTable{
\begin{tabular}{cccccccccccc}
$\abs{y(j_3)}$ & $\dd{\sigma}{\abs{y(j_3)}}$ & Tot. unc & Stat & JES & JER & Eff & Lumi & Bkg & Pileup & Unf model & Unf stat\\
 & [\text{pb]} & [\%]& [\%] & [\%] & [\%] & [\%] & [\%] & [\%] & [\%] & [\%] & [\%]\\
\hline
$0 \ldots 0.3$ & 3.14 & 9.9 & 2.5 & 9.0 & 0.26 & 1.7 & 2.3 & 0.27 & 0.28 & 0.15 & 1.1 \\
$0.3 \ldots 0.6$ & 3.02 & 10. & 2.6 & 9.4 & 0.13 & 1.7 & 2.3 & 0.27 & 0.31 & 0.088 & 1.1 \\
$0.6 \ldots 0.9$ & 3.06 & 9.6 & 2.6 & 8.7 & 0.20 & 1.6 & 2.3 & 0.25 & 0.20 & 0.012 & 1.2 \\
$0.9 \ldots 1.2$ & 2.70 & 9.5 & 2.7 & 8.5 & 0.22 & 1.7 & 2.3 & 0.25 & 0.22 & 0.34 & 1.2 \\
$1.2 \ldots 1.5$ & 2.51 & 12. & 2.8 & 11. & 0.14 & 1.6 & 2.3 & 0.23 & 0.59 & 0.78 & 1.3 \\
$1.5 \ldots 1.8$ & 2.21 & 11. & 3.1 & 10. & 0.17 & 1.6 & 2.3 & 0.22 & 0.13 & 0.62 & 1.4 \\
$1.8 \ldots 2.1$ & 1.89 & 13. & 3.1 & 12. & 0.13 & 1.7 & 2.3 & 0.22 & 0.57 & 1.8 & 1.4 \\
$2.1 \ldots 2.4$ & 1.70 & 11. & 3.4 & 10. & 0.66 & 1.7 & 2.3 & 0.21 & 0.87 & 2.4 & 1.6 \\
\end{tabular}}
\label{tab:combThirdJetAbsRapidity_Zinc3jet}
\end{table*}

\begin{table*}
\centering
\topcaption{Differential cross section in \HT ($\Nj \geq 1$) for the combination of both decay channels and breakdown of the uncertainties.}
\cmsTable{
\begin{tabular}{cccccccccccc}
\HT & $\dd{\sigma}{\HT}$ & Tot. unc & Stat & JES & JER & Eff & Lumi & Bkg & Pileup & Unf model & Unf stat\\
{[{\GeV}]} & ${\scriptstyle [\frac{\text{pb}}{{\GeV}}]}$ & [\%]& [\%] & [\%] & [\%] & [\%] & [\%] & [\%] & [\%] & [\%] & [\%]\\
\hline
$30 \ldots 41$ & 3.71 & 5.9 & 0.41 & 5.1 & 0.18 & 1.5 & 2.3 & $<$0.01 & 0.38 & 0.92 & 0.19 \\
$41 \ldots 59$ & 1.678 & 4.7 & 0.53 & 3.6 & 0.16 & 1.5 & 2.3 & $<$0.01 & 0.26 & 1.1 & 0.21 \\
$59 \ldots 83$ & 0.852 & 5.3 & 0.66 & 4.4 & 0.23 & 1.5 & 2.3 & $<$0.01 & 0.30 & 0.62 & 0.26 \\
$83 \ldots 118$ & 0.449 & 6.0 & 0.74 & 5.3 & 0.13 & 1.6 & 2.3 & 0.015 & 0.34 & 0.54 & 0.30 \\
$118 \ldots 168$ & 0.199 & 5.9 & 0.92 & 5.1 & 0.20 & 1.6 & 2.3 & 0.040 & 0.18 & 0.41 & 0.38 \\
$168 \ldots 220$ & 0.0886 & 6.3 & 1.5 & 5.4 & 0.36 & 1.6 & 2.3 & 0.078 & 0.35 & 0.33 & 0.61 \\
$220 \ldots 300$ & 0.0373 & 6.9 & 1.6 & 6.0 & 0.10 & 1.7 & 2.3 & 0.14 & 0.20 & 0.17 & 0.66 \\
$300 \ldots 400$ & 0.0148 & 6.8 & 2.3 & 5.6 & 0.21 & 1.6 & 2.3 & 0.20 & 0.18 & 0.21 & 0.98 \\
$400 \ldots 550$ & 0.00449 & 7.3 & 3.2 & 5.7 & 0.20 & 1.8 & 2.3 & 0.36 & 0.63 & 0.28 & 1.3 \\
$550 \ldots 780$ & 0.00133 & 8.1 & 5.3 & 4.8 & 0.13 & 1.6 & 2.3 & 0.40 & 1.2 & 0.24 & 2.1 \\
$780 \ldots 1100$ & 0.000306 & 12. & 8.2 & 7.5 & 0.22 & 1.8 & 2.3 & 0.59 & 0.69 & 0.56 & 3.2 \\
\end{tabular}}
\label{tab:combJetsHT_Zinc1jet}
\end{table*}

\begin{table*}
\centering
\topcaption{Differential cross section in \HT ($\Nj \geq 2$) for the combination of both decay channels and breakdown of the uncertainties.}
\cmsTable{
\begin{tabular}{cccccccccccc}
\HT & $\dd{\sigma}{\HT}$ & Tot. unc & Stat & JES & JER & Eff & Lumi & Bkg & Pileup & Unf model & Unf stat\\
{[{\GeV}]} & ${\scriptstyle [\frac{\text{pb}}{{\GeV}}]}$ & [\%]& [\%] & [\%] & [\%] & [\%] & [\%] & [\%] & [\%] & [\%] & [\%]\\
\hline
$60 \ldots 83$ & 0.208 & 9.5 & 1.1 & 8.9 & 0.25 & 1.5 & 2.3 & 0.023 & 0.63 & 1.0 & 0.67 \\
$83 \ldots 118$ & 0.228 & 7.9 & 0.89 & 7.3 & 0.15 & 1.6 & 2.3 & 0.027 & 0.45 & 0.59 & 0.42 \\
$118 \ldots 168$ & 0.1371 & 6.8 & 0.96 & 6.0 & 0.18 & 1.6 & 2.3 & 0.030 & 0.32 & 0.58 & 0.42 \\
$168 \ldots 220$ & 0.0705 & 7.3 & 1.4 & 6.6 & 0.29 & 1.6 & 2.3 & 0.10 & 0.36 & 0.31 & 0.57 \\
$220 \ldots 300$ & 0.0329 & 7.1 & 1.6 & 6.2 & 0.11 & 1.7 & 2.3 & 0.16 & 0.18 & 0.29 & 0.64 \\
$300 \ldots 400$ & 0.01360 & 6.8 & 2.2 & 5.7 & 0.20 & 1.6 & 2.3 & 0.22 & 0.33 & 0.29 & 0.90 \\
$400 \ldots 550$ & 0.00436 & 7.3 & 3.1 & 5.8 & 0.18 & 1.8 & 2.3 & 0.36 & 0.56 & 0.28 & 1.2 \\
$550 \ldots 780$ & 0.00129 & 8.1 & 5.0 & 5.1 & 0.17 & 1.6 & 2.3 & 0.41 & 1.1 & 0.21 & 1.9 \\
$780 \ldots 1100$ & 0.000304 & 12. & 7.9 & 7.2 & 0.25 & 1.7 & 2.3 & 0.58 & 0.65 & 0.41 & 3.1 \\
\end{tabular}}
\label{tab:combJetsHT_Zinc2jet}
\end{table*}

\begin{table*}
\centering
\topcaption{Differential cross section in \HT ($\Nj \geq 3$) for the combination of both decay channels and breakdown of the uncertainties.}
\cmsTable{
\begin{tabular}{cccccccccccc}
\HT & $\dd{\sigma}{\HT}$ & Tot. unc & Stat & JES & JER & Eff & Lumi & Bkg & Pileup & Unf model & Unf stat\\
{[{\GeV}]} & ${\scriptstyle [\frac{\text{pb}}{{\GeV}}]}$ & [\%]& [\%] & [\%] & [\%] & [\%] & [\%] & [\%] & [\%] & [\%] & [\%]\\
\hline
$90 \ldots 130$ & 0.0166 & 17. & 3.5 & 15. & 0.64 & 1.6 & 2.3 & 0.013 & 0.61 & 5.4 & 2.3 \\
$130 \ldots 168$ & 0.0300 & 12. & 2.5 & 11. & 0.10 & 1.7 & 2.3 & 0.097 & 0.35 & 1.8 & 1.2 \\
$168 \ldots 220$ & 0.0254 & 11. & 2.8 & 9.7 & 0.088 & 1.7 & 2.3 & 0.20 & 0.46 & 0.75 & 1.2 \\
$220 \ldots 300$ & 0.0163 & 9.3 & 2.4 & 8.4 & 0.27 & 1.7 & 2.3 & 0.28 & 0.21 & 0.73 & 1.0 \\
$300 \ldots 400$ & 0.00841 & 8.4 & 3.1 & 7.2 & 0.13 & 1.7 & 2.3 & 0.36 & 0.26 & 0.43 & 1.3 \\
$400 \ldots 550$ & 0.00307 & 8.9 & 3.9 & 7.2 & 0.22 & 1.8 & 2.3 & 0.53 & 0.72 & 0.40 & 1.5 \\
$550 \ldots 780$ & 0.00103 & 10. & 6.3 & 6.8 & 0.33 & 1.7 & 2.3 & 0.53 & 1.1 & 0.22 & 2.5 \\
$780 \ldots 1100$ & 0.000246 & 12. & 9.1 & 6.5 & 0.17 & 1.7 & 2.3 & 0.67 & 0.88 & 2.7 & 3.5 \\
\end{tabular}}
\label{tab:combJetsHT_Zinc3jet}
\end{table*}

\begin{table*}
\centering
\topcaption{Differential cross section in \ptbal ($\Nj \geq 1$) for the combination of both decay channels and breakdown of the uncertainties.}
\cmsTable{
\begin{tabular}{cccccccccccc}
\ptbal & $\dd{\sigma}{\ptbal}$ & Tot. unc & Stat & JES & JER & Eff & Lumi & Bkg & Pileup & Unf model & Unf stat\\
{[{\GeV}]} & ${\scriptstyle [\frac{\text{pb}}{{\GeV}}]}$ & [\%]& [\%] & [\%] & [\%] & [\%] & [\%] & [\%] & [\%] & [\%] & [\%]\\
\hline
$0 \ldots 10$ & 2.65 & 6.0 & 0.45 & 5.2 & 0.42 & 1.5 & 2.3 & $<$0.01 & 0.45 & 1.1 & 0.18 \\
$10 \ldots 20$ & 3.53 & 6.1 & 0.36 & 5.3 & 0.28 & 1.5 & 2.3 & $<$0.01 & 0.40 & 1.2 & 0.14 \\
$20 \ldots 35$ & 2.35 & 6.3 & 0.37 & 5.1 & 0.38 & 1.6 & 2.3 & $<$0.01 & 0.31 & 2.2 & 0.15 \\
$35 \ldots 50$ & 1.116 & 6.0 & 0.53 & 4.1 & 0.69 & 1.6 & 2.3 & 0.023 & 0.30 & 3.2 & 0.23 \\
$50 \ldots 65$ & 0.467 & 4.4 & 0.87 & 2.2 & 0.77 & 1.6 & 2.3 & 0.053 & 0.092 & 2.0 & 0.39 \\
$65 \ldots 80$ & 0.208 & 5.0 & 1.2 & 1.0 & 0.85 & 1.9 & 2.3 & 0.17 & 0.33 & 3.5 & 0.54 \\
$80 \ldots 100$ & 0.0883 & 5.1 & 1.8 & 1.6 & 0.81 & 2.0 & 2.4 & 0.37 & 0.62 & 2.9 & 0.75 \\
$100 \ldots 125$ & 0.0344 & 6.9 & 2.7 & 2.9 & 0.66 & 2.2 & 2.4 & 0.62 & 0.42 & 4.2 & 1.1 \\
$125 \ldots 150$ & 0.0154 & 7.5 & 4.1 & 4.3 & 0.57 & 2.1 & 2.4 & 0.69 & 0.54 & 2.6 & 1.6 \\
$150 \ldots 175$ & 0.00686 & 12. & 6.1 & 7.7 & 0.23 & 2.2 & 2.4 & 0.76 & 0.67 & 4.5 & 2.3 \\
$175 \ldots 200$ & 0.00357 & 12. & 8.0 & 5.2 & 0.82 & 2.3 & 2.5 & 0.71 & 0.51 & 4.7 & 2.9 \\
\end{tabular}}
\label{tab:combVisPt_Zinc1jetQun}
\end{table*}

\begin{table*}
\centering
\topcaption{Differential cross section in \ptbal ($\Nj \geq 2$) for the combination of both decay channels and breakdown of the uncertainties.}
\cmsTable{
\begin{tabular}{cccccccccccc}
\ptbal & $\dd{\sigma}{\ptbal}$ & Tot. unc & Stat & JES & JER & Eff & Lumi & Bkg & Pileup & Unf model & Unf stat\\
{[{\GeV}]} & ${\scriptstyle [\frac{\text{pb}}{{\GeV}}]}$ & [\%]& [\%] & [\%] & [\%] & [\%] & [\%] & [\%] & [\%] & [\%] & [\%]\\
\hline
$0 \ldots 15$ & 0.522 & 8.7 & 0.70 & 8.2 & 0.38 & 1.5 & 2.3 & 0.027 & 0.56 & 0.40 & 0.32 \\
$15 \ldots 30$ & 0.635 & 8.1 & 0.56 & 7.5 & 0.29 & 1.6 & 2.3 & 0.023 & 0.48 & 0.97 & 0.26 \\
$30 \ldots 45$ & 0.372 & 6.6 & 0.75 & 5.7 & 0.48 & 1.6 & 2.3 & 0.040 & 0.38 & 1.4 & 0.35 \\
$45 \ldots 60$ & 0.178 & 6.3 & 1.0 & 5.4 & 0.94 & 1.6 & 2.3 & 0.14 & 0.24 & 0.87 & 0.47 \\
$60 \ldots 80$ & 0.0738 & 6.7 & 1.4 & 5.0 & 1.2 & 1.9 & 2.3 & 0.35 & 0.29 & 2.6 & 0.60 \\
$80 \ldots 100$ & 0.0308 & 7.3 & 2.3 & 5.2 & 1.3 & 2.2 & 2.4 & 0.75 & 0.34 & 2.7 & 0.91 \\
$100 \ldots 125$ & 0.0133 & 8.7 & 3.7 & 5.4 & 1.3 & 2.2 & 2.4 & 1.1 & 0.60 & 4.0 & 1.4 \\
$125 \ldots 150$ & 0.00682 & 12. & 5.1 & 9.0 & 0.98 & 2.5 & 2.4 & 1.3 & 0.59 & 4.1 & 1.9 \\
$150 \ldots 175$ & 0.00352 & 14. & 7.3 & 10. & 0.15 & 2.6 & 2.4 & 1.4 & 0.22 & 5.1 & 2.6 \\
$175 \ldots 200$ & 0.00182 & 15. & 9.5 & 10. & 0.33 & 2.3 & 2.5 & 1.2 & 0.78 & 4.3 & 3.0 \\
\end{tabular}}
\label{tab:combVisPt_Zinc2jetQun}
\end{table*}

\begin{table*}
\centering
\topcaption{Differential cross section in \ptbal ($\Nj \geq 3$) for the combination of both decay channels and breakdown of the uncertainties.}
\cmsTable{
\begin{tabular}{cccccccccccc}
\ptbal & $\dd{\sigma}{\ptbal}$ & Tot. unc & Stat & JES & JER & Eff & Lumi & Bkg & Pileup & Unf model & Unf stat\\
{[{\GeV}]} & ${\scriptstyle [\frac{\text{pb}}{{\GeV}}]}$ & [\%]& [\%] & [\%] & [\%] & [\%] & [\%] & [\%] & [\%] & [\%] & [\%]\\
\hline
$0 \ldots 20$ & 0.102 & 12. & 1.8 & 11. & 0.71 & 1.5 & 2.3 & 0.044 & 0.57 & 4.4 & 0.78 \\
$20 \ldots 40$ & 0.106 & 11. & 1.4 & 9.9 & 0.61 & 1.6 & 2.3 & 0.095 & 0.29 & 2.8 & 0.66 \\
$40 \ldots 65$ & 0.0483 & 9.3 & 2.2 & 7.8 & 1.2 & 1.7 & 2.3 & 0.33 & 0.32 & 3.0 & 1.0 \\
$65 \ldots 90$ & 0.0160 & 8.5 & 4.0 & 4.8 & 1.4 & 2.1 & 2.4 & 1.1 & 0.16 & 4.1 & 1.7 \\
$90 \ldots 120$ & 0.00580 & 13. & 7.1 & 8.3 & 1.9 & 2.3 & 2.4 & 2.0 & 0.61 & 4.6 & 2.9 \\
$120 \ldots 150$ & 0.00243 & 23. & 13. & 16. & 0.81 & 2.6 & 2.4 & 2.8 & 1.5 & 6.8 & 5.0 \\
$150 \ldots 175$ & 0.00127 & 26. & 18. & 16. & 1.3 & 2.6 & 2.4 & 2.9 & 0.96 & 4.3 & 6.7 \\
$175 \ldots 200$ & 0.00079 & 26. & 20. & 9.9 & 1.8 & 2.8 & 2.5 & 3.1 & 0.41 & 8.5 & 7.4 \\
\end{tabular}}
\label{tab:combVisPt_Zinc3jetQun}
\end{table*}

\begin{table*}
\centering
\topcaption{Differential cross section in \JZB (full phase space) for the combination of both decay channels and breakdown of the uncertainties.}
\cmsTable{
\begin{tabular}{cccccccccccc}
\JZB & $\dd{\sigma}{\JZB}$ & Tot. unc & Stat & JES & JER & Eff & Lumi & Bkg & Pileup & Unf model & Unf stat\\
{[{\GeV}]} & ${\scriptstyle [\frac{\text{pb}}{{\GeV}}]}$ & [\%]& [\%] & [\%] & [\%] & [\%] & [\%] & [\%] & [\%] & [\%] & [\%]\\
\hline
$-140 \ldots -105$ & 0.00274 & 17. & 11. & 9.8 & 1.3 & 1.6 & 2.4 & 0.10 & 1.6 & 6.4 & 4.8 \\
$-105 \ldots -80$ & 0.0115 & 11. & 6.3 & 7.0 & 0.66 & 1.7 & 2.4 & 0.12 & 0.64 & 2.5 & 2.9 \\
$-80 \ldots -60$ & 0.0388 & 15. & 3.7 & 11. & 0.73 & 1.7 & 2.4 & 0.061 & 0.82 & 5.7 & 1.7 \\
$-60 \ldots -40$ & 0.153 & 14. & 2.0 & 11. & 0.73 & 1.7 & 2.3 & 0.047 & 0.59 & 7.0 & 0.90 \\
$-40 \ldots -20$ & 0.658 & 9.0 & 0.96 & 6.7 & 1.3 & 1.7 & 2.3 & 0.012 & 0.53 & 4.7 & 0.40 \\
$-20 \ldots 0$ & 2.45 & 8.0 & 0.43 & 6.9 & 0.54 & 1.6 & 2.3 & $<$0.01 & 0.46 & 2.8 & 0.17 \\
$0 \ldots 20$ & 2.16 & 5.1 & 0.58 & 3.6 & 0.64 & 2.1 & 2.3 & $<$0.01 & 0.17 & 1.3 & 0.24 \\
$20 \ldots 40$ & 0.69 & 15. & 0.89 & 14. & 1.5 & 1.6 & 2.3 & 0.027 & 0.41 & 5.4 & 0.38 \\
$40 \ldots 60$ & 0.142 & 11. & 2.1 & 9.5 & 1.4 & 1.7 & 2.3 & 0.18 & 0.34 & 3.9 & 0.92 \\
$60 \ldots 85$ & 0.0356 & 13. & 3.9 & 11. & 1.9 & 1.9 & 2.4 & 0.55 & 1.0 & 2.6 & 1.6 \\
$85 \ldots 110$ & 0.0114 & 14. & 7.3 & 9.1 & 0.83 & 2.1 & 2.4 & 0.93 & 2.0 & 5.7 & 3.0 \\
$110 \ldots 140$ & 0.0053 & 19. & 11. & 12. & 0.66 & 2.4 & 2.5 & 1.1 & 1.5 & 8.0 & 4.4 \\
\end{tabular}}
\label{tab:combJZB}
\end{table*}

\begin{table*}
\centering
\topcaption{Differential cross section in \JZB ($\pt(\cPZ)<50$ GeV) for the combination of both decay channels and breakdown of the uncertainties.}
\cmsTable{
\begin{tabular}{cccccccccccc}
\JZB & $\dd{\sigma}{\JZB}$ & Tot. unc & Stat & JES & JER & Eff & Lumi & Bkg & Pileup & Unf model & Unf stat\\
{[{\GeV}]} & ${\scriptstyle [\frac{\text{pb}}{{\GeV}}]}$ & [\%]& [\%] & [\%] & [\%] & [\%] & [\%] & [\%] & [\%] & [\%] & [\%]\\
\hline
$-50 \ldots -30$ & 0.00859 & 7.8 & 5.8 & 1.9 & 1.2 & 1.8 & 2.3 & 0.042 & 0.92 & 2.5 & 2.6 \\
$-30 \ldots -15$ & 0.1212 & 5.1 & 2.1 & 1.5 & 2.6 & 2.3 & 2.3 & 0.042 & 0.19 & 0.61 & 1.0 \\
$-15 \ldots 0$ & 1.30 & 8.0 & 0.52 & 6.9 & 0.26 & 1.7 & 2.3 & $<$0.01 & 0.55 & 2.9 & 0.23 \\
$0 \ldots 15$ & 1.63 & 12. & 0.44 & 11. & 0.51 & 1.6 & 2.3 & $<$0.01 & 0.32 & 3.1 & 0.19 \\
$15 \ldots 30$ & 0.83 & 14. & 0.65 & 13. & 1.3 & 1.6 & 2.3 & 0.013 & 0.34 & 3.4 & 0.29 \\
$30 \ldots 50$ & 0.219 & 11. & 1.2 & 11. & 1.4 & 1.6 & 2.3 & 0.036 & 0.15 & 1.2 & 0.50 \\
$50 \ldots 75$ & 0.0410 & 11. & 2.6 & 9.2 & 1.4 & 1.8 & 2.3 & 0.29 & 0.39 & 4.6 & 1.1 \\
$75 \ldots 105$ & 0.0097 & 13. & 5.4 & 9.6 & 0.63 & 2.3 & 2.4 & 0.89 & 1.0 & 6.1 & 2.2 \\
$105 \ldots 150$ & 0.00241 & 14. & 10. & 6.3 & 1.4 & 2.4 & 2.4 & 1.3 & 0.87 & 5.1 & 3.8 \\
\end{tabular}}
\label{tab:combJZB_ptLow}
\end{table*}

\begin{table*}
\centering
\topcaption{Differential cross section in \JZB ($\pt(\cPZ)>50$ GeV) for the combination of both decay channels and breakdown of the uncertainties.}
\cmsTable{
\begin{tabular}{cccccccccccc}
\JZB & $\dd{\sigma}{\JZB}$ & Tot. unc & Stat & JES & JER & Eff & Lumi & Bkg & Pileup & Unf model & Unf stat\\
{[{\GeV}]} & ${\scriptstyle [\frac{\text{pb}}{{\GeV}}]}$ & [\%]& [\%] & [\%] & [\%] & [\%] & [\%] & [\%] & [\%] & [\%] & [\%]\\
\hline
$-165 \ldots -125$ & 0.00165 & 11. & 8.8 & 1.6 & 0.32 & 1.7 & 2.4 & 0.15 & 0.87 & 3.3 & 5.0 \\
$-125 \ldots -95$ & 0.00475 & 8.8 & 6.2 & 2.8 & 1.3 & 1.9 & 2.4 & 0.14 & 0.46 & 2.0 & 3.4 \\
$-95 \ldots -70$ & 0.0182 & 19. & 3.6 & 16. & 0.64 & 1.8 & 2.4 & 0.12 & 0.30 & 5.2 & 2.0 \\
$-70 \ldots -45$ & 0.091 & 14. & 1.4 & 13. & 0.36 & 1.6 & 2.3 & 0.052 & 0.58 & 3.5 & 0.78 \\
$-45 \ldots -20$ & 0.551 & 6.1 & 0.63 & 3.8 & 0.71 & 1.6 & 2.3 & 0.011 & 0.28 & 1.0 & 0.33 \\
$-20 \ldots 0$ & 1.404 & 5.3 & 0.38 & 4.4 & 0.13 & 1.5 & 2.3 & $<$0.01 & 0.43 & 0.33 & 0.18 \\
$0 \ldots 25$ & 0.607 & 4.9 & 0.62 & 3.4 & 0.92 & 2.1 & 2.3 & 0.021 & 0.30 & 1.1 & 0.30 \\
$25 \ldots 55$ & 0.090 & 19. & 1.3 & 18. & 2.3 & 1.7 & 2.3 & 0.14 & 0.43 & 3.5 & 0.68 \\
$55 \ldots 85$ & 0.0162 & 19. & 3.5 & 14. & 2.4 & 2.0 & 2.4 & 0.52 & 0.93 & 11. & 1.8 \\
$85 \ldots 120$ & 0.00454 & 18. & 6.9 & 14. & 3.2 & 2.0 & 2.4 & 0.79 & 1.8 & 8.1 & 3.3 \\
$120 \ldots 150$ & 0.00195 & 21. & 11. & 14. & 1.2 & 2.3 & 2.6 & 1.3 & 1.8 & 9.4 & 5.0 \\
\end{tabular}}
\label{tab:combJZB_ptHigh}
\end{table*}

\section{Results}
\label{results}

The measurements from the electron and muon channels are found to be consistent and are combined using a weighted average as described in Ref.~\cite{Khachatryan:2016crw}. For each bin of the measured differential cross sections, the results of each of the two measurements are weighted by the inverse of the squared total uncertainty. The covariance matrix of the combination, the diagonal elements of which are used to extract the measurement uncertainties, is computed assuming full correlation between the two channels for all the sources of uncertainty sources except the statistical uncertainties and those associated with lepton reconstruction and identification, which are taken to be uncorrelated. The integrated cross section is measured for different exclusive and inclusive multiplicities and the results are shown in Tables~\ref{tab:combZNGoodJets_Zexc} and~\ref{tab:combZNGoodJets_Zinc}.

\begin{figure*}
  \centering
  \includegraphics[width=0.48\textwidth]{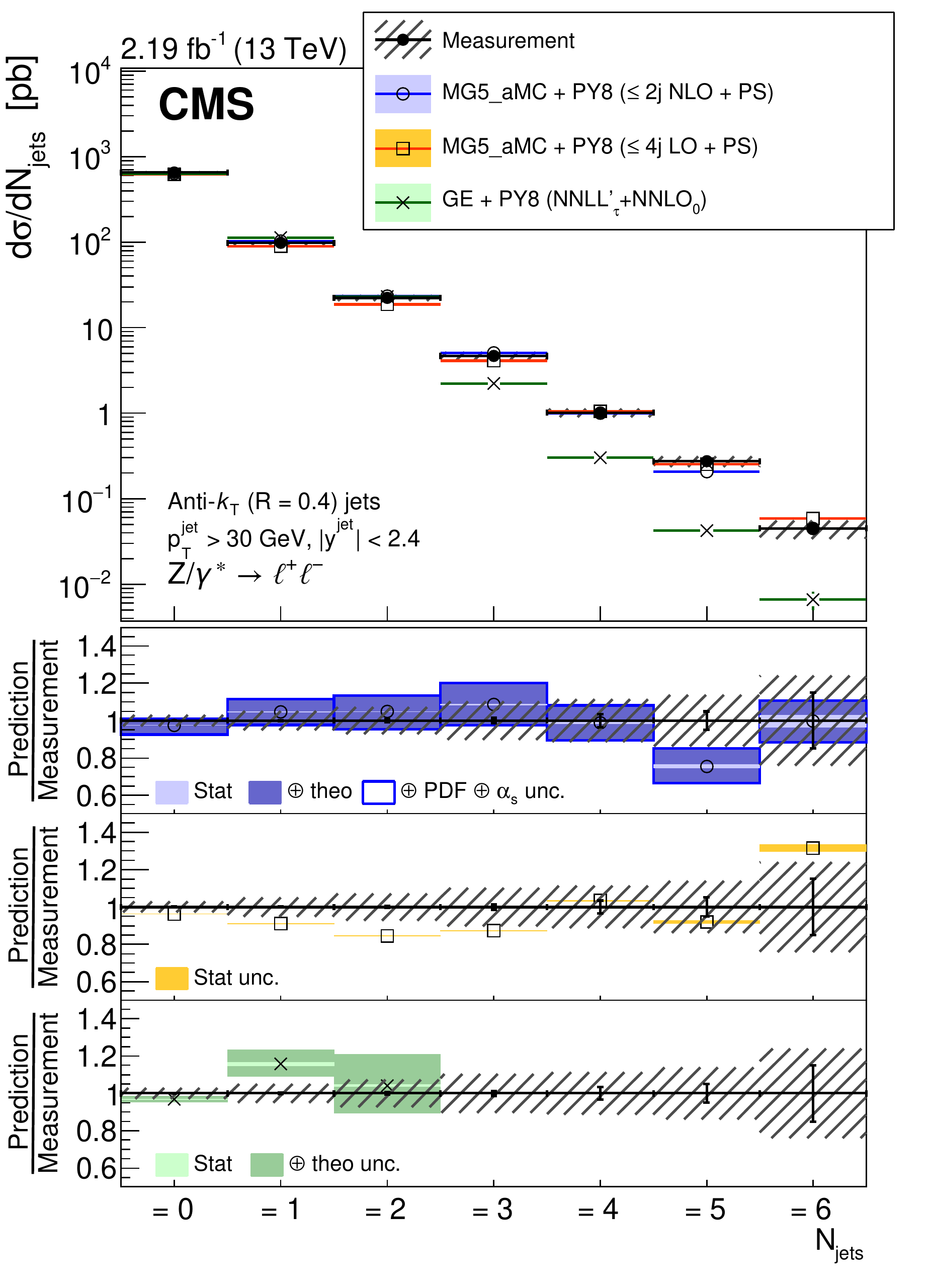}
  \includegraphics[width=0.48\textwidth]{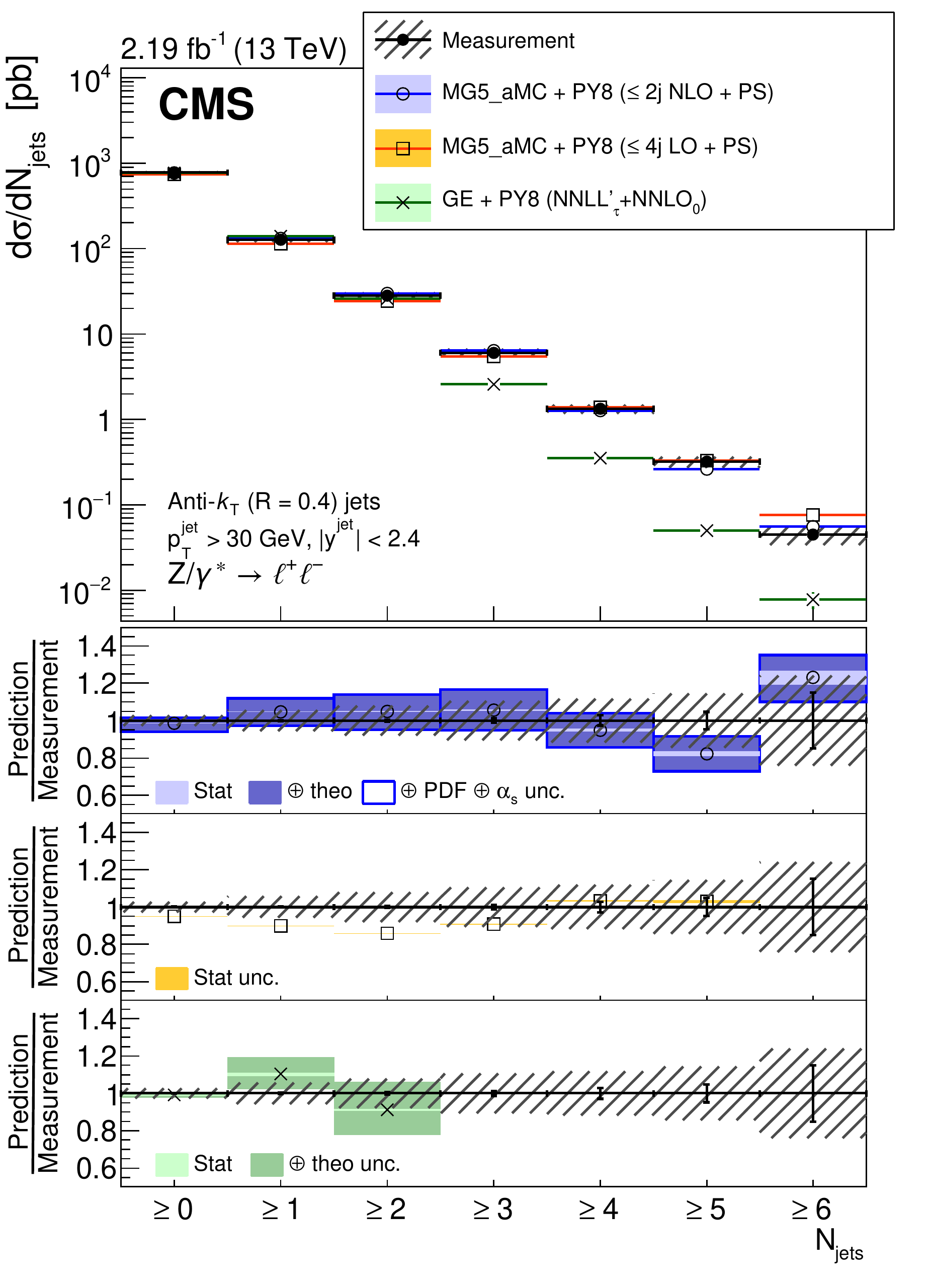}
  \caption{Measured cross section for $\cPZ+\text{ jets}$ as a function of the jet exclusive (left) and inclusive (right) multiplicity. The error bars represent the statistical uncertainty and the grey hatched bands represent the total uncertainty, including the systematic and statistical components. The measurement is compared with different predictions, which are described in the text. The ratio of each prediction to the measurement is shown together with the measurement statistical (black bars) and total (black hatched bands) uncertainties and the prediction (coloured bands) uncertainties. Different uncertainties were considered for the predictions: statistical (stat), ME calculation (theo), and PDF together with the strong coupling constant (\alpS). The complete set was computed for one of the predictions. These uncertainties were added together in quadrature (represented by the $\oplus$ sign in the legend).}
  \label{fig:sigNjet}
\end{figure*}

\begin{figure}
  \centering
  \includegraphics[width=0.48\textwidth]{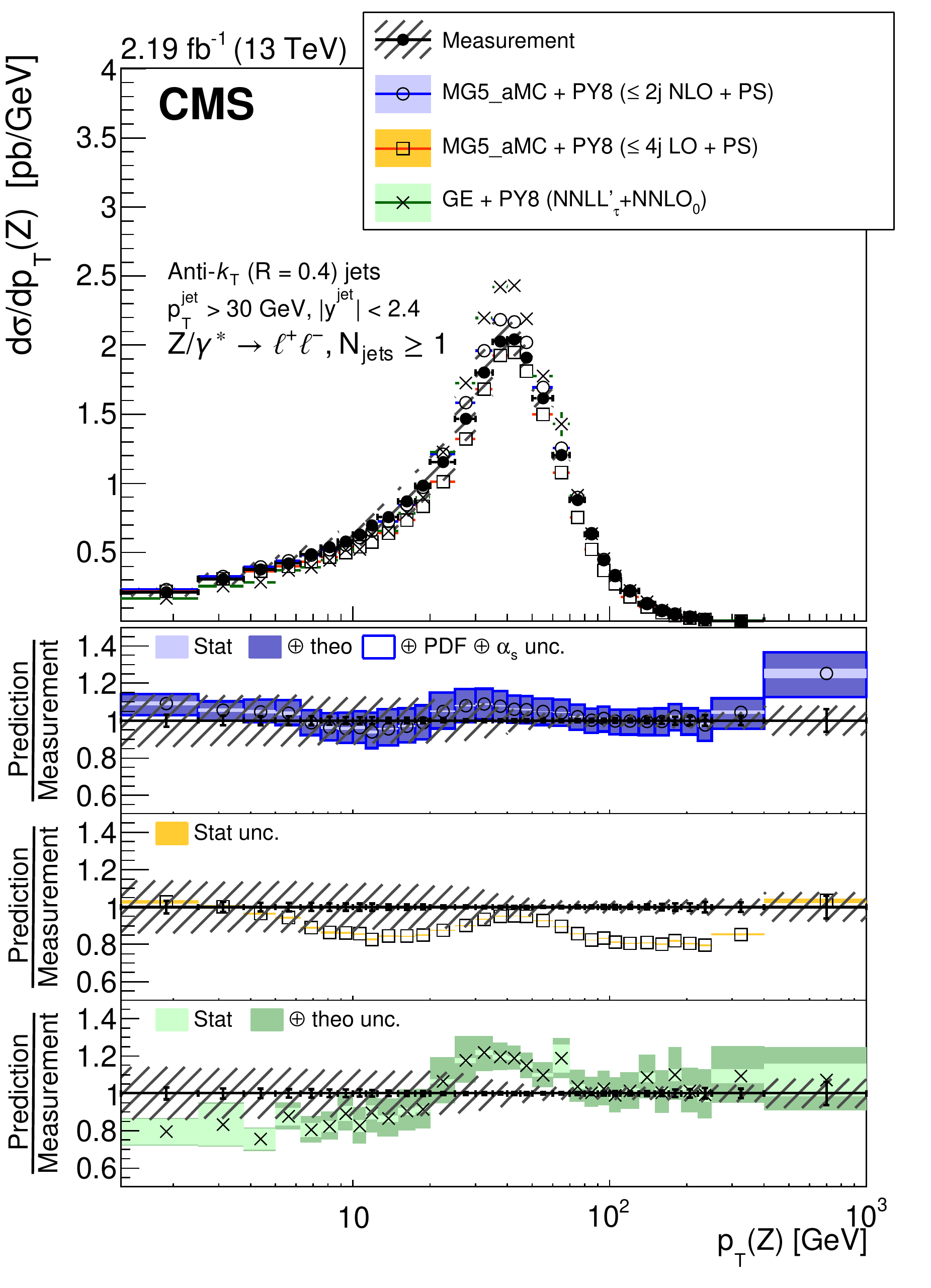}
  \caption{Measured cross section for $\cPZ+\text{ jets}$ as a function of the transverse momentum of the $\cPZ$ boson for events with at least one jet. \citefigfour .}
  \label{fig:sigZPt1j}
\end{figure}

\begin{figure}
  \centering
  \includegraphics[width=0.48\textwidth]{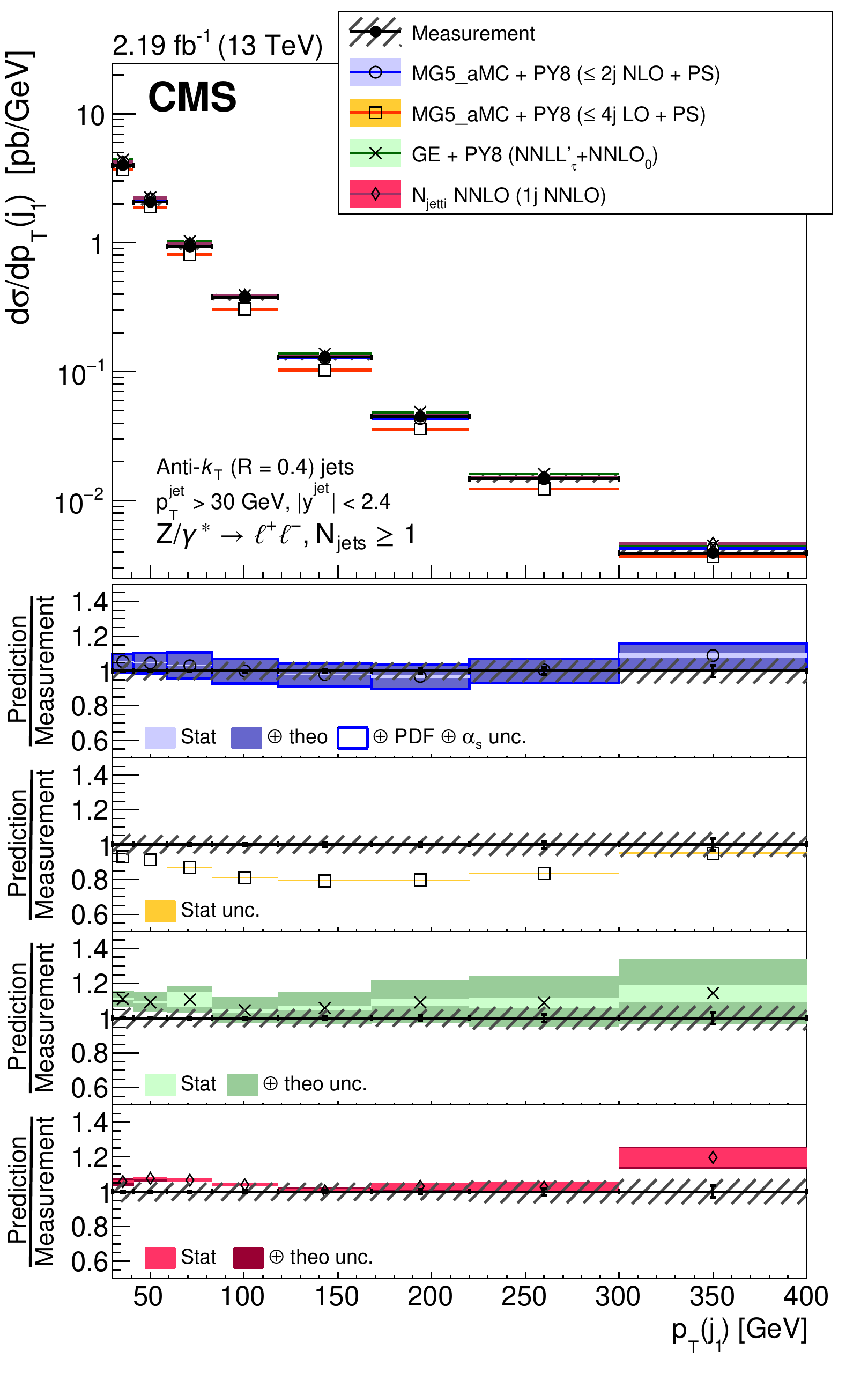}
  \caption{Measured cross section for $\cPZ+\text{ jets}$ as a function of the transverse momentum of the first jet. \citefigfour .}
  \label{fig:sigPtjet_a}
\end{figure}

\begin{figure}
  \centering
  \raisebox{-\height}{\includegraphics[width=0.48\textwidth]{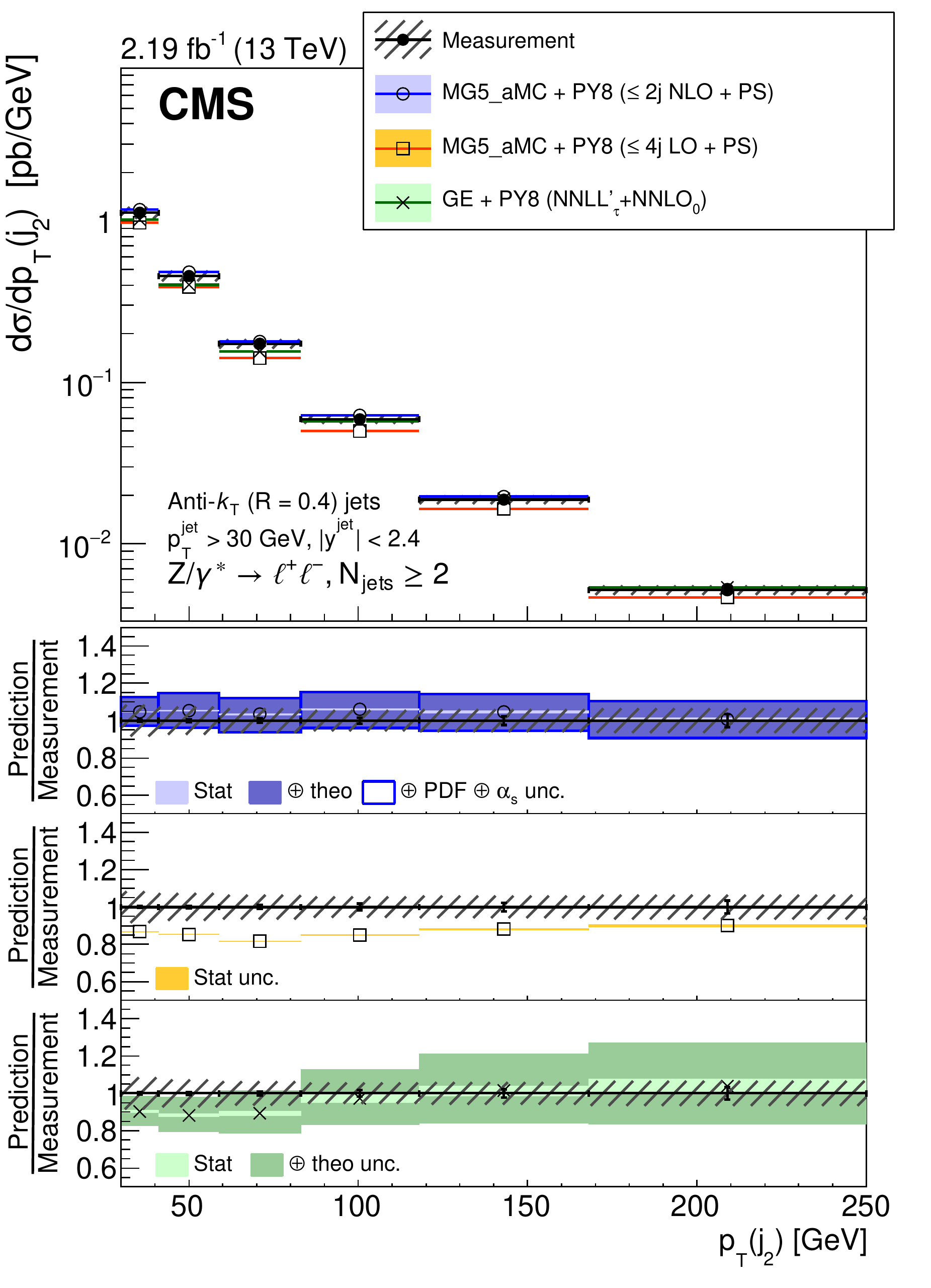}}
  \raisebox{-\height}{\includegraphics[width=0.48\textwidth]{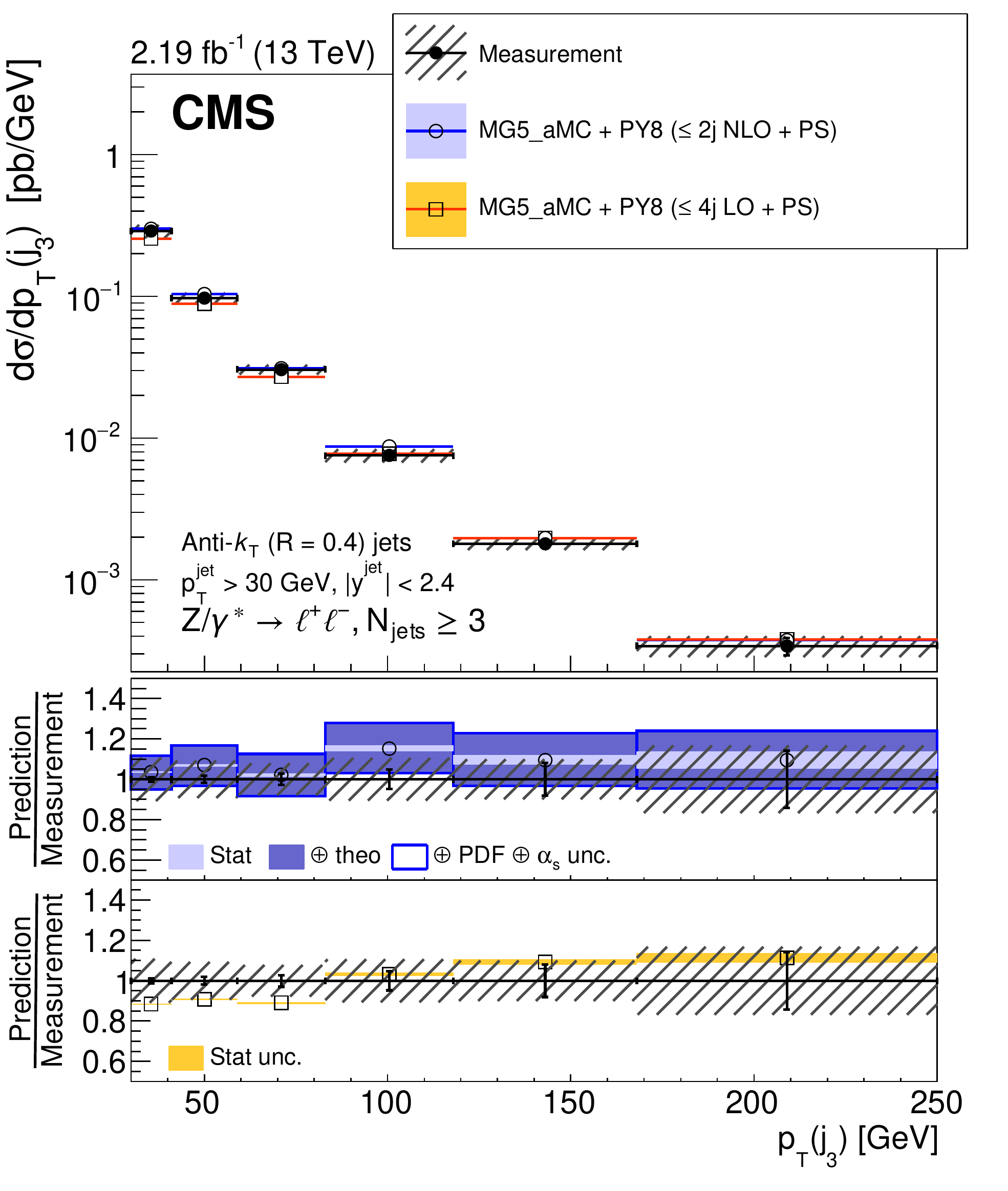}}
  \caption{Measured cross section for $\cPZ+\text{ jets}$ as a function of the transverse momentum of the second (\cmsLeft) and third (\cmsRight) jet. \citefigfour .}
  \label{fig:sigPtjet_b}
\end{figure}

\begin{figure}
  \centering
  \raisebox{-\height}{\includegraphics[width=0.43\textwidth]{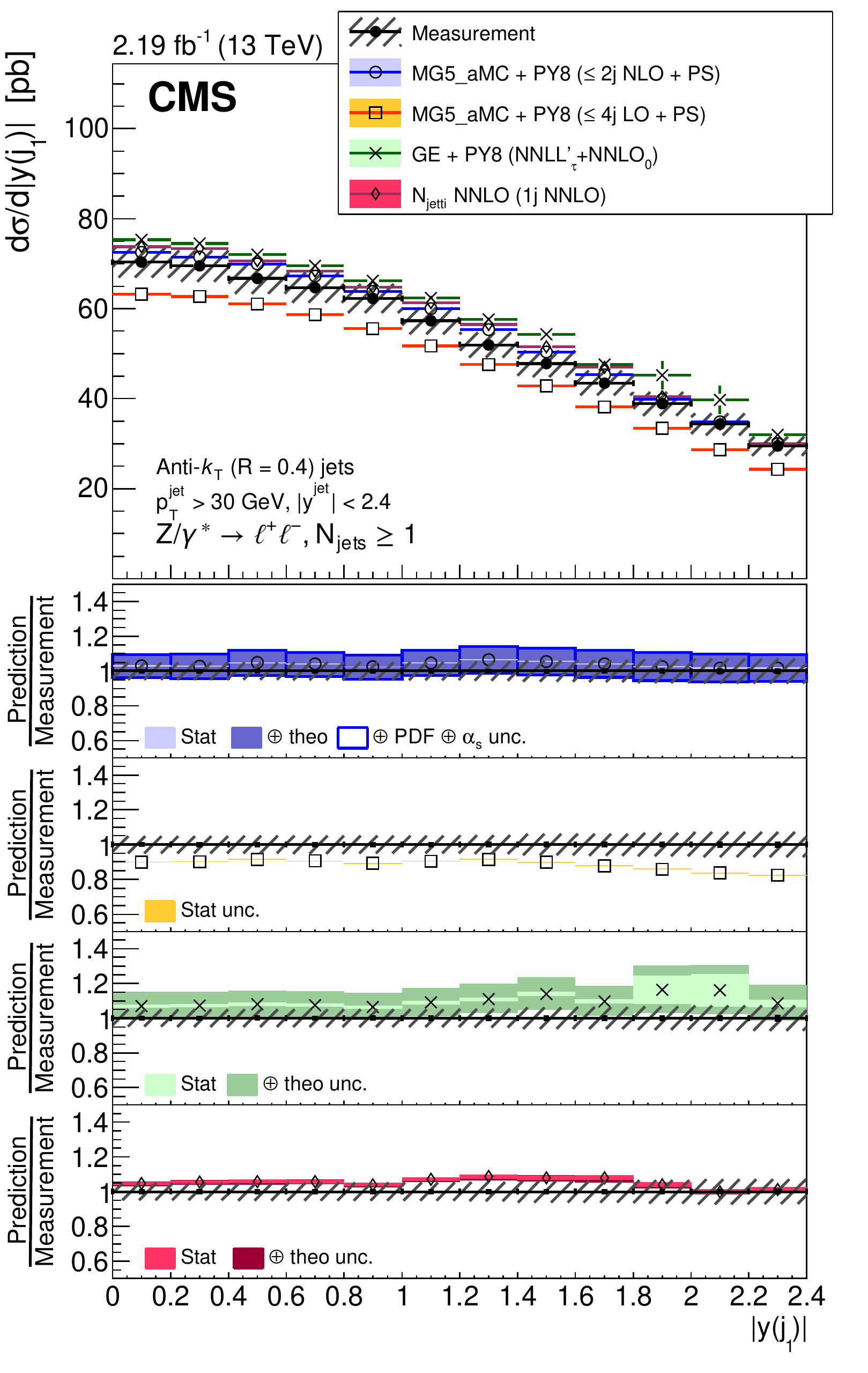}}
  \raisebox{-\height}{\includegraphics[width=0.43\textwidth]{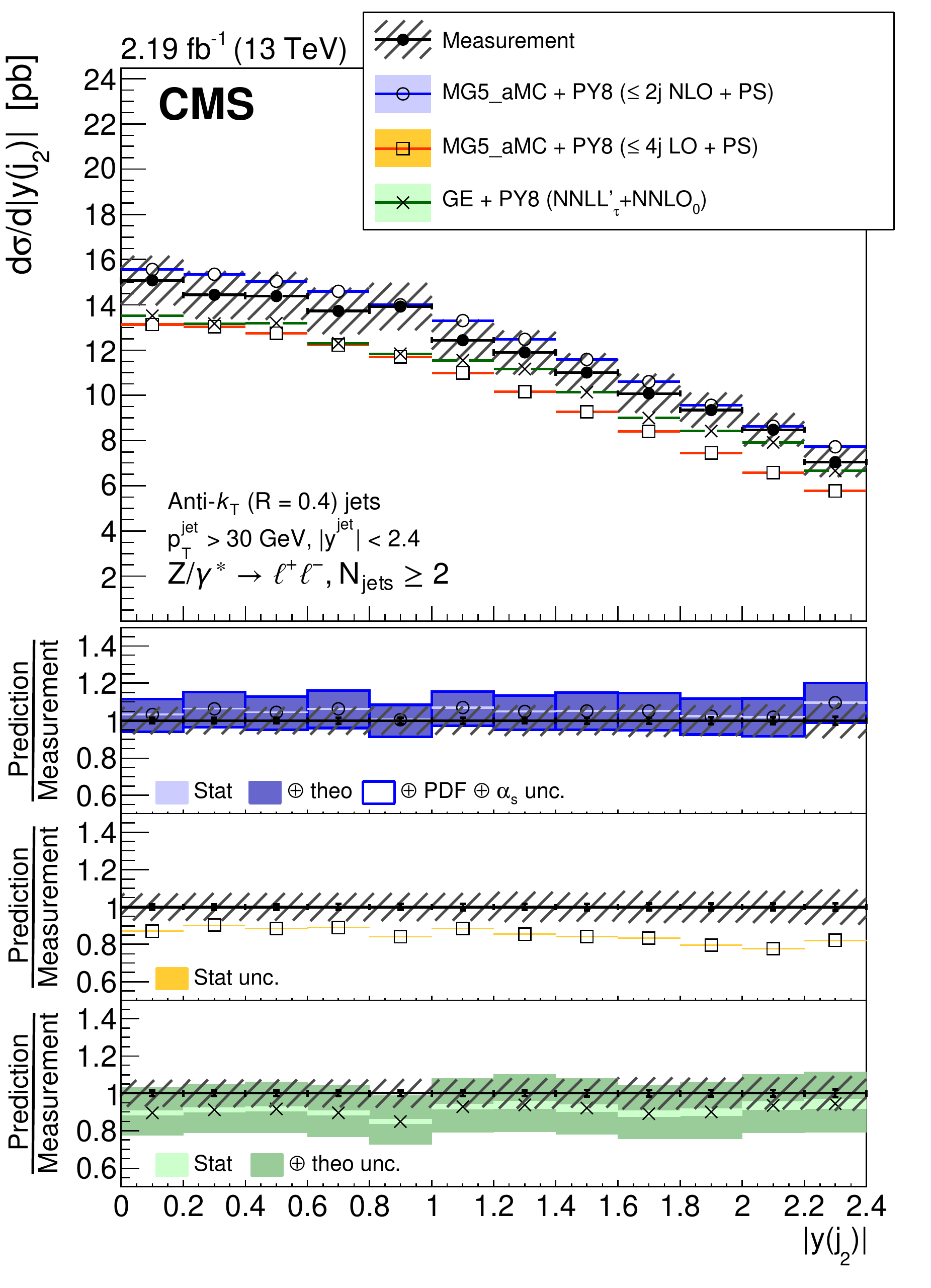}}
  \caption{Measured cross section for $\cPZ+\text{ jets}$ as a function of the absolute rapidity of the first (\cmsLeft) and second (\cmsRight) jet. \citefigfour .}
  \label{fig:sigEtajet_a}
\end{figure}

\begin{figure}
  \centering
  \includegraphics[width=0.43\textwidth]{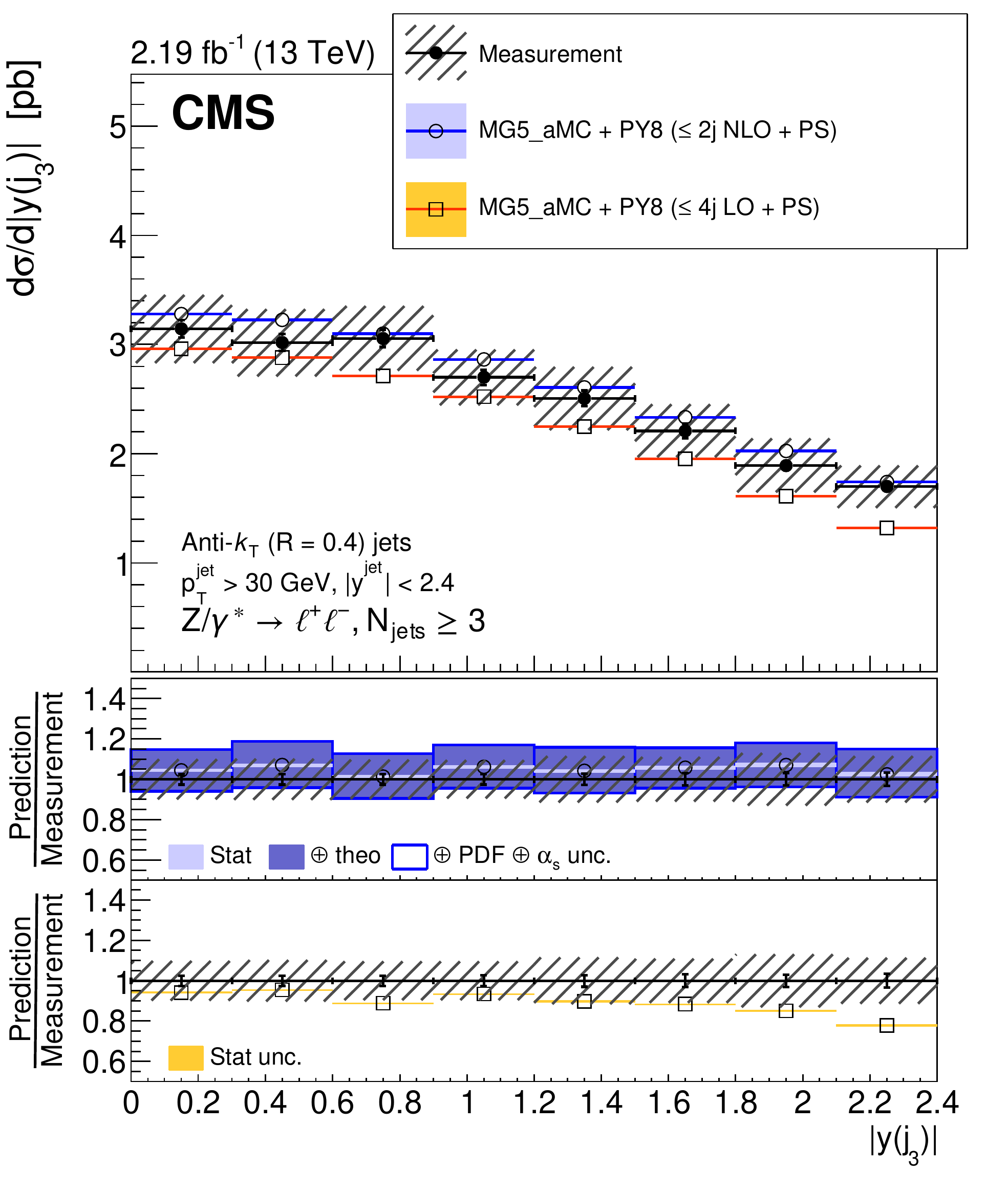}
  \caption{Measured cross section for $\cPZ+\text{ jets}$ as a function of the absolute rapidity of the third jet. \citefigfour .}
  \label{fig:sigEtajet_b}
\end{figure}

\begin{figure}
  \centering
  \includegraphics[width=0.4\textwidth]{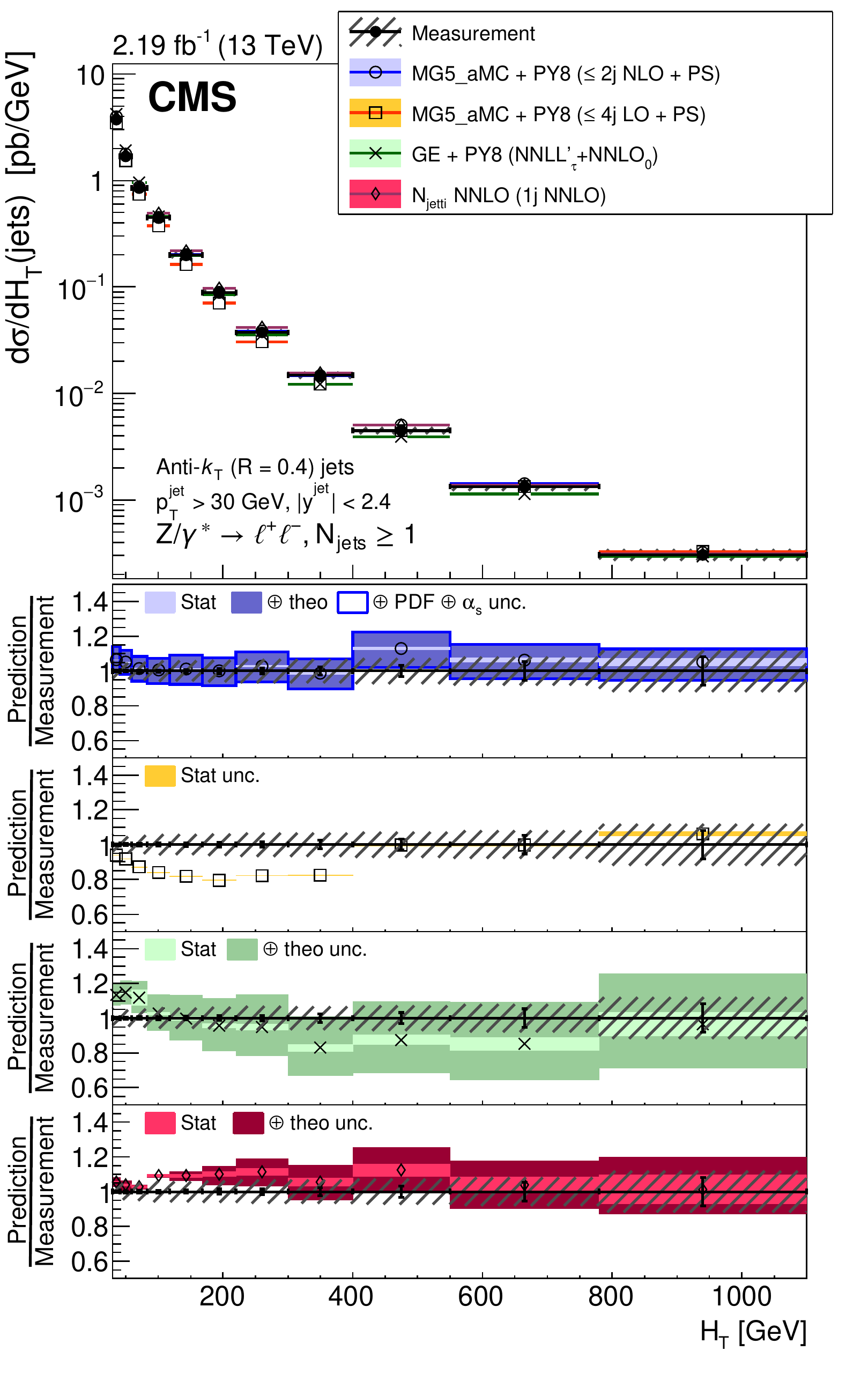}
  \caption{Measured cross section for $\cPZ+\text{ jets}$ as a function of the \HT observable for events with at least one jet. \citefigfour .}
  \label{fig:sigHtjet_a}
\end{figure}

\begin{figure}
  \centering
  \raisebox{-\height}{\includegraphics[width=0.4\textwidth]{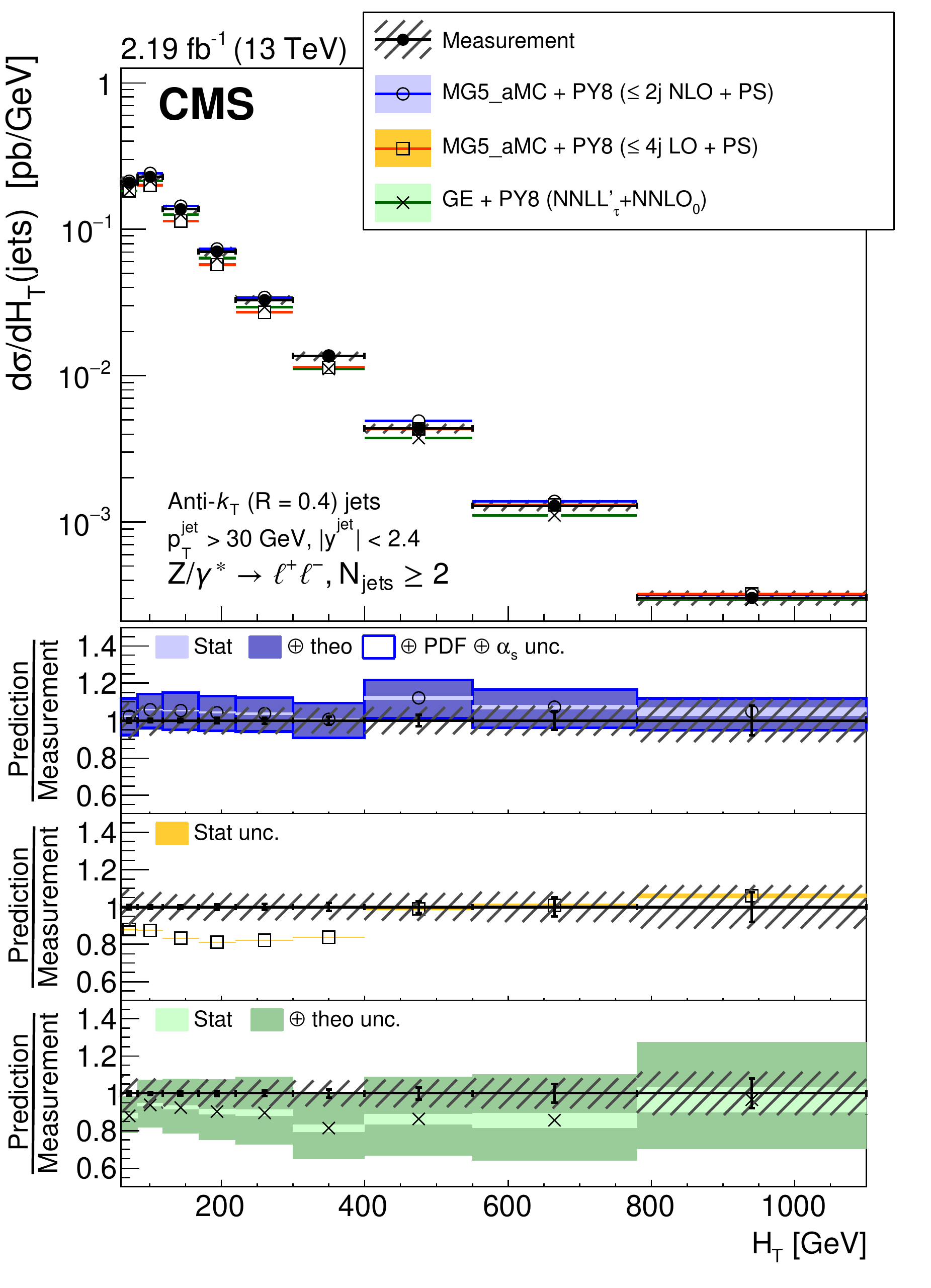}}
  \raisebox{-\height}{\includegraphics[width=0.4\textwidth]{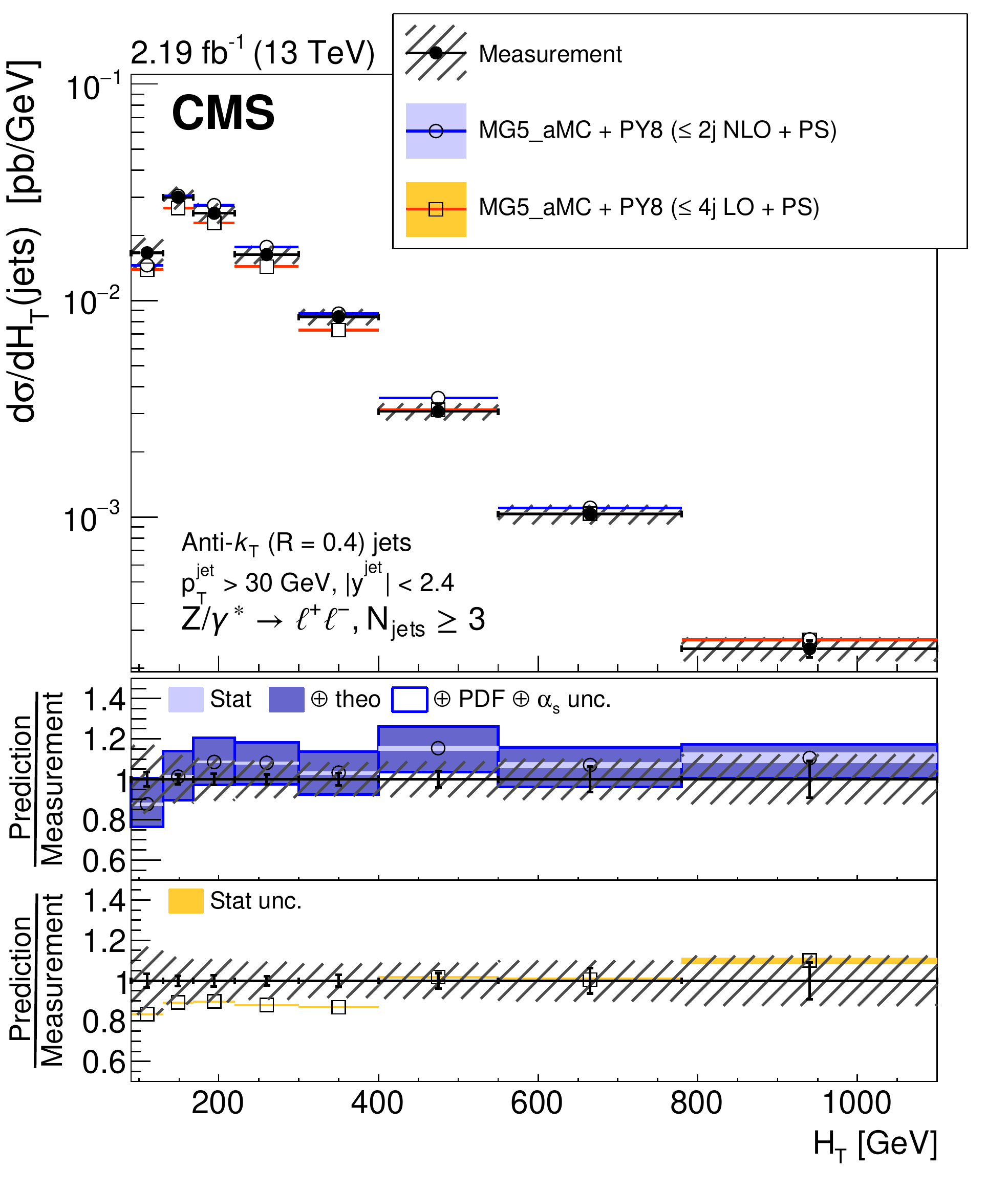}}
  \caption{Measured cross section for $\cPZ+\text{ jets}$ as a function of the \HT observable of jets for events with at least two (\cmsLeft) and three (\cmsRight) jets. \citefigfour .}
  \label{fig:sigHtjet_b}
\end{figure}

\begin{figure}
  \centering
  \raisebox{-\height}{\includegraphics[width=0.4\textwidth]{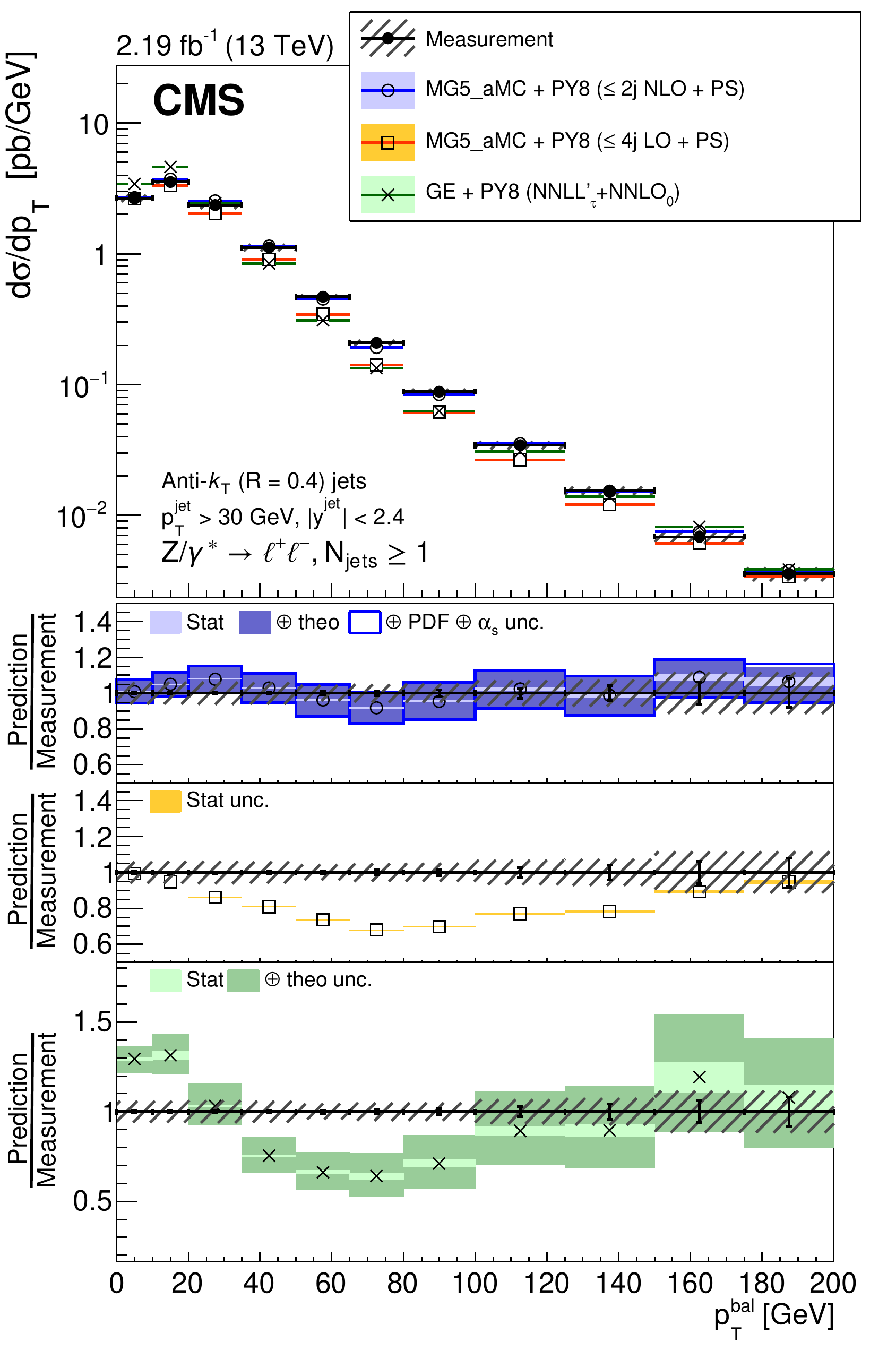}}
  \raisebox{-\height}{\includegraphics[width=0.4\textwidth]{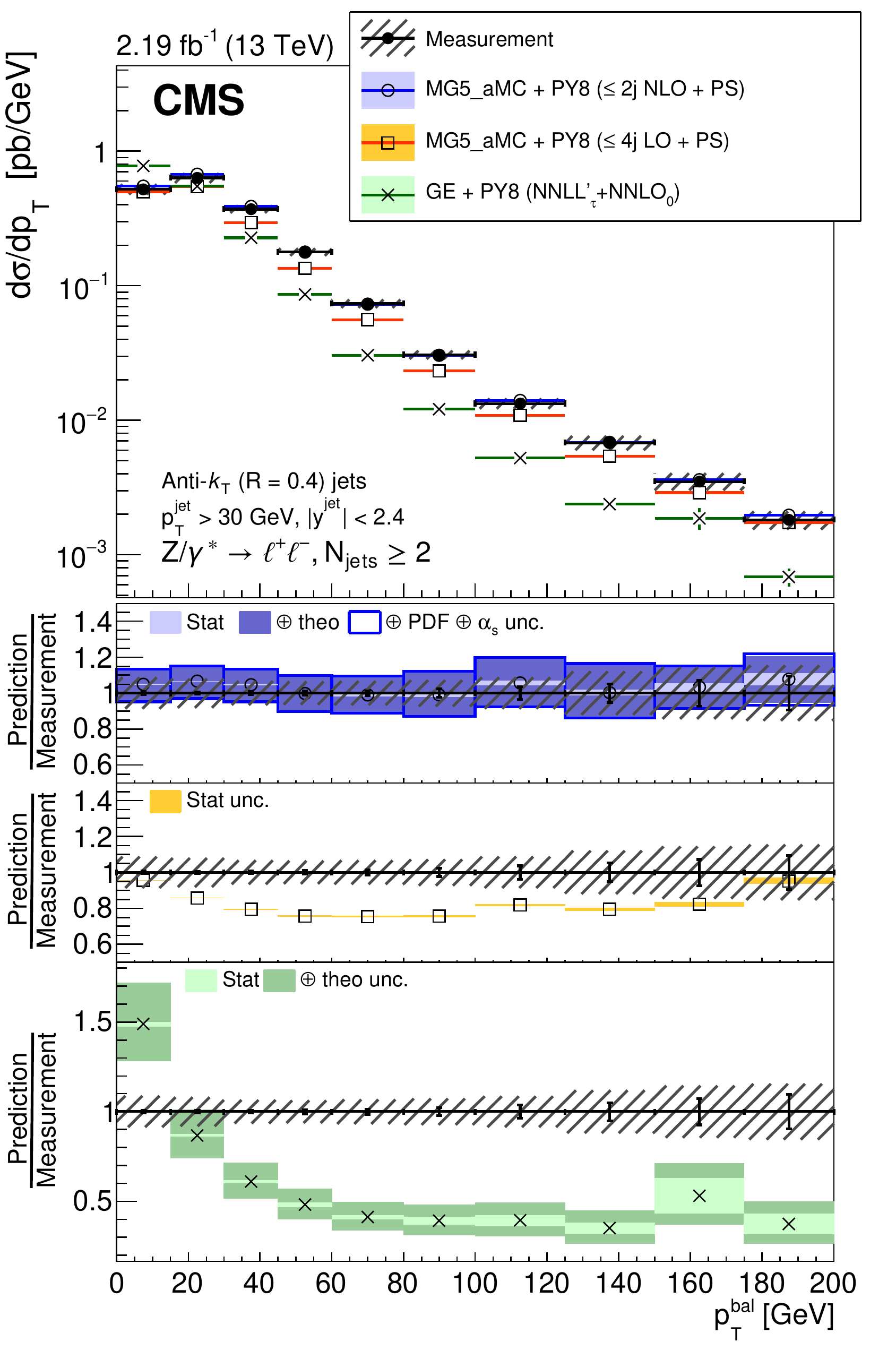}}
  \caption{Measured cross section for $\cPZ+\text{ jets}$ as a function of the transverse momentum balance between the $\cPZ$ boson and the accompanying jets for events with at least one (\cmsLeft) and two (\cmsRight) jets. \citefigfour .}
  \label{fig:ptbalmumu_a}
\end{figure}

\begin{figure}
  \centering
  \raisebox{-\height}{\includegraphics[width=0.4\textwidth]{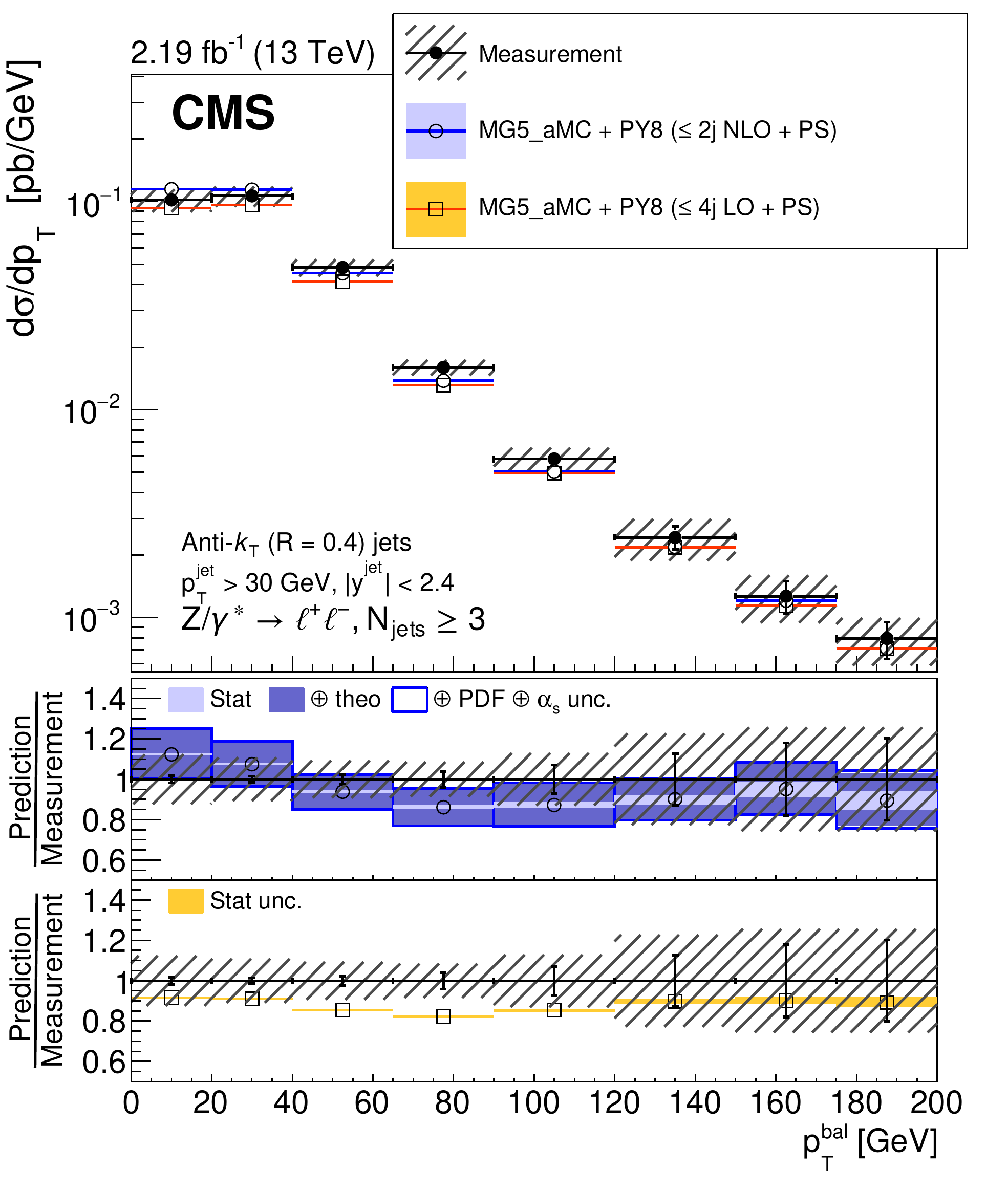}}
  \caption{Measured cross section for $\cPZ+\text{ jets}$ as a function of the transverse momentum balance between the $\cPZ$ boson and the accompanying jets for events with at least three jets. \citefigfour .}
  \label{fig:ptbalmumu_b}
\end{figure}

\begin{figure}
  \centering
  \includegraphics[width=0.4\textwidth]{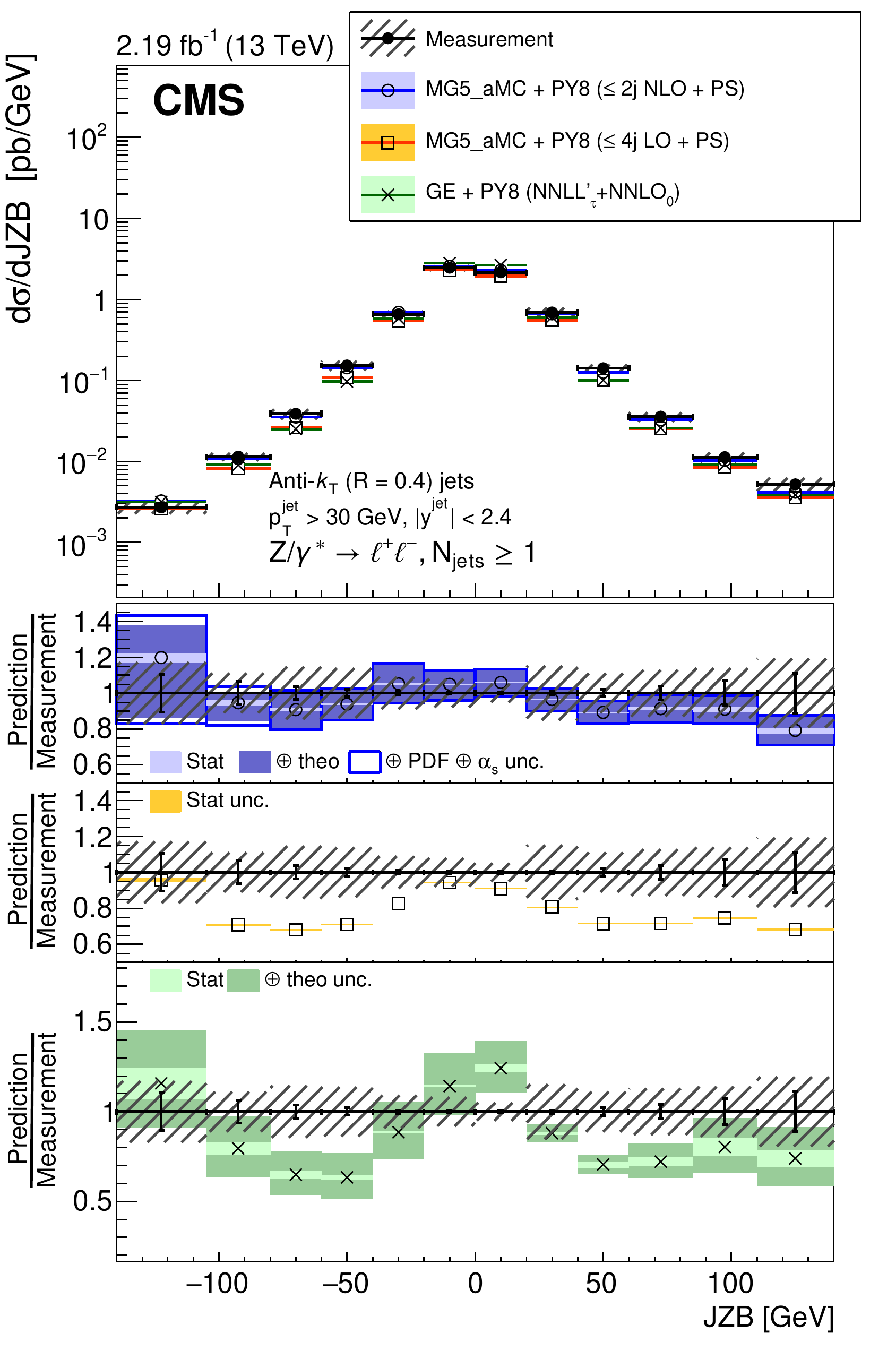}
  \caption{Measured cross section for $\cPZ+\text{ jets}$ as a function of the \JZB variable (see text), with no restriction on $\pt(\cPZ)$. \citefigfour .}
  \label{fig:JZBmumu_a}
\end{figure}

\begin{figure*}
  \centering
  \raisebox{-\height}{\includegraphics[width=0.4\textwidth]{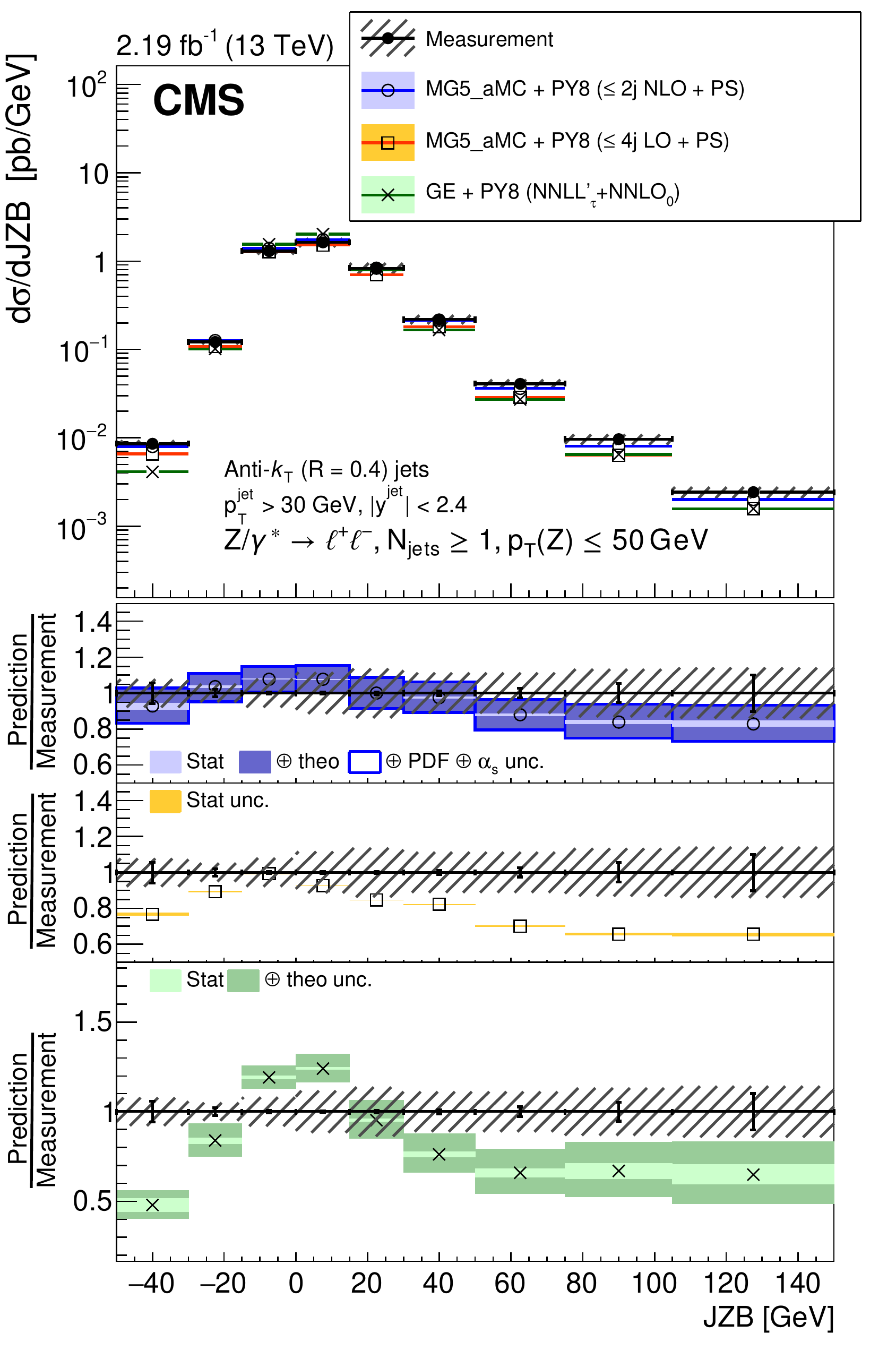}}
  \raisebox{-\height}{\includegraphics[width=0.4\textwidth]{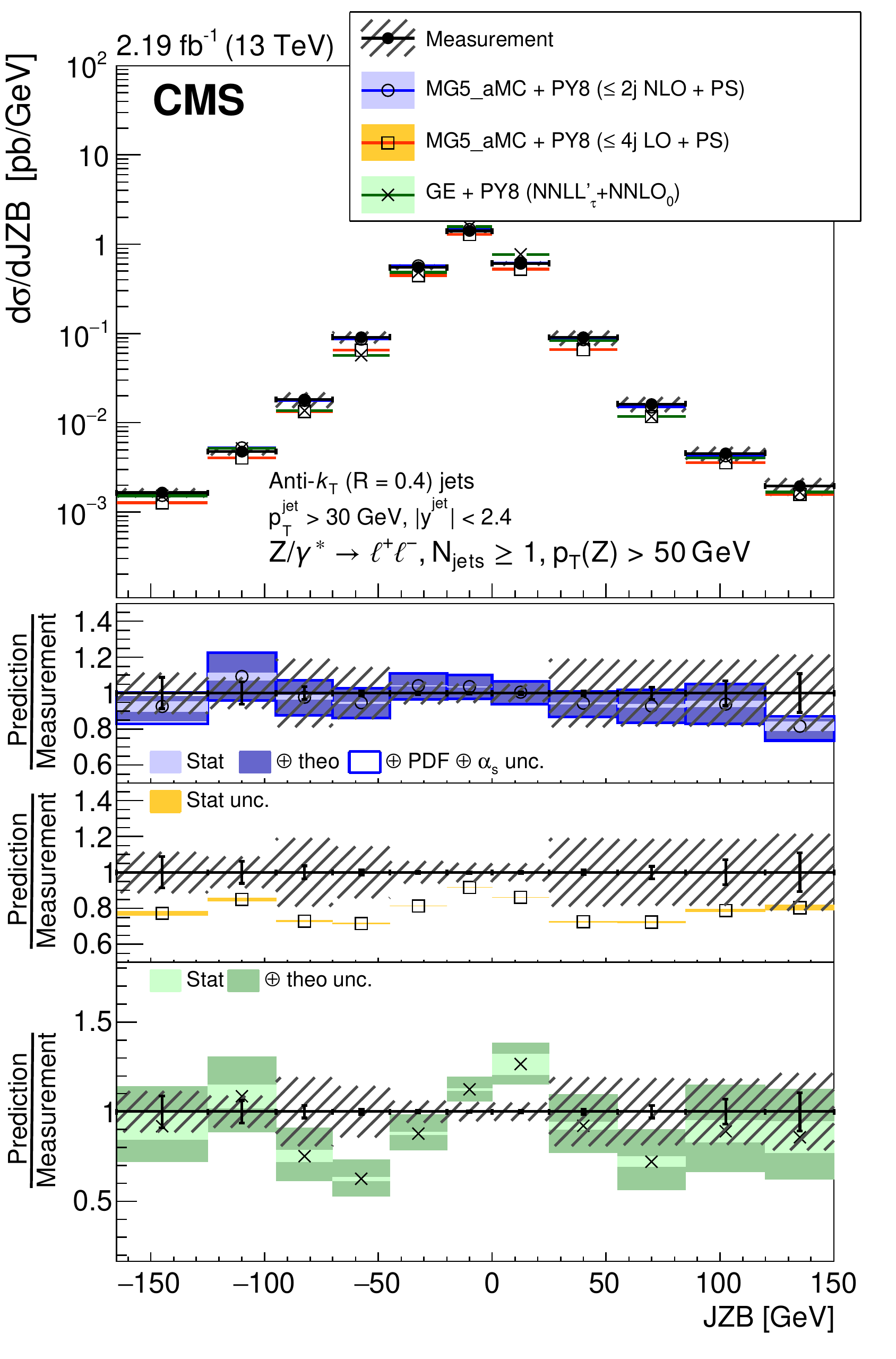}}
  \caption{Measured cross section for $\cPZ+\text{ jets}$ as a function of the \JZB variable (see text), for $\pt(\cPZ)<50\,$GeV (left) and $\pt(\cPZ)>50\,$GeV (right). \citefigfour .}
  \label{fig:JZBmumu_b}
\end{figure*}

The results for the differential cross sections are shown in Figs.~\ref{fig:sigNjet} to~\ref{fig:JZBmumu_b} and are compared to the predictions described in Section~\ref{theory}. For the two predictions obtained from \MGaMC and \PYTHIAeight the number of partons included in the ME calculation and the order of the calculation is indicated by distinctive labels (``${\leq} 4$j LO'' for up to four partons at LO and ``${\leq} 2$j NLO'' for up to two partons at NLO). The prediction of \GENEVA is denoted as ``GE''. The label ``PY8'' indicates that \PYTHIAeight is used in these calculations for the parton showering and the hadronisation. The NNLO $\cPZ + 1 \text{ jet}$ calculation is denoted as \Njetti NNLO in the legends. The measured cross section values along with the uncertainties discussed in Section~\ref{systematics} are given in Tables~\ref{tab:combZNGoodJets_Zexc} to~\ref{tab:combJZB_ptHigh}.

Fig.~\ref{fig:sigNjet} shows the measured cross section as a function of the exclusive (Table~\ref{tab:combZNGoodJets_Zexc}) and the inclusive (Table~\ref{tab:combZNGoodJets_Zinc}) jet multiplicities. Agreement between the measurement and the \MGaMC prediction is observed. The cross section obtained from LO \MGaMC tends to be lower than NLO \MGaMC up to a jet multiplicity of 3. The total cross section for $\cPZ (\to \ell^+\ell^-)+\ge 0 \text{ jet}, m_{\ell^+\ell^-}>50\GeV$ computed at NNLO and used to normalise the cross section of the LO prediction is similar to the NLO cross section as seen in Table~\ref{tab:theory_xsec}. The smaller cross section seen when requiring at least one jet is explained by a steeply falling \pt spectrum of the leading jet in the LO prediction. The \GENEVA prediction describes the measured cross section up to a jet multiplicity of 2, but fails to describe the data for higher jet multiplicities, where one or more jets arise from the parton shower. This effect is not seen in the NLO (LO) \MGaMC predictions, which give a fair description of the data for multiplicities above three (four).

The measured cross section as a function of the transverse momentum of the $\cPZ$ boson for events with at least one jet is presented in Fig.~\ref{fig:sigZPt1j} and Table~\ref{tab:combZPt_Zinc1jet}. The best model for describing the measurement at low \pt, below the peak, is NLO \MGaMC, showing a better agreement than the NNLL$_{\tau}$' calculation from \GENEVA. The shape of the distribution in the region below 10\GeV is better described by \GENEVA than by the other predictions, as shown by the flat ratio plot. This kinematic region is covered by events with extra hadronic activity in addition to the jet required by the event selection. The estimation of the uncertainty in the shape in this region shows that it is dominated by the statistical uncertainty, represented by error bars on the plot since the systematic uncertainties are negligible. In the intermediate region, \GENEVA predicts a steeper rise for the distribution than the other two predictions and than the measurement. The high-\pt region, where \GENEVA and NLO \MGaMC are expected to have similar accuracy (NLO), is equally well described by the two. The LO predictions undershoot the measurement in this region despite the normalisation of the total $\cPZ + \ge 0 \text{ jet}$ cross section to its NNLO value.

The jet transverse momenta for the 1$^{\text{st}}$, 2$^{\text{nd}}$ and~3$^{\text{rd}}$ leading jets can be seen in Figs.~\ref{fig:sigPtjet_a} and~\ref{fig:sigPtjet_b} (Tables~\ref{tab:combFirstJetPt_Zinc1jet}--\ref{tab:combThirdJetPt_Zinc3jet}). The LO \MGaMC predicted spectrum differs from the measurement, showing a steeper slope in the low $\pt$ region. The same feature was observed in the previous measurements~\cite{Chatrchyan:2011ne,Khachatryan:2014zya}. The comparison with NLO \MGaMC and  \Njetti NNLO calculation shows that adding NLO terms cures this discrepancy. The \GENEVA prediction shows good agreement for the measured \pt of the first jet, while it undershoots the data at low \pt for the second jet. The jet rapidities for the first three leading jets have also been measured and the distributions are shown in Figs.~\ref{fig:sigEtajet_a} and~\ref{fig:sigEtajet_b} (Tables~\ref{tab:combFirstJetAbsRapidity_Zinc1jet}--\ref{tab:combThirdJetAbsRapidity_Zinc3jet}). All the predictions are in agreement with data.

The total jet activity has been measured via the \HT variable. The differential cross section as a function of this observable is presented in Figs.~\ref{fig:sigHtjet_a} and~\ref{fig:sigHtjet_b} (Tables~\ref{tab:combJetsHT_Zinc1jet}--\ref{tab:combJetsHT_Zinc3jet}) for inclusive jet multiplicities of 1, 2, and~3. The LO \MGaMC calculation predicts fewer events than found in the data for the region $\HT < 400\GeV$. For higher jet multiplicities both LO and NLO \MGaMC are compatible with the measurement, although the contribution in the region $\HT < 400\GeV$ is smaller for LO than for NLO \MGaMC. 
The contribution at lower values of \HT is slightly overestimated, but the discrepancy is compatible with the theoretical and experimental uncertainties. The \GENEVA generator predicts a steeper spectrum than measured. For jet multiplicities of at least one, we also compare with \Njetti NNLO, and the level of agreement is similar to that found with NLO \MGaMC. The uncertainty for \Njetti NNLO is larger than in the jet transverse momentum distribution because of the contribution from the additional jets.

The balance in transverse momentum between the jets and the $\cPZ$ boson, \ptbal, is shown in Figs.~\ref{fig:ptbalmumu_a} and~\ref{fig:ptbalmumu_b} (Tables~\ref{tab:combVisPt_Zinc1jetQun}--\ref{tab:combVisPt_Zinc3jetQun}) for inclusive jet multiplicities of 1, 2, and 3. When more jets are included, the peak of \ptbal is shifted to larger values. The measurement is in good agreement with NLO \MGaMC predictions. The slopes of the distributions for the first two jet multiplicities predicted by LO \MGaMC do not fully describe the data. This observation indicates that the NLO correction is important for the description of hadronic activity beyond the jet acceptance used in this analysis, $\pt>30\GeV$ and $\abs{y}>2.4$. An imbalance in the event, \ie \ptbal not equal to zero, requires two partons in the final state with one of the two out of the acceptance. Such events are described with NLO accuracy for the NLO \MGaMC sample and LO accuracy for the two other samples. In the case of the \GENEVA simulation, when at least two jets are required, as in the second plot of Fig.~\ref{fig:ptbalmumu_a}, the additional jet must come from parton showering and this leads to an underestimation of the cross section, as in the case of the jet multiplicity distribution. When requiring two jets within the acceptance, the NLO \MGaMC prediction, which has an effective  LO accuracy for this observable, starts to show discrepancies with the measurement. The estimated theoretical uncertainties cover the observed discrepancies.

The \JZB distribution is shown in Figs.~\ref{fig:JZBmumu_a} and~\ref{fig:JZBmumu_b} (Tables~\ref{tab:combJZB}--\ref{tab:combJZB_ptHigh}) for the inclusive one-jet events, in the full phase space, and separately for $\pt(\cPZ)$ below and above 50\GeV. As expected in the high-$\pt(\cPZ)$ region, \ie in the high jet multiplicity sample, the distribution is more symmetric. The NLO \MGaMC prediction provides a better description of the \JZB distribution than \GENEVA and LO \MGaMC. This applies to both configurations, $\JZB<0$ and ${}>0$. This observation indicates that the NLO correction is important for the description of hadronic activity beyond the jet acceptance used in this analysis.

\FloatBarrier

\section{Summary}
\label{summary}

We have measured differential cross sections for the production of a $\cPZ$ boson in association with jets, where the $\cPZ$ boson decays into two charged leptons with $\pt > 20\GeV$ and $\abs{\eta}<2.4$. The data sample corresponds to an integrated luminosity of 2.19\fbinv collected with the CMS detector during the 2015 proton-proton LHC run at a centre-of-mass energy of 13\TeV .

The cross section has been measured as functions of the exclusive and inclusive jet multiplicities up to 6, of the transverse momentum of the $\cPZ$ boson, jet kinematic variables including jet transverse momentum (\pt), the scalar sum of jet transverse momenta (\HT), and the jet rapidity ($y$) for inclusive jet multiplicities of 1, 2, and~3. The balance in transverse momentum between the reconstructed jet recoil and the $\cPZ$ boson has been measured for different jet multiplicities. This balance has also been measured separating events with a recoil smaller and larger than the boson \pt using the \JZB variable. Jets with $\pt>30\GeV$ and $\abs{y} < 2.4$ are used in the definition of the different jet quantities.

The results are compared to the predictions of four different calculations.  The first  two merge matrix elements with different final-state parton multiplicities.  The first is LO for multiplicities up to 4, the second NLO for multiplicities up to 2 and LO for a jet multiplicity of 3, and both are based on \MGaMC. The third is a combination of NNLO calculation with NNLL resummation, based on \GENEVA. The fourth is a fixed order NNLO calculation of one $\cPZ$ boson and one jet.  The first three calculations include parton showering, based on \PYTHIAeight.

The measurements are in good agreement with the results of the NLO multiparton calculation. Even the measurements for events with more than 2 jets agree within the $\approx 10\%$ measurement and 10\% theoretical uncertainties, although this part of the calculation is only LO. The multiparton LO prediction does not agree as well as the NLO multiparton one. It exhibits significant discrepancies with data in jet multiplicity and in both transverse momentum and rapidity distributions of the leading jet.

The transverse momentum balance between the $\cPZ$ boson and the hadronic recoil, which is expected to be sensitive to soft-gluon radiation, has been measured for the first time at the LHC. The multiparton LO prediction fails to describe the measurement, while the multiparton NLO prediction provides a very good description for jet multiplicities computed with NLO accuracy.

Inclusive measurement for events with at least one jet are compared with the NNLO $\cPZ+\ge 1 \text{ jet}$ fixed order calculation. The agreement is good, even for the \HT observable, which is sensitive to events of different jet multiplicities.

The NNLO+NNLL predictions provide similar agreement for the measurements of the kinematic variables of the two leading jets, but fail to describe observables sensitive to extra jets. At low transverse momentum of the $\cPZ$ boson, the NLO multiparton calculation provides a better description than the NNLO+NNLL calculation, whereas both calculations provide a similar description at high transverse momentum.

The results suggest using multiparton NLO predictions for the estimation of the $\cPZ+\text{ jets}$ contribution at the LHC in measurements and searches, and its associated uncertainty.

\begin{acknowledgments}

\hyphenation{Bundes-ministerium Forschungs-gemeinschaft Forschungs-zentren Rachada-pisek} We congratulate our colleagues in the CERN accelerator departments for the excellent performance of the LHC and thank the technical and administrative staffs at CERN and at other CMS institutes for their contributions to the success of the CMS effort. In addition, we gratefully acknowledge the computing centres and personnel of the Worldwide LHC Computing Grid for delivering so effectively the computing infrastructure essential to our analyses. Finally, we acknowledge the enduring support for the construction and operation of the LHC and the CMS detector provided by the following funding agencies: the Austrian Federal Ministry of Science, Research and Economy and the Austrian Science Fund; the Belgian Fonds de la Recherche Scientifique, and Fonds voor Wetenschappelijk Onderzoek; the Brazilian Funding Agencies (CNPq, CAPES, FAPERJ, and FAPESP); the Bulgarian Ministry of Education and Science; CERN; the Chinese Academy of Sciences, Ministry of Science and Technology, and National Natural Science Foundation of China; the Colombian Funding Agency (COLCIENCIAS); the Croatian Ministry of Science, Education and Sport, and the Croatian Science Foundation; the Research Promotion Foundation, Cyprus; the Secretariat for Higher Education, Science, Technology and Innovation, Ecuador; the Ministry of Education and Research, Estonian Research Council via IUT23-4 and IUT23-6 and European Regional Development Fund, Estonia; the Academy of Finland, Finnish Ministry of Education and Culture, and Helsinki Institute of Physics; the Institut National de Physique Nucl\'eaire et de Physique des Particules~/~CNRS, and Commissariat \`a l'\'Energie Atomique et aux \'Energies Alternatives~/~CEA, France; the Bundesministerium f\"ur Bildung und Forschung, Deutsche Forschungsgemeinschaft, and Helmholtz-Gemeinschaft Deutscher Forschungszentren, Germany; the General Secretariat for Research and Technology, Greece; the National Research, Development and Innovation Fund, Hungary; the Department of Atomic Energy and the Department of Science and Technology, India; the Institute for Studies in Theoretical Physics and Mathematics, Iran; the Science Foundation, Ireland; the Istituto Nazionale di Fisica Nucleare, Italy; the Ministry of Science, ICT and Future Planning, and National Research Foundation (NRF), Republic of Korea; the Lithuanian Academy of Sciences; the Ministry of Education, and University of Malaya (Malaysia); the Mexican Funding Agencies (BUAP, CINVESTAV, CONACYT, LNS, SEP, and UASLP-FAI); the Ministry of Business, Innovation and Employment, New Zealand; the Pakistan Atomic Energy Commission; the Ministry of Science and Higher Education and the National Science Centre, Poland; the Funda\c{c}\~ao para a Ci\^encia e a Tecnologia, Portugal; JINR, Dubna; the Ministry of Education and Science of the Russian Federation, the Federal Agency of Atomic Energy of the Russian Federation, Russian Academy of Sciences, the Russian Foundation for Basic Research and the Russian Competitiveness Program of NRNU ``MEPhI"; the Ministry of Education, Science and Technological Development of Serbia; the Secretar\'{\i}a de Estado de Investigaci\'on, Desarrollo e Innovaci\'on, Programa Consolider-Ingenio 2010, Plan Estatal de Investigaci\'on Cient\'{\i}fica y T\'ecnica y de Innovaci\'on 2013-2016, Plan de Ciencia, Tecnolog\'{i}a e Innovaci\'on 2013-2017 del Principado de Asturias and Fondo Europeo de Desarrollo Regional, Spain; the Swiss Funding Agencies (ETH Board, ETH Zurich, PSI, SNF, UniZH, Canton Zurich, and SER); the Ministry of Science and Technology, Taipei; the Thailand Center of Excellence in Physics, the Institute for the Promotion of Teaching Science and Technology of Thailand, Special Task Force for Activating Research and the National Science and Technology Development Agency of Thailand; the Scientific and Technical Research Council of Turkey, and Turkish Atomic Energy Authority; the National Academy of Sciences of Ukraine, and State Fund for Fundamental Researches, Ukraine; the Science and Technology Facilities Council, UK; the US Department of Energy, and the US National Science Foundation.

Individuals have received support from the Marie-Curie programme and the European Research Council and Horizon 2020 Grant, contract No. 675440 (European Union); the Leventis Foundation; the A. P. Sloan Foundation; the Alexander von Humboldt Foundation; the Belgian Federal Science Policy Office; the Fonds pour la Formation \`a la Recherche dans l'Industrie et dans l'Agriculture (FRIA-Belgium); the Agentschap voor Innovatie door Wetenschap en Technologie (IWT-Belgium); the F.R.S.-FNRS and FWO (Belgium) under the ``Excellence of Science - EOS" - be.h project n. 30820817; the Ministry of Education, Youth and Sports (MEYS) of the Czech Republic; the Lend\"ulet (``Momentum") Programme and the J\'anos Bolyai Research Scholarship of the Hungarian Academy of Sciences, the New National Excellence Program \'UNKP, the NKFIA research grants 123842, 123959, 124845, 124850 and 125105 (Hungary); the Council of Scientific and Industrial Research, India; the HOMING PLUS programme of the Foundation for Polish Science, cofinanced from European Union, Regional Development Fund, the Mobility Plus programme of the Ministry of Science and Higher Education, the National Science Center (Poland), contracts Harmonia 2014/14/M/ST2/00428, Opus 2014/13/B/ST2/02543, 2014/15/B/ST2/03998, and 2015/19/B/ST2/02861, Sonata-bis 2012/07/E/ST2/01406; the National Priorities Research Program by Qatar National Research Fund; the Programa de Excelencia Mar\'{i}a de Maeztu and the Programa Severo Ochoa del Principado de Asturias; the Thalis and Aristeia programmes cofinanced by EU-ESF and the Greek NSRF; the Rachadapisek Sompot Fund for Postdoctoral Fellowship, Chulalongkorn University and the Chulalongkorn Academic into Its 2nd Century Project Advancement Project (Thailand); the Welch Foundation, contract C-1845; and the Weston Havens Foundation (USA).
\end{acknowledgments}

\clearpage

\bibliography{auto_generated}
\cleardoublepage \appendix\section{The CMS Collaboration \label{app:collab}}\begin{sloppypar}\hyphenpenalty=5000\widowpenalty=500\clubpenalty=5000\textbf{Yerevan Physics Institute,  Yerevan,  Armenia}\\*[0pt]
A.M.~Sirunyan, A.~Tumasyan
\vskip\cmsinstskip
\textbf{Institut f\"{u}r Hochenergiephysik,  Wien,  Austria}\\*[0pt]
W.~Adam, F.~Ambrogi, E.~Asilar, T.~Bergauer, J.~Brandstetter, E.~Brondolin, M.~Dragicevic, J.~Er\"{o}, A.~Escalante Del Valle, M.~Flechl, M.~Friedl, R.~Fr\"{u}hwirth\cmsAuthorMark{1}, V.M.~Ghete, J.~Hrubec, M.~Jeitler\cmsAuthorMark{1}, N.~Krammer, I.~Kr\"{a}tschmer, D.~Liko, T.~Madlener, I.~Mikulec, N.~Rad, H.~Rohringer, J.~Schieck\cmsAuthorMark{1}, R.~Sch\"{o}fbeck, M.~Spanring, D.~Spitzbart, A.~Taurok, W.~Waltenberger, J.~Wittmann, C.-E.~Wulz\cmsAuthorMark{1}, M.~Zarucki
% ...
\vskip\cmsinstskip
% ...
1:~~Also at Vienna University of Technology, Vienna, Austria\\
% ...

\end{sloppypar}
\end{document}